\documentclass[]{aa}
\usepackage{graphicx}
\usepackage{times}
\usepackage{natbib}

\begin{document}
  \title{Diagnostics of irradiated dense gas in galaxy nuclei} 

  \subtitle{II. A grid of XDR and PDR models}

  \author{R. Meijerink\inst{1} \and M. Spaans \inst{2} \and F.P. Israel \inst{1}}

  \institute{Sterrewacht Leiden, P.O. Box 9513, 2300 RA, Leiden, 
    The Netherlands\\
    \email{meijerin@strw.leidenuniv.nl}
    \and Kapteyn Astronomical Institute, P.O. Box 800, 9700 AV Groningen,
    The Netherlands
  }

  \offprints{R.~Meijerink}

  \date{Received ?????? / Accepted ?????}

  \abstract{The nuclei of active galaxies harbor massive young
  stars, an accreting central black hole, or both. In order to
  determine the physical conditions that pertain to molecular gas
  close to the sources of radiation, numerical models are
  constructed. These models iteratively determine the thermal and
  chemical balance of molecular gas that is exposed to X-rays (1-100
  keV) and far-ultraviolet radiation (6-13.6 eV), as a function of
  depth. We present a grid of XDR and PDR models that span ranges in
  density ($10^2-10^{6.5}$ cm$^{-3}$), irradiation ($10^{0.5}-10^5
  G_0$ and $F_X=1.6\times 10^{-2}-160$~erg~cm$^{-2}$~s$^{-1}$) and
  column density ($3\times 10^{21}-1\times
  10^{25}$~cm$^{-2}$). Predictions are made for the most important
  atomic fine-structure lines, e.g., [CII], [OI], [CI], [SiII], and
  for molecular species like HCO$^+$, HCN, HNC, CS and SiO up to
  $J=4$, CO and $^{13}$CO up to $J=16$, and column densities for CN,
  CH, CH$^+$, HCO, HOC$^+$, NO and N$_2$H$^+$. We find that surface
  temperatures are higher (lower) in PDRs compared to XDRs for
  densities $>10^4$ ($<10^4$) cm$^{-3}$.  For the atomic lines, we
  find that, largely due to the different XDR ionization balance, the
  fine-structure line ratios of [SiII] 35~$\mu$m/[CII] 158~$\mu$m,
  [OI] 63~$\mu$m/[CII] 158~$\mu$m, [FeII] 26~$\mu$m/[CII] 158~$\mu$m
  and [CI] 369~$\mu$m/[CI] 609~$\mu$m are larger in XDRs than in PDRs,
  for a given density, column and irradiation strength.  Similarly,
  for the molecular lines, we find that the line ratios HCN/HCO$^+$
  and HNC/HCN, as well as the column density ratio CN/HCN,
  discriminate between PDRs and XDRs. In particular, the HCN/HCO$^+$
  1-0 ratio is $<1$ ($>1$) for XDRs (PDRs) if the density exceeds
  $10^5$ cm$^{-3}$ and if the column density is larger than
  $10^{23}$~cm$^{-2}$.  For columns less than $10^{22.5}$ cm$^{-2}$
  the XDR HCN/HCO$^+$ 1-0 ratio becomes larger than one, although the
  individual HCN 1-0 and HCO$^+$ 1-0 line intensities are weaker.  For
  modest densities, $n=10^4-10^5$~cm$^{-3}$, and strong radiation
  fields ($>100$~erg~s$^{-1}$~cm$^{-2}$), HCN/HCO$^+$ ratios can
  become larger in XDRs than PDRs as well. Also, the HCN/CO 1-0 ratio
  is typically smaller in XDRs, and the HCN emission in XDRs is
  boosted with respect to CO only for high (column) density gas, with
  columns in excess of $10^{23}$ cm$^{-2}$ and densities larger than
  $10^4$ cm$^{-3}$. Furthermore, CO is typically warmer in XDRs than
  in PDRs, for the same total energy input. This leads to higher CO
  J=N+1-N/CO 1-0, $N\ge 1$, line ratios in XDRs. In particular, lines
  with $N\ge 10$, like CO(16-15) and CO(10-9) observable with
  HIFI/Herschel, discriminate very well between XDRs and PDRs.  This
  is crucial since the XDR/AGN contribution will typically be of a
  much smaller (possibly beam diluted) angular scale and a 10-25$\%$
  PDR contribution can already suppress XDR distinguishing features
  involving HCN/HCO$+$ and HNC/HCN. For possible future observations,
  column density ratios indicate that CH, CH$^+$, NO, HOC$^+$ and HCO
  are good PDR/XDR discriminators.}

  \maketitle
%

\section{Introduction}

The radiation that emanates from galaxy nuclei, such as those of
NGC~253 and NGC~1068 or even more extreme (ultra-)luminous infrared
galaxies, is believed to originate from regions with active star
formation, an accretion disk around a central super-massive black hole
or both (e.g.\ Silk \citeyear{Silk2005}; Maloney
\citeyear{Maloney1999}; Sanders \& Mirabel
\citeyear{Sanders1996}). The unambiguous identification of the central
energy source, or the relative contributions from stars and an active
galaxy nucleus (AGN), remains a major challenge in the study of active
galaxy centers. The general aim of this paper is to determine how the
properties of the irradiated interstellar medium (ISM) may further our
understanding.

Unlike emission at optical wavelengths, atomic, molecular and dust
emission in the far-infrared and (sub-)mm range allow one to probe
deeply into the large column densities of gas and dust that occupy the
centers of active galaxies. Observational studies of the ISM in galaxy
centers have been presented by various authors (c.f. Aalto
\citeyear{Aalto2005}; Baan \citeyear{Baan2005}; Ott et
al. \citeyear{Ott2005}; Israel \citeyear{Israel2005}; Spoon et
al. \citeyear{Spoon2001, Spoon2003, Spoon2005}; Kl\"ockner et
al. \citeyear{Klockner2003}; Israel and Baas \citeyear{Israel2002};
Garrett et al. \citeyear{Garrett2001}; H\"uttemeister and Aalto
\citeyear{Huttemeister2001}; Curran et al. \citeyear{Curran2000}).
Theoretical models show that the spectral energy distribution of the
radiation representing star formation (peaking in the ultraviolet) and
AGN (peaking in the X-ray regime) activity respectively influences the
thermal and chemical balance of the ambient ISM in fundamentally
different ways (Meijerink \& Spaans \citeyear{Meijerink2005}; Maloney,
Hollenbach \& Tielens \citeyear{Maloney1996}; Lepp \& Dalgarno
\citeyear{Lepp1996}). The specific aim of this paper is thus to study
the extent to which emission from commonly observed molecular and
atomic line transitions may be used as a diagnostic tool in the study
of external galaxy centers to determine the ambient conditions in
general, and the type of irradiation in particular.

To this effect, we have extended the chemical calculations described
by Meijerink \& Spaans (2005; hereafter Paper I) for ultraviolet and
X-ray irradiated gas to a much larger parameter space of ambient
conditions and we have performed detailed radiative transfer
calculations to compute the line intensities of many atomic and
molecular transitions. We refer the interested reader to Paper I for a
detailed description of the combined photon-dominated region
(PDR)/X-ray dominated region (XDR) code that we have used to compute
the impact of ultraviolet (PDR) and X-ray (XDR) photons on nearby ISM.
The results described here will be applied to observations of external
galaxy centers in a subsequent paper (Meijerink et al., in
preparation). \citet{Stauber2005} also developed PDR and XDR codes,
for applications to Young Stellar Objects. The main difference is that
this code has a time dependent chemistry and that it includes ice
evaporation at $t=0$. This affects the chemistry compared with
traditional steady-state models of pure gas-phase PDRs (e.g.,
CN/HCN). This is not relevant for our cloud models. However,
freeze-out and evaporation start to become important for clouds with
density $n_{\rm H}\ge 10^5$~cm$^{-3}$.

\section{'Standard' clouds}

\begin{table*}[!ht] 
\centering
\caption[c]{'Standard' clouds}
\begin{tabular}{llllll} 
\hline 
\noalign{\smallskip} 
Type & $r$(pc)	 & $n({\rm cm}^{-3})$ & $N({\rm cm}^{-2})$         & $G_0$        &  $F_X$ [erg~s$^{-1}$~cm$^{-2}$]  \\
\noalign{\smallskip}
\hline
\noalign{\smallskip}
A  &  1    & $10^4 - 10^{6.5}$  & $3 \times 10^{22}$ - $1 \times 10^{25}$   & $10^2$ - $10^5$ & $1.6$ - $160$      \\
B  &  10   & $10^3 - 10^4$      & $3 \times 10^{22}$ - $3 \times 10^{23}$   & $10^1$ - $10^4$     & $1.6\times10^{-1}$ - $16$ \\
C  &  10   & $10^2 - 10^3$      & $3 \times 10^{21}$ - $3 \times 10^{22}$   & $10^{0.5}$ - $10^3$     & $1.6\times10^{-2}$ - $1.6$   \\
\noalign{\smallskip}
\hline
\end{tabular}
\label{standard_clouds}
\end{table*}

The current spatial resolution of sub-millimeter telescopes, such as
the James Clerk Maxwell Telescope (JCMT), the Institute de Radio
Astronomie Millim\'etrique (IRAM) telescope and even the Combined
Array for Research in Submillimeter Astronomy (CARMA) or the
Submillimeter Array (SMA), is insufficient to resolve individual
clouds in extragalactic sources. By using these telescopes, each
resolution element thus measures the combined emission from a large
ensemble of molecular clouds. As a consequence, it is frequently
impossible to use a single model cloud solution to describe the
observed molecular lines.  Instead, more complicated solutions
involving two or more model clouds, with differing densities and
incident radiation fields, are needed. This contrasts with the study
of Galactic objects, where usually a single model cloud solution is
sufficient to fit the measurements of single resolution elements.

In this paper, we calculate a grid of 'standard' clouds sampling the
different physical conditions believed to be relevant for the centers
of active galaxy nuclei. In order to sample both the hierarchical size
and (column) density properties of the ISM, we have chosen to
construct models for a number of fixed sizes as well as
densities. Note that the column densities are not the same for each
model, since we use fixed cloud sizes. In these clouds, we investigate
the detailed column density dependence for the line ratios HCN/CO,
HNC/HCN, HCO$^+$/HCN, SiO/CO and CS/HCN, which include line ratios
observed in several galaxies. From our computational grid, one or
more, properly weighted, models can be chosen to reproduce observed
atomic and molecular lines.

We distinguish three different 'standard' clouds, each with their own
characteristic combinations of size and volume density range, hence
also column density (cf. Table \ref{standard_clouds}), for which we
calculate a set of models for different incident radiation fields, and
where a distinction between irradiation by far-ultraviolet (FUV) and
X-ray photons is made. The X-ray radiation field is a power-law
distribution $F(E)=F(0)(E/1 {\rm keV})^{-\alpha}$ integrated between 1
and 100 keV, where $\alpha=-0.9$. A low energy cut-off of 1 keV is
chosen, since most of the emitted radiation below 1~keV is absorbed in
the hot, highly ionized medium close to the black hole.  This power
law spectrum is generally believed to be representative for accreting
black holes. Please note that this differs from Paper I, where we used
a thermal distribution at $10^6$ K instead. The ultraviolet radiation
field (6-13.6 eV) is expressed in multiples of the Habing flux, where
$G_0=1$ corresponds to $1.6\times 10^{-3}$ erg cm$^{-2}$ s$^{-1}$,
which is the local Galactic interstellar radiation field. We express
the X-ray flux in erg~s$^{-1}$~cm$^{-2}$. An accreting black hole with
an X-ray luminosity of $10^{44}$~erg~s$^{-1}$, produces an X-ray flux
of $~100$~erg$^{-1}$~cm$^{-2}$ at a distance of 100~pc, when there is
not extinction. We use a line width $\delta_v=2.7$~km~s$^{-1}$. Cloud
type A represents compact, high-density environments such as molecular
cloud cores, and clouds very close to active nuclei; cloud type B
corresponds to more traditional molecular cloud environments, and
cloud type C is representative of the more diffuse extended
(molecular) medium in which clouds of type B are usually embedded.

Late-type galaxies frequently have radial metallicity gradients, with
the highest metallicity in the center \citep{VilaCostas1992,
Zaritsky1994}. For this reason, we have adjusted the metallicity used
in Table 2 of Paper I. The published metallicity gradients and
suggestions of a gradient flattening in the very center have led us to
adopt a twice Solar metallicity as a reasonable value. Since [C]/[O]
abundance ratios decrease at higher metallicities, we have taken the
carbon abundance equal to the oxygen abundance (see, for instance,
Garnett et al. \citeyear{Garnett1999}; Kobulnicky \& Skillman
\citeyear{Kobulnicky1998}). Note that the [C]/[O] ratio affects the
abundances of O$_2$ and H$_2$O. See for example \citet{Spaans2001} and
especially Fig.\ 2 in \citet{Bergin2000}.

From the models, we have calculated the intensities of the molecular
rotational lines of HCN, HNC, HCO$^+$, CS and SiO (upto $J=4$), [CI],
[CII], [OI], [SiII] and other fine-structure lines. For CO and
$^{13}$CO we calculated the intensities of the rotational lines up to
$J=16$, in order to make predictions for future observing facilities
such as the ESO Herschel/HIFI space mission. We use the Leiden Atomic
and Molecular Database (LAMDA) as described in \citet{Schoier2005} to
retrieve the collisional data needed for the calculations. Where no
collisional data are available for commonly observed molecules such as
CN, CH, CH$^+$, HCO, HOC$^+$, NO and N$_2$H$^+$, we only give column
densities.

\begin{figure}[!hb]
\unitlength1cm
\begin{minipage}[b]{8.8cm}
\resizebox{8.8cm}{!}{\includegraphics*[angle=0]{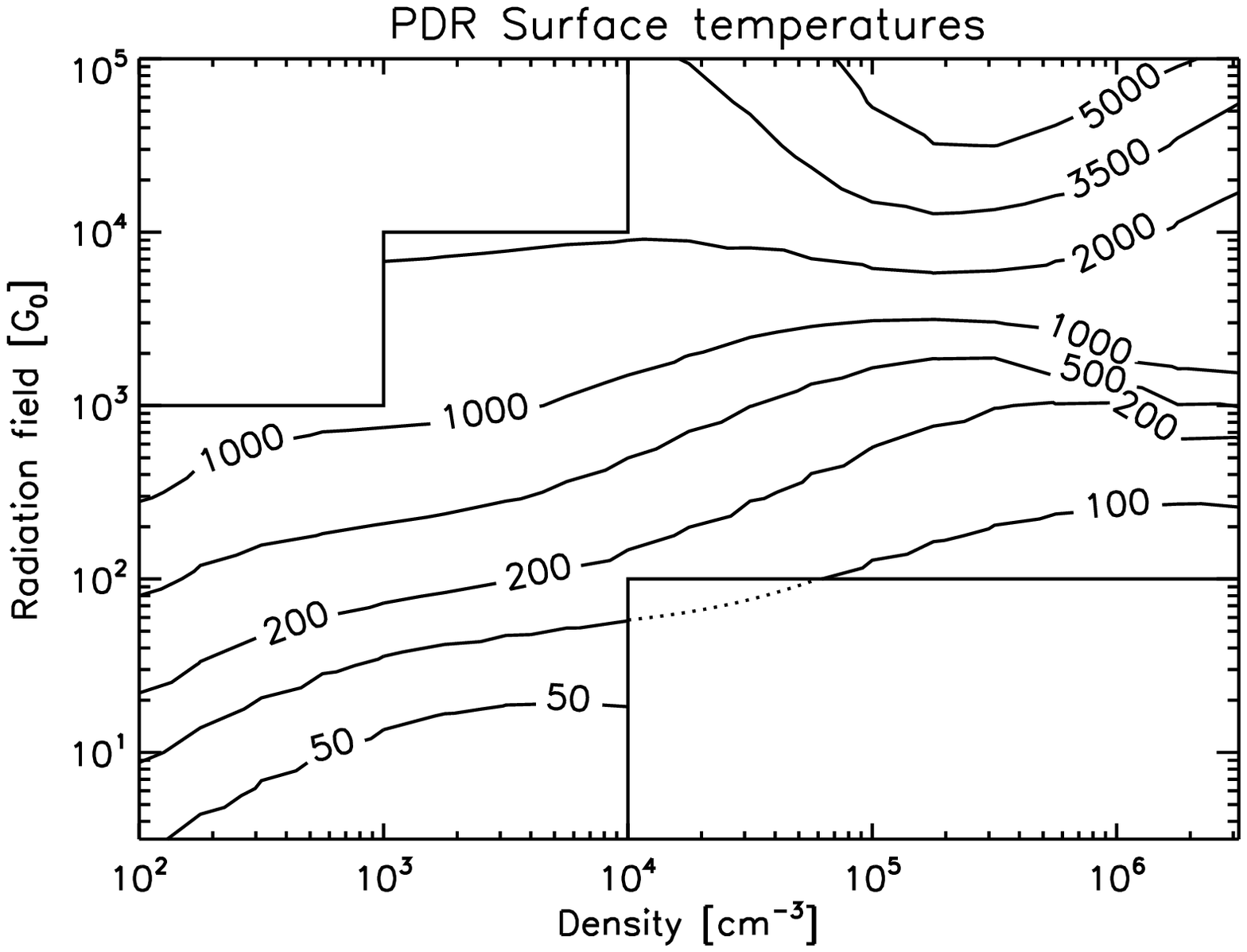}}
\end{minipage}
\begin{minipage}[b]{8.8cm}
\resizebox{8.8cm}{!}{\includegraphics*[angle=0]{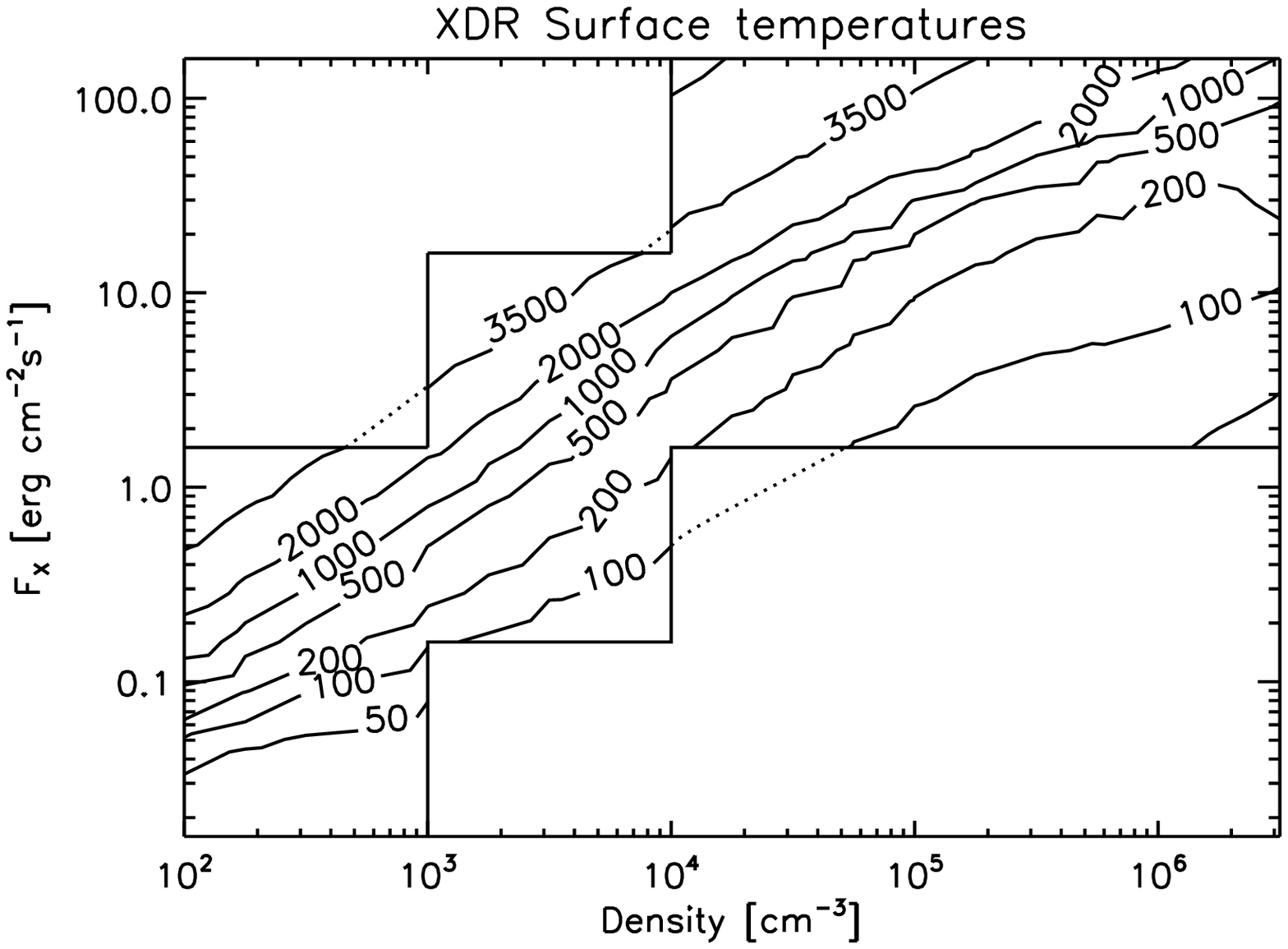}}
\end{minipage}
\caption[] {Surface temperatures for PDR (top) and XDR (bottom) models.}
\label{surface_temperatures}
\end{figure}

\section{Surface temperatures}

To illustrate the coupling differences of FUV and X-ray photons to the
gas, we first discuss the surface temperatures of the low, mid, and
high density models respectively over the parameter space given for
the PDR and XDR models in Table \ref{standard_clouds}. We calculated a
larger range of radiation fields for the PDR models than for the XDR
models, for reasons related to the heating efficiency which are
discussed below. The resulting surface temperatures as a function of
gas density and incident radiation intensity for both PDR and XDR
models are shown in Fig. \ref{surface_temperatures}.

The most important heating mechanism at the edge of a PDR is
photo-electric heating. FUV photons are absorbed by dust grains, which
release electrons that lose their surplus kinetic energy to the gas by
Coulomb interactions. The efficiency of this process increases, when
the grains are more negatively charged, which is determined by a
complex interplay of the impinging radiation field $G_0$, electron
density $n_e$ and gas temperature $T$. The absolute efficiency is very
low, since about 0.5-3 percent of the photon energy goes into gas
heating. This sharply contrasts with direct X-ray heating, important
at the edge of XDRs. Direct ionization of an atom yields a kinetic
electron with an energy higher than 1~keV. This electron heats,
ionizes and excites the gas. Depending on the H, H$_2$, He and
electron abundances, the heating efficiency can be up to 70 percent,
much higher than for photo-electric heating. However, there is an
opposing effect, namely the much lower absorption cross section for
X-rays. Since the cross sections scale roughly as $E^{-3}$, there are
many fewer X-ray photons absorbed than FUV photons.

We find that at high densities ($n > 10^4$~cm$^{-3}$) PDR models
produce higher surface temperatures than the XDR models. At low
densities, however, we find the opposite, especially in the case of
high radiation fields. This is explained by the drop in the efficiency
of photo-electric heating due to grain charging at densities $n <
10^5$~cm$^{-3}$ when the same impinging radiation field is
considered. This is nicely illustrated in Fig. 6 of
\citet{Kaufman1999}, where the ratios of the intensity of the
[CII]~158~$\mu$m and [OI]~63~$\mu$m lines to the total far-infrared
intensity emitted from the surface of the clouds are plotted as a
function of density and radiation field. This ratio is a measure of
the heating efficiency, since [CII]~158~$\mu$m and [OI]~63~$\mu$m are
the most important coolants in PDRs.

In the regime discussed here, the surface temperatures of the XDR
models are quite well correlated with $H_X/n$, where $H_X$ is the
energy deposition per particle and $n$ the total hydrogen density.
This means that the highest surface temperatures are found at the
lowest densities and highest impinging radiation fields.
Consequently, the contours of equal surface temperature are almost
straight lines in the XDR plot of Fig. \ref{surface_temperatures}. In
the PDR models their behavior is more complicated, as already
discussed by \citet{Kaufman1999}, since grains are involved in heating
the gas. At the edge of the cloud, the cooling is dominated by [CII]
158~$\mu$m and [OI] 63~$\mu$m, which have critical densities of
$n_{cr}({\rm CII})\approx3\times10^3$~cm$^{-3}$ and $n_{cr}({\rm
OI})\approx5\times10^5$~cm$^{-3}$ for collisions with atomic and
molecular hydrogen. The cooling rate in this regime is more or less
proportional to $n^2$. The heating rate is at least proportional to
$n$, because the grain density is proportional to $n$.  It can be
larger, because grains become less positively charged at increasing
electron densities making it easier for electrons to escape the
grains, so that the heating efficiency increases. Nevertheless, the
density dependence of heating remains less steep than $n^2$, i.e.,
less than that of cooling, which causes a drop in the temperature at
$n < 10^4$~cm$^{-3}$ and for fixed $G_0$ with increasing density.
Between $n\approx1-3\times10^3$~cm$^{-3}$ where the [CII]~158~$\mu$m
line thermalizes, the drop in temperature stagnates. For densities
between $n=10^4-10^5$~cm$^{-3}$ and $G_0 < 10^4$, we also find that
the surface temperature drops with increasing $n$. The [OI] 63$\mu$m
line thermalizes at higher densities $n=10^5-10^6$~cm$^{-3}$ and,
therefore, cooling will be proportional to $n$. Heating increases
faster with density, which results in a higher surface temperature.
For $G_0>10^4$, we find that the surface temperature rises up to
$n\approx5\times10^5$~cm$^{-3}$ due to the increase in the heating
efficiency with density at fixed $G_0$. Above this density and at
these high temperatures, coolants with high critical densities and
excitation energies such as the [OI] 6300\AA\ line become important,
causing the surface temperature to drop again.

When we compare the surface temperatures with those derived by
\citet{Kaufman1999}, we find that our model surface temperatures are
higher. A possible explanation for this has been discussed recently by
\citet{Roellig2006}, who present scaling relations for heating and
cooling as a function of metallicity $Z$. They state that the
photo-electric heating rate is $\propto Z$ for $n<10^3$~cm$^{-3}$
increasing to $Z^2$ when $n>10^6$~cm$^{-3}$. The cooling rate is
always proportional to $Z$, and, therefore, higher metallicities
result in higher surface temperatures. Note however that the
temperature differences found are very likely not only because of a
change in metallicity. In the PDR comparison test
\citep{Roellig2006a}, we found significant scatter between different
PDR codes in the thermal balance. Therefore, one should not take the
absolute values of line intensities too literally in the
interpretation of data.

\begin{figure*}
\centerline{\includegraphics[height=150mm,clip=]{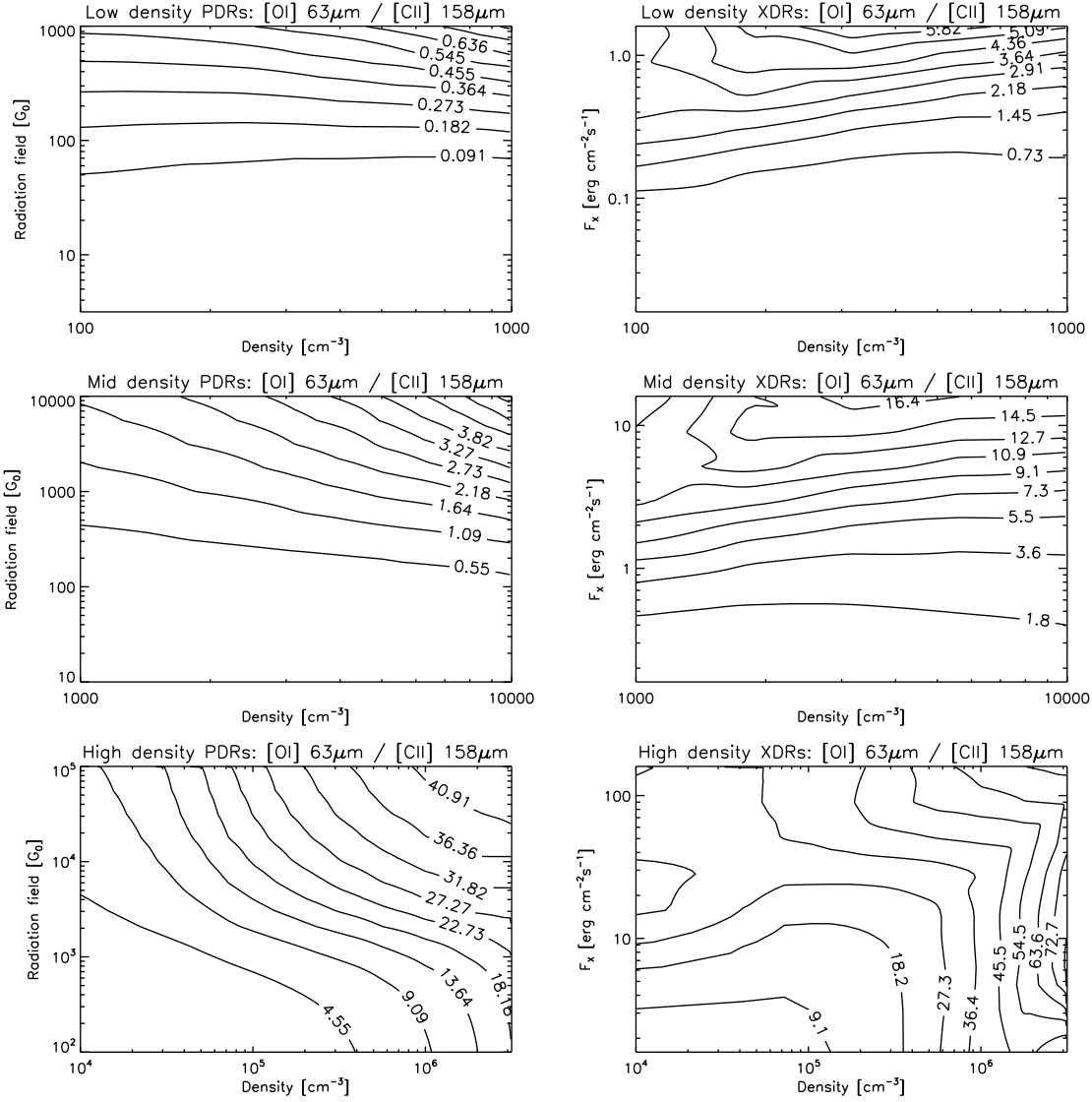}}
\caption{[OI] 63~$\mu$m / [CII] 158~$\mu$m ratio for PDR (left) and
XDR (right) models.}
\label{fine_ratio_OI_CII}
\end{figure*}

\begin{figure*}
\centerline{\includegraphics[height=150mm,clip=]{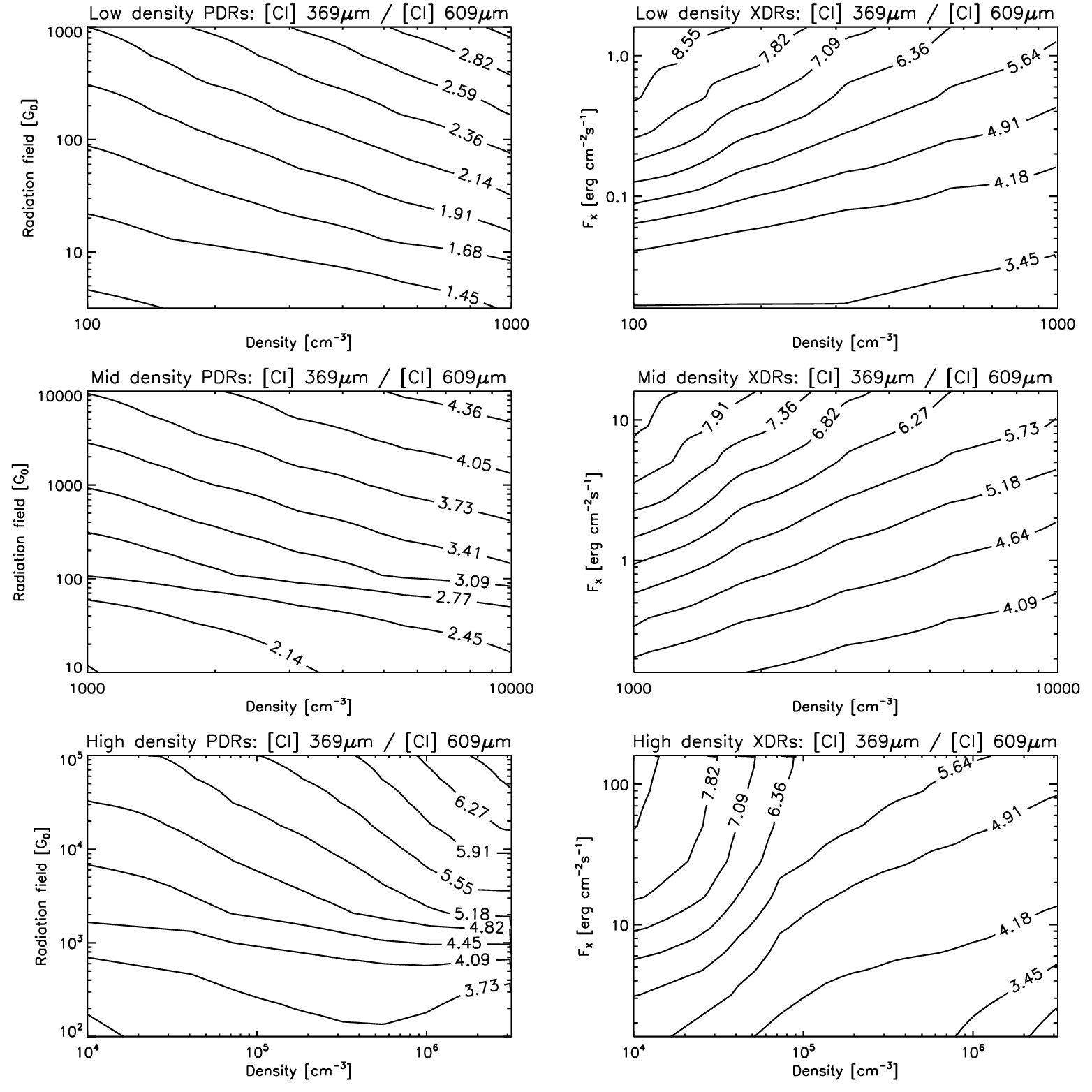}}
\caption{[CI] 369~$\mu$m / [CI] 609~$\mu$m ratio for PDR (left) and
XDR (right) models.}
\label{fine_ratio_CI_CI}
\end{figure*}

\section{Fine-structure lines}

In principle, we can use combinations of fine-structure line
intensities to constrain densities and incident radiation fields.
Here we discuss how such line ratios depend on gas density, ambient
radiation field strength, and cloud column density, and compare the
results for the PDR and XDR models. In previous studies only results
for PDRs were shown. \citet{Wolfire1990} were the first to show
[CII]/[SiII] contour plots for PDRs. Recently, \citet{Kaufman2006}
showed updated plots for [SiII] and [FeII] for PDR gas. The
fine-structure lines are selfconsistently calculated in the PDR and
XDR codes, and we take line-trapping into account for ALL lines.

\subsection{[SiII] 35 $\mu$m/[CII] 158 $\mu$m intensity ratio}

On the website, we show the [SiII] 34.8~$\mu$m / [CII] 158~$\mu$m
fine-structure line ratio for both the PDR and the XDR models. The
[SiII] 34.8~$\mu$m line has an energy of $E/k=414$~K and a critical
density of $n_{cr}=3.4\times10^5$~cm$^{-3}$, while this is $E/k=92$~K
and $n_{cr}=2.8\times10^3$~cm$^{-3}$ for the [CII] 158~$\mu$m. Very
high radiation fields produce gas temperatures in the PDR models that
are sufficiently high to excite the [SiII] and [CII] upper states and
the ratio depends mostly on density. With lower radiation fields, the
surface temperature drops. The upper level energy of [SiII]
34.8~$\mu$m is reached first and the [SiII]/[CII] line ratio drops.
As the density increases at a given FUV radiation field, the ratios
limit to roughly constant values that are set by the corresponding
surface temperatures.

In the XDR models, the ratio is not only determined by the temperature
and density, but also by the fractional abundances and column
densities. In PDRs clearly defined layers occur, in which carbon and
silicon are both almost fully ionized, but throughout XDRs neutral and
ionized species co-exist. Despite the fact that surface temperatures
in the XDR models are lower, we find much higher [SiII]/[CII] ratios,
since silicon is much easier to ionize than carbon. The dominant
source for ionization is not the direct absorption of an X-ray photon
(primary ionization), but the produced kinetic electron. This electron
can ionize a species either directly by collisions (secondary
ionizations) or indirectly by first exciting H and H$_2$ and producing
Lyman $\alpha$ and Lyman-Werner photons, which then may ionize species
in turn. The cross section of Si for secondary ionization is about
twice that of C. This does not, however, fully explain the order of
magnitude difference between the calculated PDR and XDR ratios. This
difference also reflects the fact that ionization of C can only be
done by Lyman-Werner photons, whereas both Lyman $\alpha$ and
Lyman-Werner photons are capable of ionizing Si. It is thus harder to
ionize C than Si in regions where the gas is mostly atomic, which
results in a large increase of the ratio for all densities at a given
irradiation strength.

This also explains the fairly constant ratio below
$F_X\approx10$~erg~s$^{-1}$~cm$^{-2}$ for $n\sim 2\times 10^3-3\times
10^6$ cm$^{-3}$, which results from an interplay between various
effects. With ambient radiation fields constant, we find that the
surface temperatures drop with increasing density in the XDR. At lower
temperatures and higher densities, H$_2$ is more easily formed. Both
the temperature drop itself and the enhanced H$_2$ (leading to more
carbon ionizations) thus suppresses the [SiII]/[CII] ratio, but this
is compensated by the relatively high critical density of the [SiII]
34.8~$\mu$m line.

For the highest, $>16$~erg~s$^{-1}$~cm$^{-2}$, radiation fields, where
carbon is highly ionized, we find the same trend as seen for the PDR,
i.e., the ratio is mostly dependent on density for the highest
radiation fields, and then there is a decrease in the ratio when the
temperature drops below the upper-state energy of [SiII].

At the lowest densities and highest X-ray radiation fields, we find
that the effect of column density become important as well. Since we
fixed cloud sizes, in each cloud type the lowest density models imply
also the lowest column densities. In the high irradiation models,
carbon is almost fully ionized at the XDR edge. The fractional
abundance of ionized carbon drops toward the H/H$_2$ transition and
then increases again for the reasons discussed earlier. In the lowest
(column) density models, we only produce the highly ionized part,
suppressing the ratio even more. This is best seen in the diagram for
the high density XDR models, at densities $n=10^4$~cm$^{-3}$. The
ratio increases from $F_X=1.6$ to $30$~erg~s$^{-1}$~cm$^{-2}$ and then
drops again. Hence, if we increase the column densities of these
models by a factor of $\sim 30$, the ratios only depend on radiation
field strength and not on density.


\subsection{[OI] 63 $\mu$m/[CII] 158 $\mu$m intensity ratio}

In Fig. \ref{fine_ratio_OI_CII}, we show the [OI] 63~$\mu$m / [CII]
158~$\mu$m ratio. The critical density is
$n_{cr}=5\times10^5$~cm$^{-3}$ and the upper-state energy is
$E/k=228$~K for [OI] 63~$\mu$m. In the PDR models and at very high
incident radiation fields, this ratio depends mostly on density as in
the case for the [SiII]/[CII] ratio, since once again temperatures are
sufficiently high to excite both upper-state levels. When the surface
temperature drops for lower radiation fields, the upper-state energy
of [OI] is reached first. The [OI]/[CII] ratio then drops, with a flat
density dependence for $n<10^4$ cm$^{-3}$ and $G_0<10^3$. The decrease
in this line ratio is, however, less pronounced than that in the
[SiII]/[CII] ratio, because the upper-state energy of [OI] is lower
than that of [SiII]. When $G_0<10^3$ and $n^>10^{5.5}$ cm$^{-3}$, the
ratio is roughly constant at the same density for all values of $G_0$.
Here, the more difficult excitation of [OI] at lower temperatures is
counteracted by the C$^+$ layer becoming thinner at lower radiation
fields.

In the XDR, the [OI]/[CII] ratio is again more complex due to its
dependence on density, radiation field, ionized carbon fraction and
total column density. The ratios are overall much higher than in the
PDR models, since carbon does not become fully ionized. At the largest
$H_X/n$, temperatures are high enough to create a dependence on
density only. At lower radiation fields, the [OI]/[CII] ratio drops as
temperatures approach the upper-state energy of the [OI] 63 $\mu$m
line. When we decrease the radiation field even more, we find that the
C$^+$ fraction is rapidly reduced at high ($>10^5$ cm$^{-3}$)
densities and the ratio increases again. At high
($>16$~erg~s$^{-1}$~cm$^{-2}$) radiation field strength and low
densities, a large $H_X/n$ is maintained throughout the whole cloud,
since the relevant type A cloud size is only one parsec. For that
reason, we find here the same effect as already seen in the
[SiII]/[CII] ratio, since column densities increase toward higher
densities. At the highest radiation fields and lowest densities, the
ratio is suppressed since carbon remains partially ionized over the
full extent of the clouds considered here. We do not find the highest
ratio at the highest radiation field in the lower density models.

\subsection{[FeII] 26 $\mu$m/[CII] 158 $\mu$m intensity ratio}

On the website, we show the [FeII] 26~$\mu$m / [CII] 158~$\mu$m
intensity ratio. [FeII] 26~$\mu$m is very difficult to excite due to
its high critical density $n_{cr}=2.2\times10^6$~cm$^{-3}$ and
upper-state energy $E/k=554$~K.  Thus, the change in ratio with
increasing density at high incident radiation fields is much larger
than that seen for the [SiII] 34.8~$\mu$m / [CII] 158~$\mu$m ratio. In
the PDR models and at high incident radiation fields, the ratio mostly
depends on the density. At lower radiation field strengths,
approaching the upper-state energy of [FeII], the ratio drops.

The same trends are seen at high radiation fields in the XDR models,
but again, we find much higher ratios than in the PDR models. It is
possible to ionize iron with Lyman~$\alpha$ photons, but not
carbon. At moderate radiation fields, we again find ratios to be more
or less independent of density, for the reason that we have already
discussed in the [SiII]/[CII] case. At the lowest densities and
highest radiation fields in each cloud type, we find that the ratios
are similarly suppressed as was the case for [SiII] 35 $\mu$m/[CII]
158 $\mu$m and [OI] 63 $\mu$m/[CII] 158 $\mu$m.

\subsection{[CI] 369 $\mu$m/[CI] 609 $\mu$m intensity ratio}

In Fig. \ref{fine_ratio_CI_CI}, we show the [CI] 369~$\mu$m / [CI]
609~$\mu$m intensity ratio. The critical densities of these lines are
$n_{cr}=3\times10^2$~cm$^{-2}$ for [CI] 609$\mu$m and
$n_{cr}=2\times10^3$~cm$^{-2}$ for [CI] 369~$\mu$m, typically lower
than the densities we are considering here. The upper-state energies
are $E/k=24$ and $63$~K for [CI] 609$\mu$m and [CI] 369~$\mu$m,
respectively. In the PDR models, the [CI] lines originate from the
C$^+$/C/CO transition layer. The temperatures in this layer slightly
rise with increasing incident radiation field strengths and range
between $T=20-100$~K (comparable to the upper-state energies). This
explains the small increase in the ratio for larger $G_0$ at the same
density. The ratio does not change much as a function of density,
since we are above the critical density. The change we do see,
however, has a temperature dependence. When densities are lowered,
recombination rates are lower as well. By consequence, at higher
densities, the transition layer is closer to the edge of the cloud and
at higher temperatures, which raises the ratio.

In the XDR, neutral carbon occurs throughout the cloud, and is also
abundant at relatively high temperatures. The spread in temperatures
is large, which was already seen in Sect. \,3, and this determines for
a large part differences in the ratios. The temperature of a cloud is
determined by $H_X/n$, resulting in the highest ratios for low $n$ and
high $F_X$, opposite to the situation in the PDR models. The three
cloud types, with their different low, mid and high density ranges
nevertheless show very similar spreads in ratios. This is caused by
the difference in column densities, which also has a very important
effect. The low, mid, and high density models have their own fixed
cloud size, and in each standard cloud type, column densities increase
toward higher densities in the same density range. The higher density
models contain larger regions of low temperature, which suppresses the
ratio at these densities even more. This can also be understood by
considering the ratio at $n=10^4$~cm$^{-3}$ in the mid (type B) and
high (type C) density range. The high density model at
$n=10^4$~cm$^{-3}$ has a smaller cloud size and therefore a higher
line ratio.

\section{Rotational lines}

Molecular rotational lines are also characteristic for the physical
condition of ISM gas and may also be used to constrain gas densities
and incident radiation fields. In the following, we discuss a number
of ratios, involving the molecular species $^{12}$CO, $^{13}$CO, HCN,
HNC, HCO$^+$, SiO and CS. Although we reproduce in this Paper only a
limited number of the diagrams showing the calculated line intensity
ratios, all model data are available on-line
\footnote{\scriptsize \tt
http://www.strw.leidenuniv.nl/$\sim$meijerin/grid/}. Hence, the reader
can determine all line ratios and integrate over all possible lines of
sight as interest dictates. Here we concentrate on molecular lines
that we consider particularly useful to distinguish between PDRs and
XDRs. We have calculated the line intensities by using a
one-dimensional version of the radiation transfer code described in
\citet{Poelman2005,Poelman2006}.

\begin{figure*}[!ht]
\centerline{\includegraphics[height=150mm,clip=]{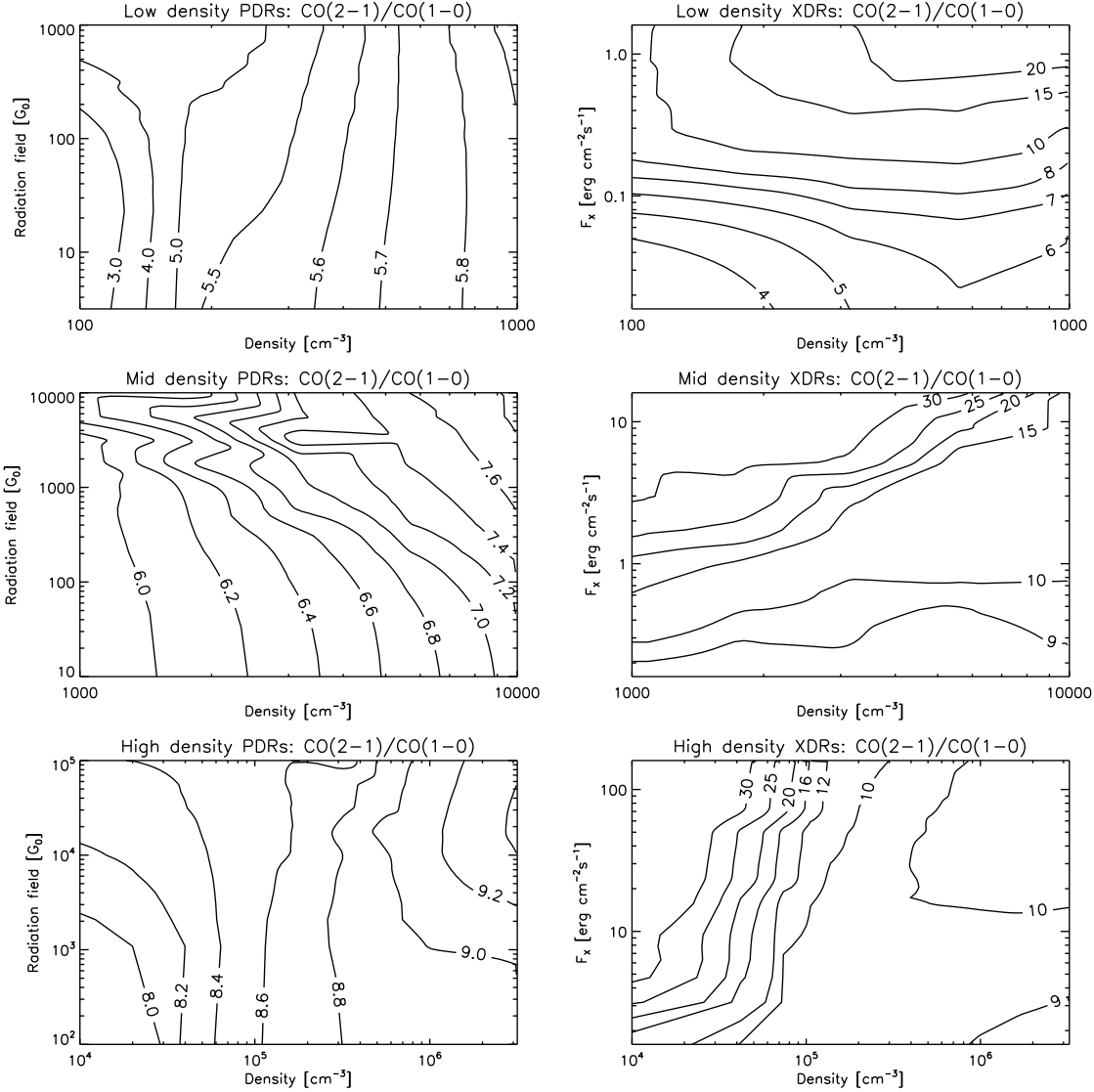}}
\caption{CO(2-1)/CO(1-0) ratio for PDR (left) and XDR (right) models.}
\label{ratio_CO21_CO10}
\end{figure*}

\begin{figure*}[!ht]
\centerline{\includegraphics[height=150mm,clip=]{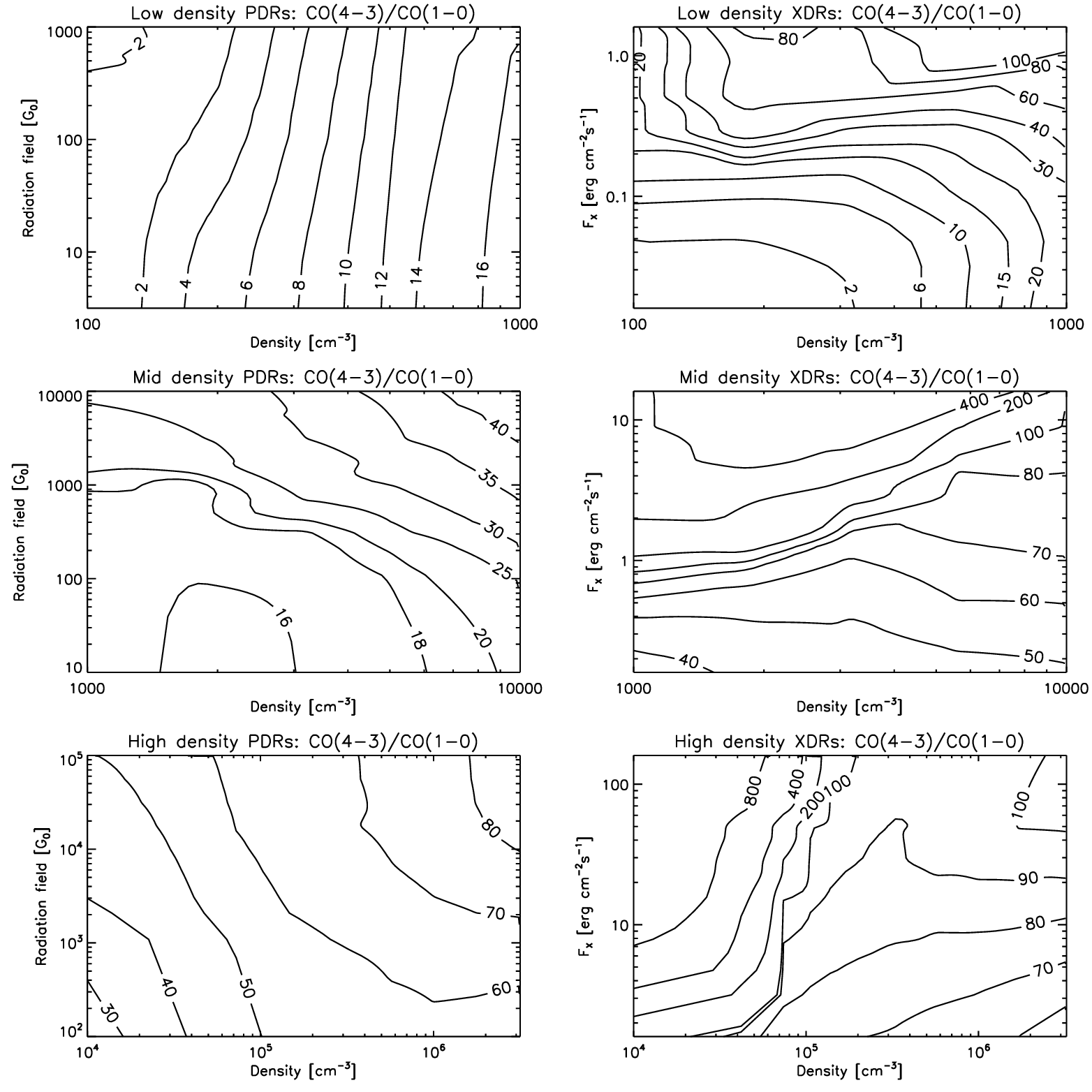}}
\caption{CO(4-3)/CO(1-0) ratio for PDR (left) and XDR (right) models.}
\label{ratio_CO43_CO10}
\end{figure*}

\subsection{CO rotational lines}

In Fig. \ref{intensity_CO10}, we show the CO(1-0) line intensity for
cloud types A (high-density), B (mid-density) and C (low-density) for
both the PDR and XDR model cases (cf. Table \ref{standard_clouds}. All
three cloud types are relevant for CO (and $^{13}$CO) since these
molecules are present ubiquitously on all galactic scales. The CO(1-0)
line has an upper state energy $E/k=5.53$~K and a critical density
$n_{cr}\sim3\times10^3$~cm$^{-3}$. In the low density PDR models
($n=10^2-10^3$~cm$^{-3}$), we find that the intensity increases with
density only. In the mid ($n=10^3-10^4$~cm$^{-3}$) and high
($n=10^4-10^{6.5}$~cm$^{-3}$) density range, we also find a small line
intensity increase at higher incident radiation fields. The emission
of the CO(1-0) line is optically thick and the emitted line intensity
mostly depends on the gas temperature near the $\tau=1$
surface. Although the upper-state energy is below the gas temperature,
we find a significant increase in the line intensity. At higher
densities, gas-grain interactions may significantly increase gas
temperatures in the highly attenuated part of the cloud. Since the
escape probability $\beta(\tau)\propto1/\tau$ when $\tau>1$, these
parts also add a small contribution to the line intensity, which is
proportional to $\sim\ln(\tau)$. For an elaborate discussion of the
CO(1-0) line intensity dependence on $G_0$ and $n$ see
\citet{Wolfire1989}. In general, the line intensities do not vary much
in the PDR models, as opposed to what is seen in the XDR models. In
all PDR models, the CO(1-0) lines are optically thick. In the XDR
models for type C clouds ($n=10^2-10^3$~cm$^{-3}$), the line
intensities vary over two orders of magnitude. Because of the fixed
cloud size, models at higher densities have larger column densities.
Even at the point farthest from the cloud edge, the low density gas in
type C clouds causes relatively little radiation attenuation, and with
high incident radiation it is at very high temperature and in a highly
ionized state throughout. Under these conditions warm CO gas present,
but only in very small amounts. Therefore, at such points in parameter
only very weak line emission is produced and the lines are optically
thin. In the high density range ($n=10^4-10^{6.5}$~cm$^{-3}$) of type
A clouds, on the other hand, the spread in intensity is much
reduced. Although at densities of $n=10^4$~cm$^{-3}$ most of the cloud
is still at a high temperature and in a highly ionized state, there is
sufficient column density to have CO abundances large enough to
produce significant line emission. At even higher densities
($n=10^6$~cm$^{-3}$), the column densities are high enough to
attenuate the radiation field in such a way that a large CO fraction
is produced ($\sim10^{-4}$), but still at a temperature of
$T\sim100$~K. Here, the CO(1-0) line emission produced in XDRs can be
two to four times stronger than that in PDRs.

For the CO(2-1) line, the upper state energy is $E/k=16.60$~K and the
critical density is $n_{cr}\sim1\times10^4$~cm$^{-3}$. Although not
shown, the line intensities exhibit a behavior as a function of
density and radiation field very similar that that of the CO(1-0)
line. Because of the higher upper state energy, we find a somewhat
stronger dependence on radiation field in the PDR models. The effect
of the larger critical density is hard to see, due to the large
optical depths, but do show up when we consider the CO(2-1)/CO(1-0)
ratio. This ratio is shown in Fig. \ref{ratio_CO21_CO10}. In the PDR
models, the ratio does not differ more than a factor of two over the
full density range considered here. In the XDR models, very large line
intensity ratios of 30 or more are found, especially at high incident
radiation fields. It is, however, very questionable whether we will
actually observe these high ratios, since the intensity of the emitted
emission is low. The CO(2-1)/CO(1-0) ratio dependence on density and
radiation field is in general weak, especially in PDRs, since the
upper state energies are not very high and the difference in critical
density is small.

The CO(4-3) line (see Fig. \ref{intensity_CO43}) has an upper state
energy $E/k=55.32$~K and critical density
$n_{cr}\sim4\times10^4$~cm$^{-3}$. As expected, the emitted intensity
shows more variation with density and radiation field. In
Fig. \ref{ratio_CO43_CO10}, we show the CO(4-3)/CO(1-0) line intensity
ratio. In the low density range (cloud type C), PDR models still
produce a line ratio increase only as function of density, but the
variation in the ratio has grown to more than a factor of two, as
opposed to only 20 percent in the corresponding CO(2-1)/CO(1-0)
ratio. The XDRs for this cloud type show a complex behavior with
density and radiation field and the ratios cover a much larger range
from about 2 to 40. At low radiation fields, only a density dependence
is seen. At high radiation field strengths, the effect of the column
density comes into play. In the mid density (type B) PDRs, the highest
line ratios are seen for the highest densities and radiation
fields. The XDRs in this range show only a dependence on radiation
field. The effect of the higher density is compensated by the fact
that at lower densities relatively more gas is at high
temperatures. The gas temperature plays a large role in the density
range applicable to this cloud type. CO is present at much higher
temperatures in the XDRs. Therefore, the XDR line ratios for the same
density and incident radiation field can be more than ten times larger
than in the PDR. This difference slowly disappears when the critical
density of the CO(4-3) line is reached and the CO(4-3) line also
thermalizes, which is seen in the high density range (cloud type A) at
densities $n > 10^5$~cm$^{-3}$.


\begin{figure*}[!ht]
\centerline{\includegraphics[height=50mm,clip=]{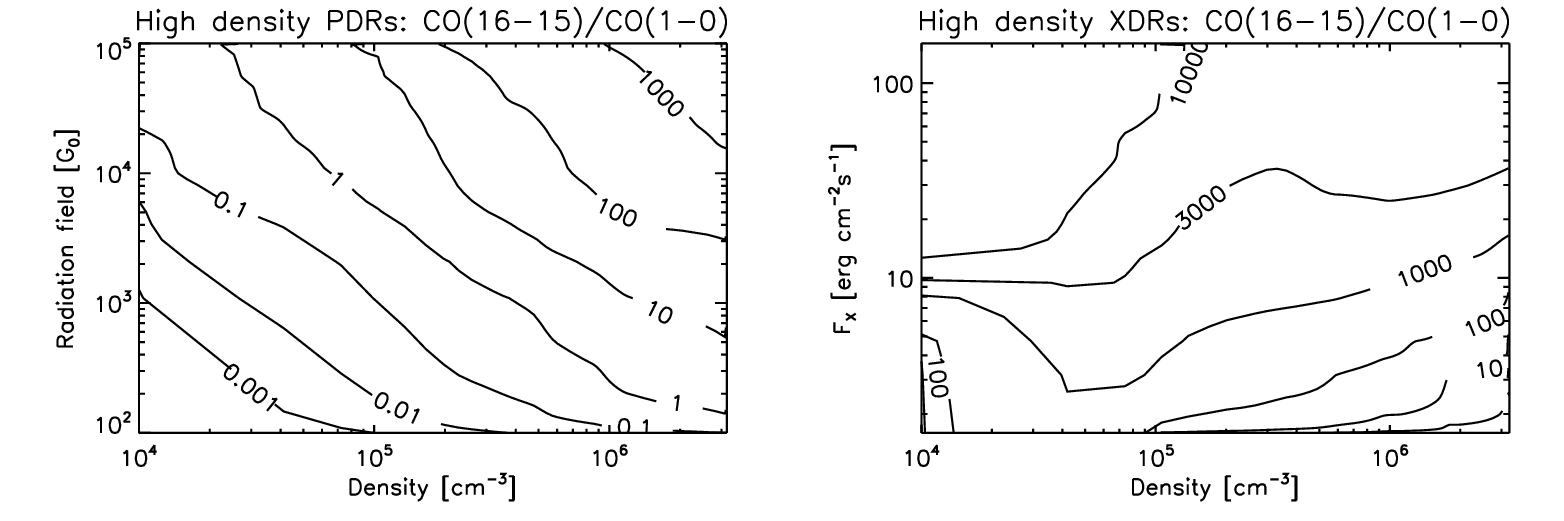}}
\centerline{\includegraphics[height=50mm,clip=]{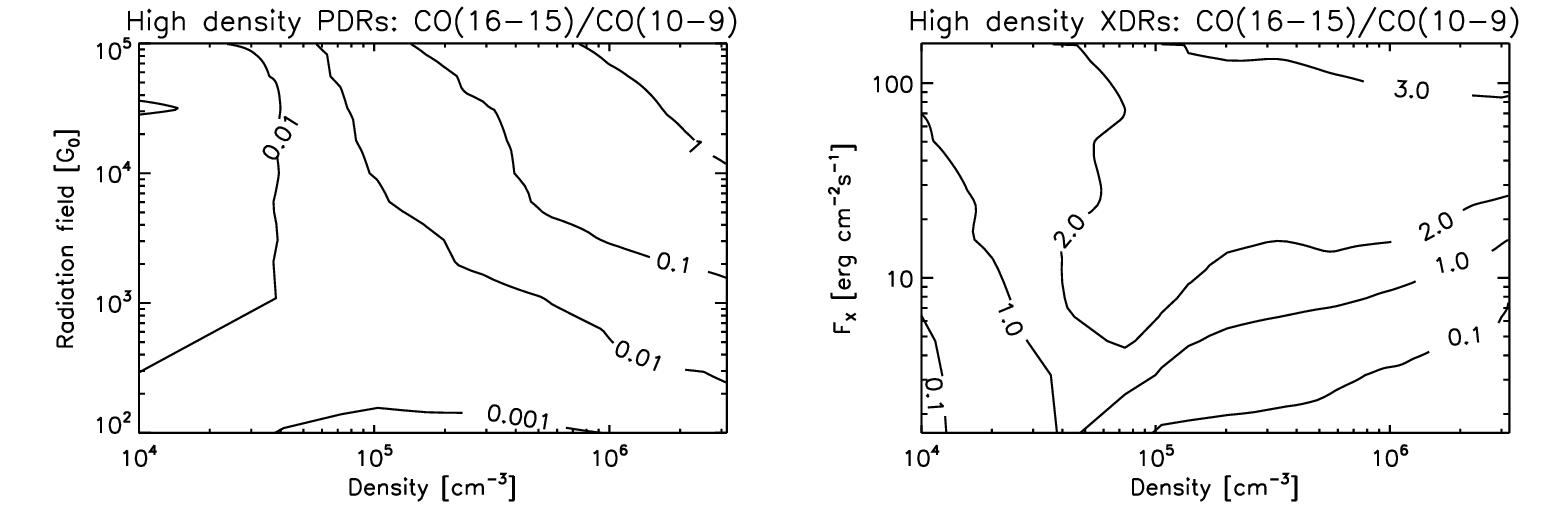}}
\centerline{\includegraphics[height=50mm,clip=]{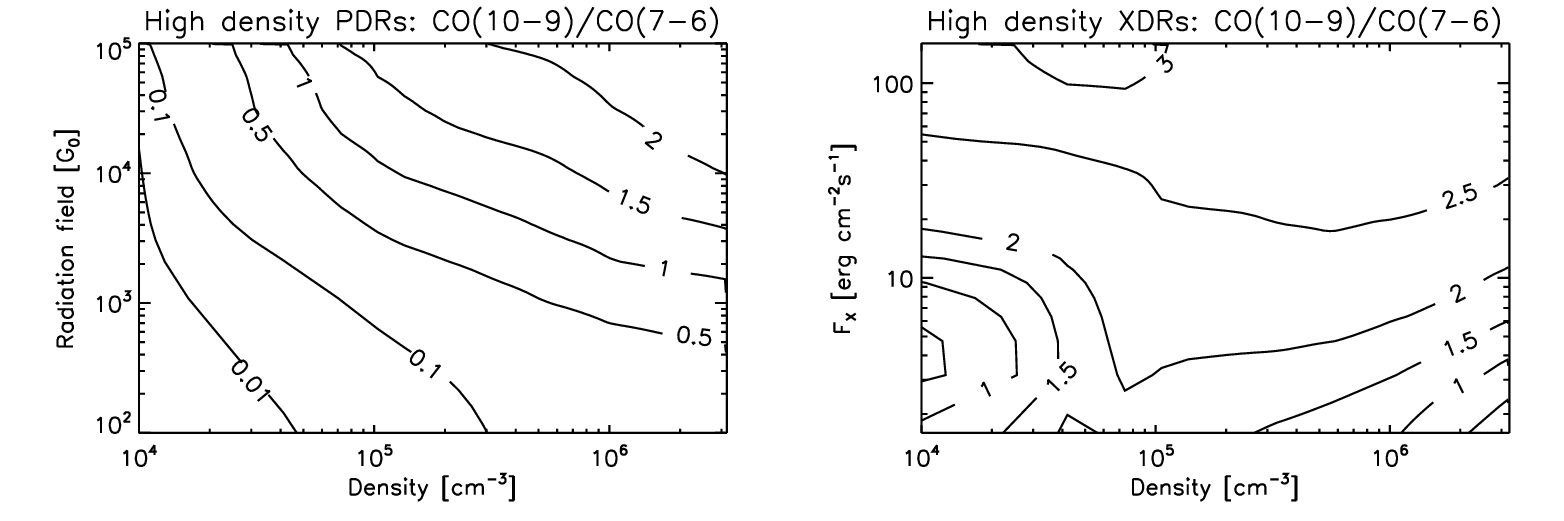}}
\centerline{\includegraphics[height=50mm,clip=]{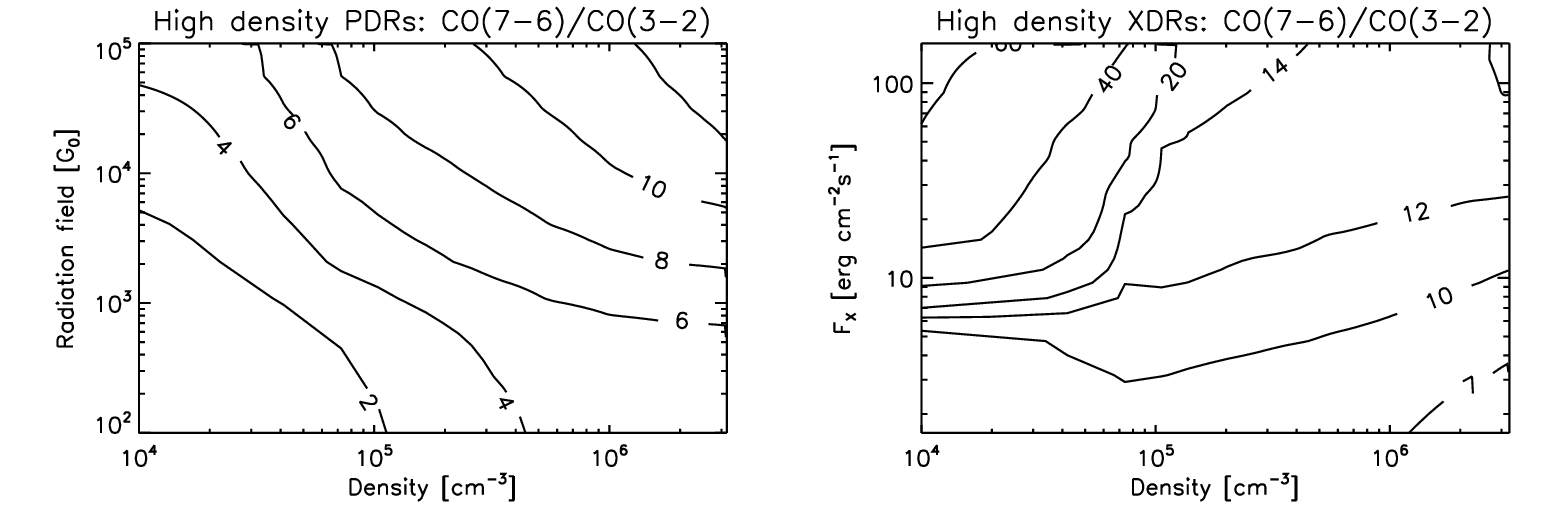}}
\caption{CO(16-15)/CO(1-0), CO(16-15)/CO(10-9), CO(10-9)/CO(7-6) and
CO(7-6)/CO(3-2) ratio for PDR (left) and XDR (right) models.}
\label{high_J_CO_ratio}
\end{figure*}


\subsection{High $J$-CO rotational transitions}

In cloud type A XDRs ($n=10^4-10^{6.5}$~cm$^{-3}$), CO is present
throughout the cloud, even when energy deposition rates $H_X/n$ are
large and temperatures are high ($T\sim200$~K). This warm CO gas
produces emission originating from high rotational transitions even
when densities are not high (e.g., $n=10^4$~cm$^{-3}$). Contrary to
the situation in XDRs, most CO in PDRs is produced beyond the H/H$_2$
transition and it has on average much lower temperatures
($T\sim20-50$~K), causing lower intensities and line
ratios. Therefore, it is very likely, that future missions such as
Herschel/HIFI will be able to distinguish between PDRs and XDRs, by
observing high rotational transitions such as CO(16-15), CO(10-9), and
CO(7-6).

In Fig. \ref{high_J_CO_intens}, we show the PDR and XDR intensities of
the CO(7-6), CO(10-9) and CO(16-15) lines for the high density range
(cloud type A). We find that for both PDRs and XDRs, the spread in
intensities increases for higher rotational lines, since the critical
densities of these transitions are higher. However, this spread is
much larger for PDRs than for XDRs. The CO(16-15) line intensity
ranges from $\sim10^{-10}$ ($n=10^4$~cm$^{-3}$ and $G_0=10^2$) to
$~\sim10^{-3}$~erg~s$^{-1}$~cm$^{-2}$~sr$^{-1}$
($n=10^{6.5}$~cm$^{-3}$ and $G_0=10^5$) for the PDR models, while this
is $\sim10^{-4}$ ($n=10^4$~cm$^{-3}$) to
$\sim10^{-2}$~erg~s$^{-1}$~cm$^{-2}$~sr$^{-1}$
($n=10^{6.5}$~cm$^{-3}$) for the XDR models. PDRs show only
significant CO(16-15) emission at very high densities and radiation
fields ($n\sim10^6$~cm$^{-3}$ and $G_0\sim10^{4}$). This very dense
and strongly irradiated gas, however, has a very small filling factor
on large (galaxy) scales, and, the probability of observing a PDR with
a very high CO(16-15) intensity is low.

The difference between the PDR and XDR models is seen even better by
considering the intensity ratios of these high rotational transitions
(see Fig. \ref{high_J_CO_ratio}). A good example is the
CO(16-15)/CO(1-0) ratio, which ranges from $~10^{-3}$ to $10^3$ for
PDRs, and from $10$ to $>10^4$ for XDRs. Especially for densities
between $10^4-10^5$~cm$^{-3}$, it is very easy to distinguish PDRs
from XDRs.


\begin{figure*}[!ht]
\centerline{\includegraphics[height=150mm,clip=]{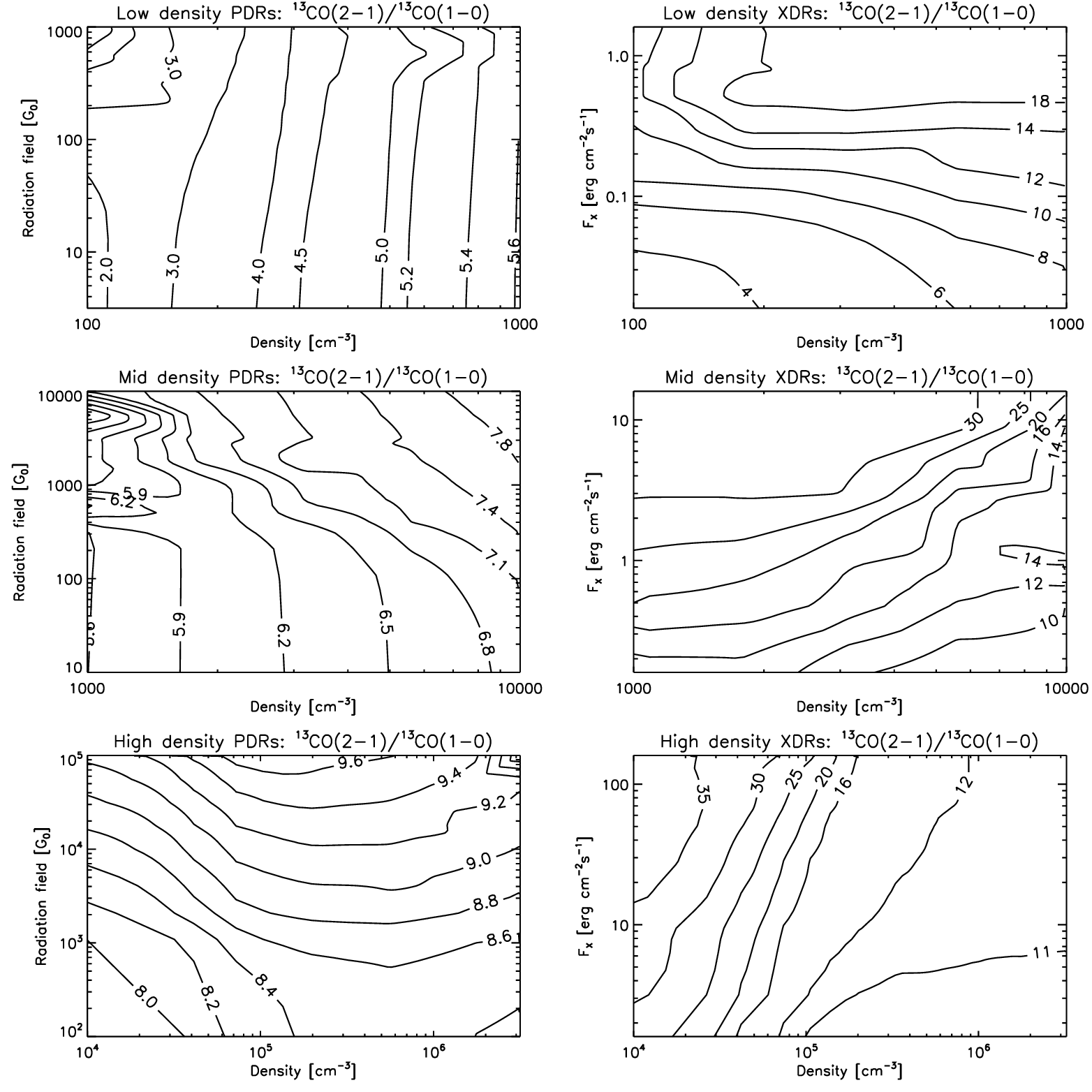}}
\caption{$^{13}$CO(2-1)/$^{13}$CO(1-0) ratio for PDR (left) and XDR (right) models.}
\label{ratio_13CO21_13CO10}
\end{figure*}

\begin{figure*}[!ht]
\centerline{\includegraphics[height=150mm,clip=]{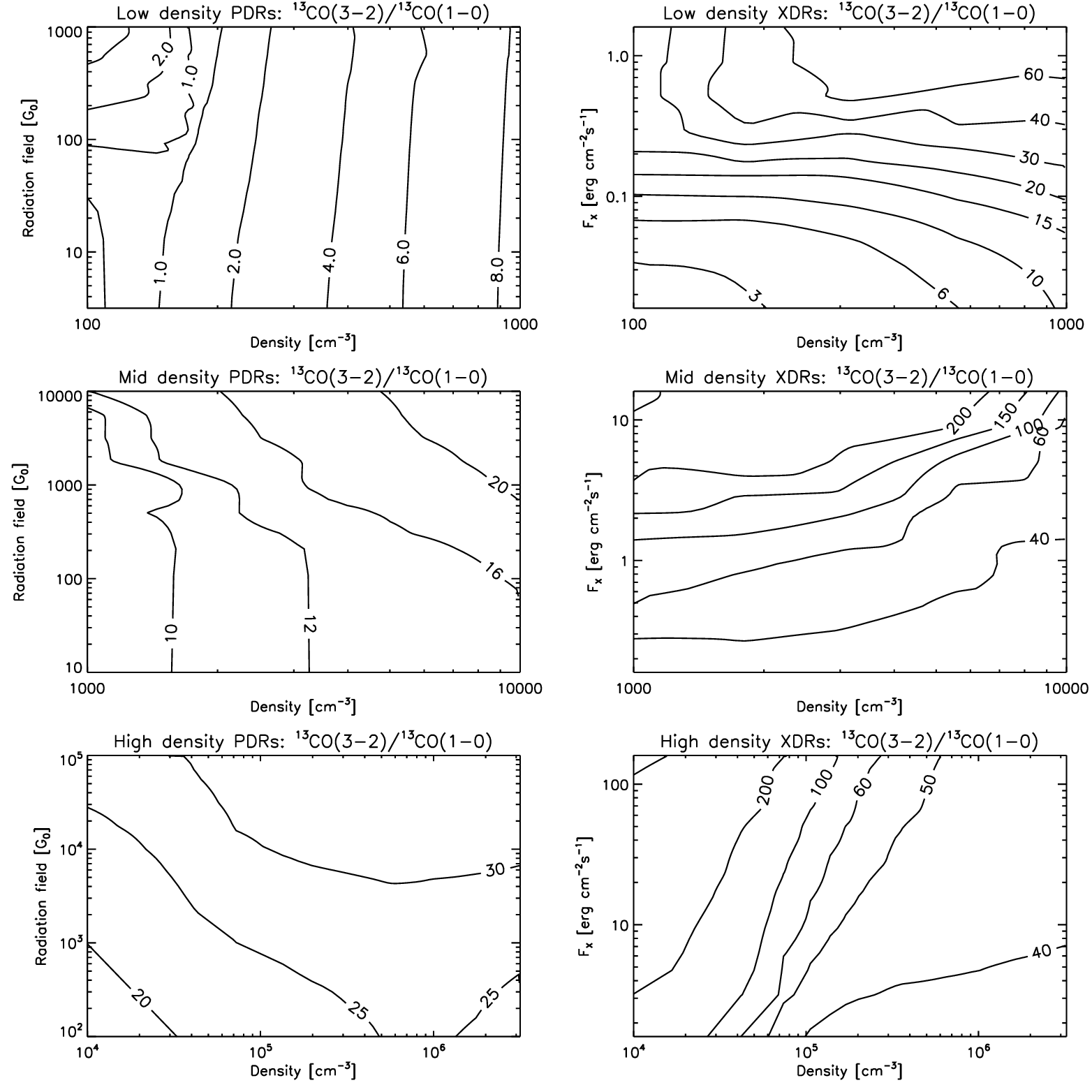}}
\caption{$^{13}$CO(3-2)/$^{13}$CO(1-0) ratio for PDR (left) and XDR (right) models.}
\label{ratio_13CO32_13CO10}
\end{figure*}

\subsection{$^{13}$CO rotational lines}

The $^{13}$CO lines have critical densities and upper state energies
for the rotational transitions almost identical to those of $^{12}$CO.
As we have adopted an abundance ratio $^{12}$C/$^{13}$C$=40$ in our
models, $^{13}$CO abundances are relatively low and the lines are much
less optically thick.

The $^{13}$CO(1-0) line intensities (Fig.\ref{intensity_13CO10}), show
the same trends with density and radiation field as the CO(1-0) line,
but there is a larger spread in intensity. For example, the difference
in the CO(1-0) line emission in the low density PDRs (cloud type C) is
a factor of two and more than a factor of 3 for the $^{13}$CO
line. This effect is even larger for the XDRs. We find similar results
for both the $^{13}$CO(2-1) (not shown) and $^{13}$CO(3-2) line
intensities as is evident from the corresponding $^{13}$CO line ratios
(see Fig. \ref{ratio_13CO21_13CO10} and \ref{ratio_13CO32_13CO10}).  A
nice illustration of the CO and $^{13}$CO behavior is supplied by the
the CO(4-3)/CO(1-0) versus the $^{13}$CO(3-2)/$^{13}$CO(1-0) intensity
ratio for the low density PDRs (cloud type C). The CO(4-3) line is
pumped due to the fact that the lower rotational lines become
optically thick. The CO(4-3)/CO(1-0) ratio changes only a factor of
two, which is a factor two and a half for the
$^{13}$CO(3-2)/$^{13}$CO(1-0) ratio, despite the lower critical
density of $^{13}$CO(3-2).


\begin{figure*}[!ht]
\centerline{\includegraphics[height=150mm,clip=]{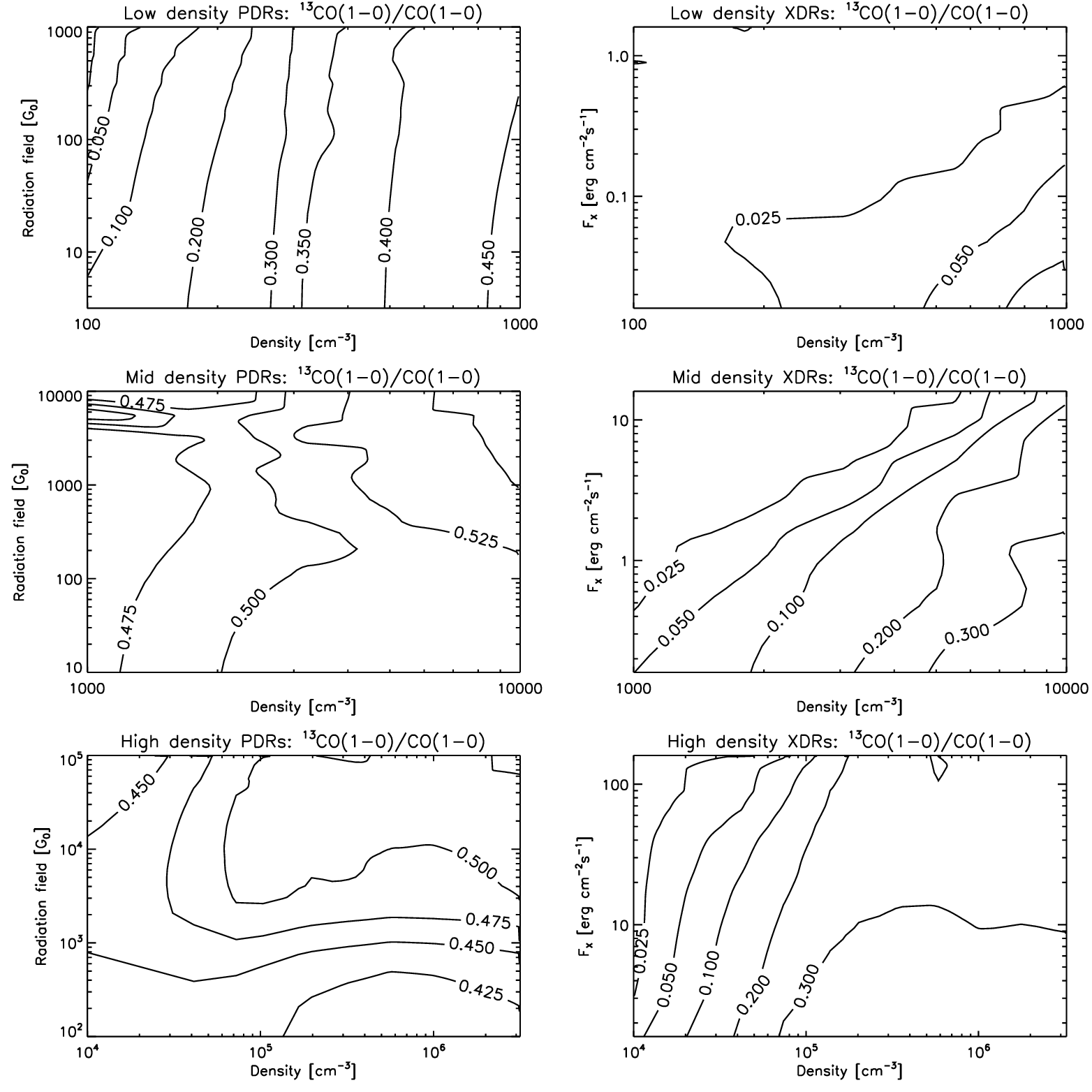}}
\caption{$^{13}$CO(1-0)/CO(1-0) ratio for PDR (left) and XDR (right) models.}
\label{ratio_13CO10_CO10}
\end{figure*}


\subsection{$^{13}$CO/CO ratios}

In Fig. \ref{ratio_13CO10_CO10} and \ref{ratio_13CO32_CO32}, the
important isotopical intensity ratios of $^{13}$CO(1-0)/CO(1-0) and
$^{13}$CO(3-2)/CO(3-2) are shown. Although for all cloud types
(density regimes) the line intensities are larger in the XDR, the PDR
isotopical ratios exceed those of XDRs, which means that the opacities
of the XDR lines are larger as well.

\subsection{[CI] 609~$\mu$m/~$^{13}$CO(2-1) ratio}

\begin{figure*}[!ht]
\centerline{\includegraphics[height=150mm,clip=]{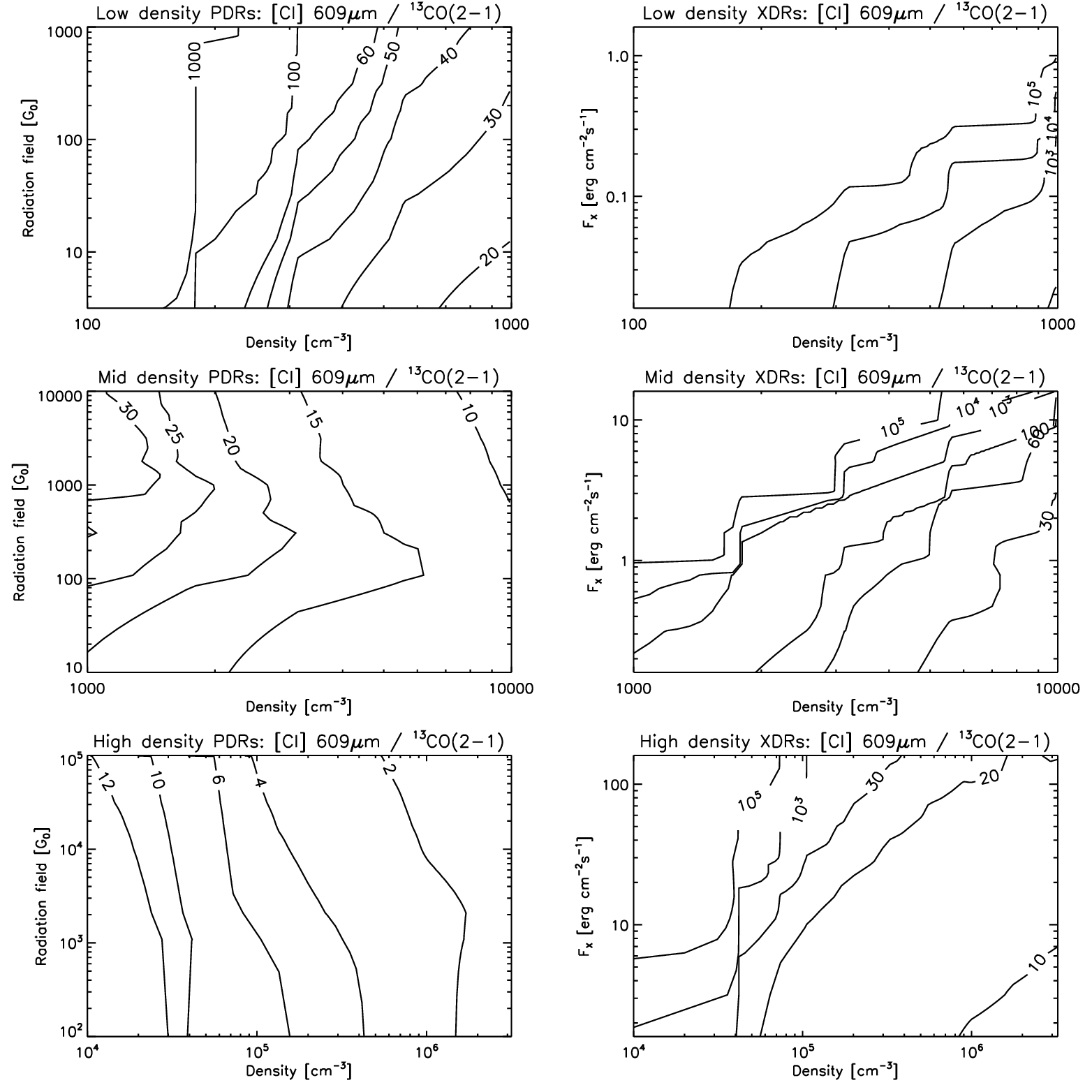}}
\caption{CI 609~$\mu$m/$^{13}$CO(2-1) ratio for PDR (left) and XDR (right) models.}
\label{ratio_CI_13CO32}
\end{figure*}

Fig. \ref{ratio_CI_13CO32} shows the [CI] 609~$\mu$m/~$^{13}$CO(2-1)
ratios. For the same gas density and incident radiation field, PDRs
have much lower ratios than XDRs. In the PDRs, the spread in the ratio
is generally not very large. At low densities, ratios rapidly decrease
from 72 ($n=10^2$~cm$^{-3}$) to 18 ($n=10^3$) and then slowly fall off
from 14.5 ($n=10^4$~cm$^{-3}$) to 2 ($n=10^6$~cm$^{-3}$). While the
PDR ratios show a more or less steady decrease with density, a rather
different picture is seen in XDRs. In each density range (cloud type),
the ratio changes by several orders of magnitude. For the lowest
density in each cloud type, the column density is too low to attenuate
the incident radiation field sufficiently to allow large amounts of CO
to be present. On the other hand, neutral carbon occurs throughout the
cloud, and, therefore, a large increase in the ratios are seen toward
high $H_X/n$.

\subsection{HCN rotational lines}

In Fig. \ref{intensity_HCN}, we show the HCN(1-0) and HCN(4-3) line
intensities for high density (cloud A type) PDR and XDR models only.
Where the CO(4-3) line has a critical density of
$n_{cr}\sim4\times10^4$~cm$^{-3}$, the HCN(1-0) line has a critical
density of $n_{cr}\sim3\times10^6$~cm$^{-3}$. Higher rotational
transitions such as HCN(2-1), $n_{cr}\sim4\times10^6$~cm$^{-3}$, and
HCN(4-3), $n_{cr}\sim2\times10^7$~cm$^{-3}$, have even higher critical
densities. The HCN rotational lines specifically trace the dense gas
component in galaxies, and the line intensities from low and medium
density gas are low. The HCN(1-0) line intensities range from
$3\times10^{-11}$ to $2\times10^{-8}$~erg~s$^{-1}$~cm$^{-2}$~sr$^{-1}$
for the low and mid density (type C en B) PDR models, and even less
for the corresponding XDR models. Because of their poor observational
prospects, we have limited ourselves to only showing the model results
in the high density range (cloud type A).  Typically, the HCN(1-0)
emission is stronger in PDRs by a factor of about two for densities
larger than $10^5$ cm$^{-3}$.  The HCN(4-3) diagrams show behavior
very similar to that of HCN(1-0), but the PDR and XDR line strengths
are now somewhat closer.

Typically, the HCN(1-0) emission is stronger in PDRs by a factor of
about two for densities larger than $10^5$ cm$^{-3}$.  Our results are
consistent with the chemical calculations of Lepp \& Dalgarno (1996)
for different ionization rates, as follows.  Our depth dependent
models cause the HCN line emissivities to be the result of a
line-of-sight integral over the HCN abundance pattern that results
from a varying (attenuated) X-ray flux. Lepp \& Dalgarno (1996, their
Fig.\ 3) find a rather narrow range of ionization rates for which the
HCN abundance is high and consequently the XDR HCN line emissivities
have difficulty competing with the PDR ones.  The HCN(4-3) contour
plots show about the same features as seen for the HCN(1-0). However,
the PDR and XDR line strengths are now somewhat closer.

\begin{figure*}[!ht]
\centering
\unitlength1cm
\begin{minipage}[b]{7cm}
\resizebox{7cm}{!}{\includegraphics*[angle=0]{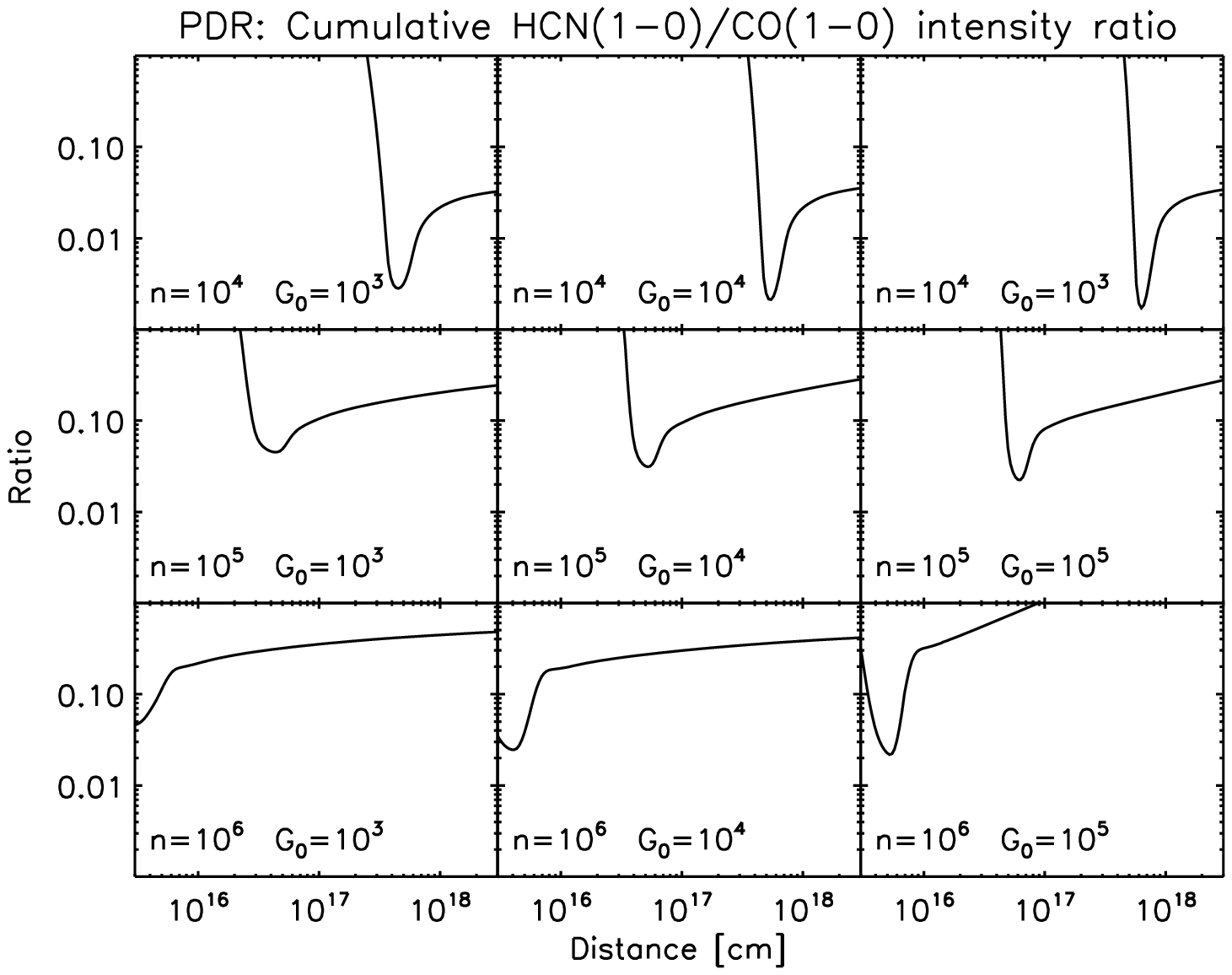}}
\end{minipage}
\begin{minipage}[b]{7cm}
\resizebox{7cm}{!}{\includegraphics*[angle=0]{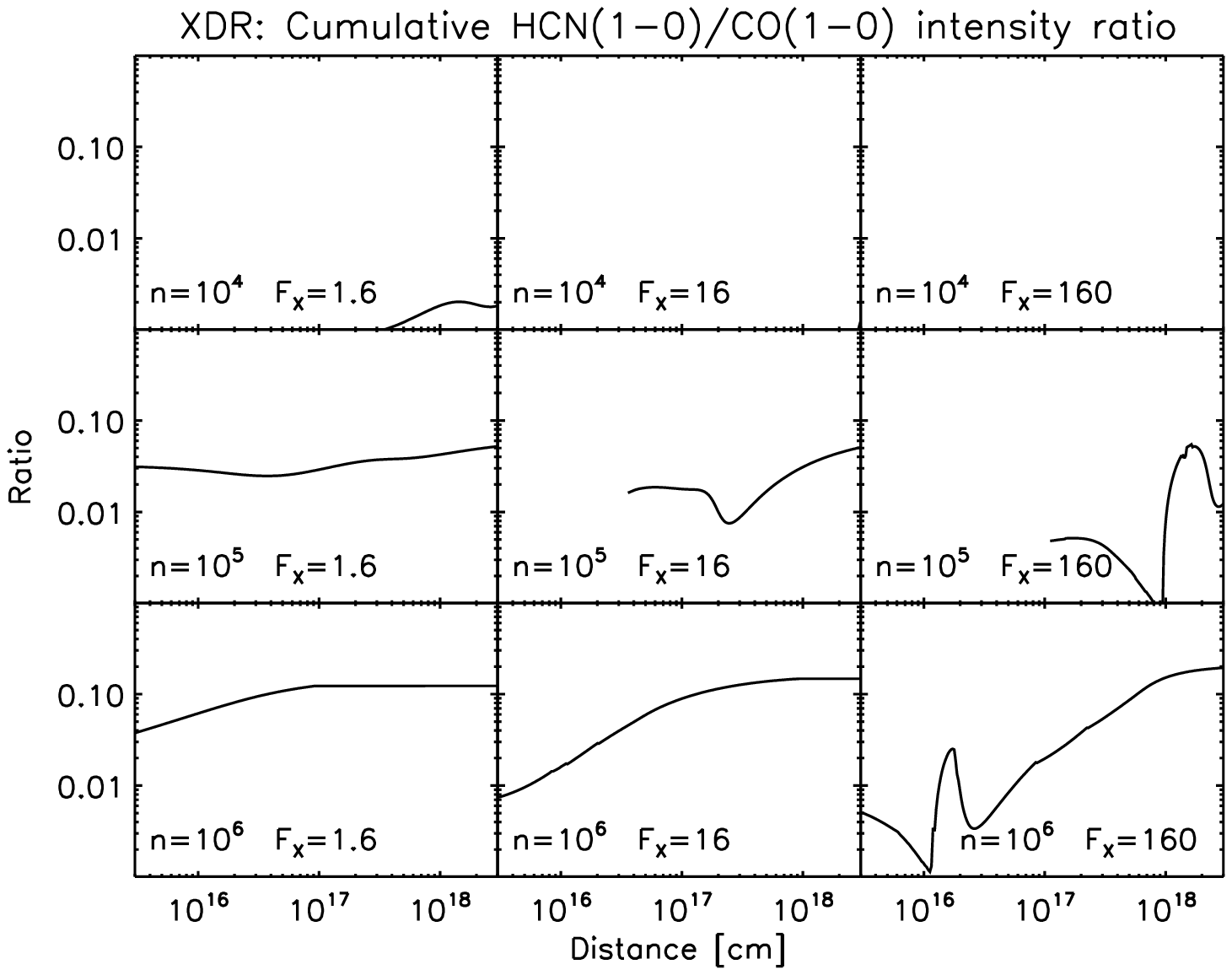}}
\end{minipage}
\caption[] {Cumulative HCN(1-0)/CO(1-0) line intensity ratios for PDR
(left) and XDR (right).}
\label{Cum_dens_HCN10_CO10}
\end{figure*}

\begin{figure*}[!ht]
\centerline{\includegraphics[height=50mm,clip=]{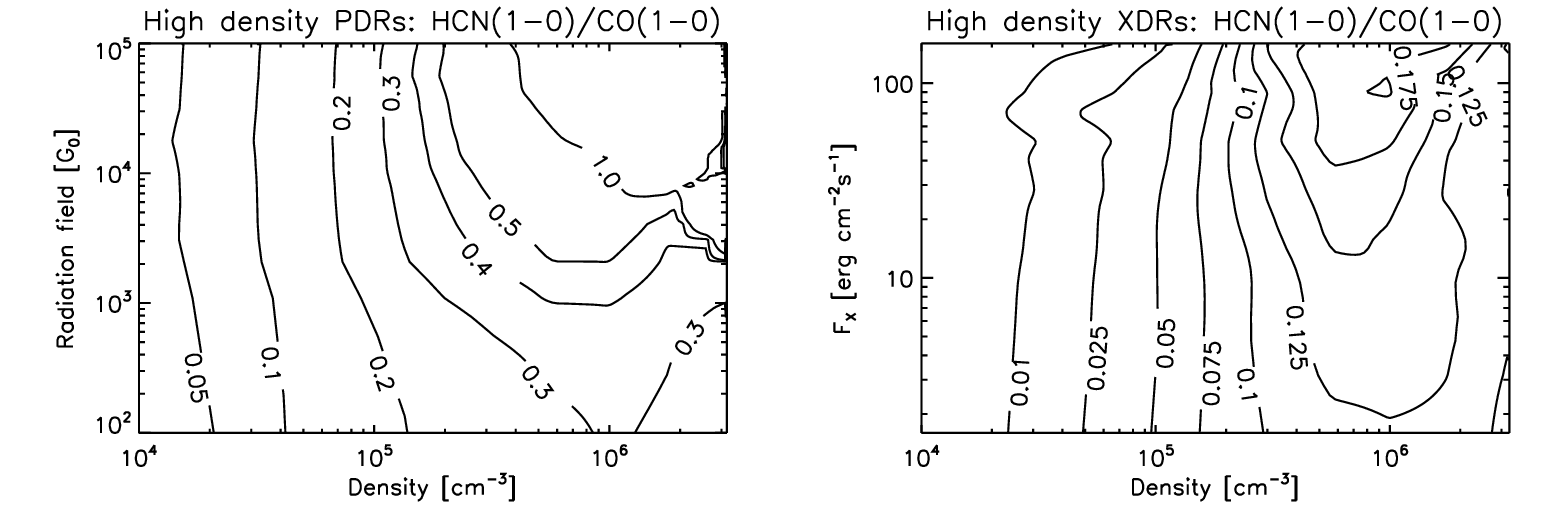}}
\centerline{\includegraphics[height=50mm,clip=]{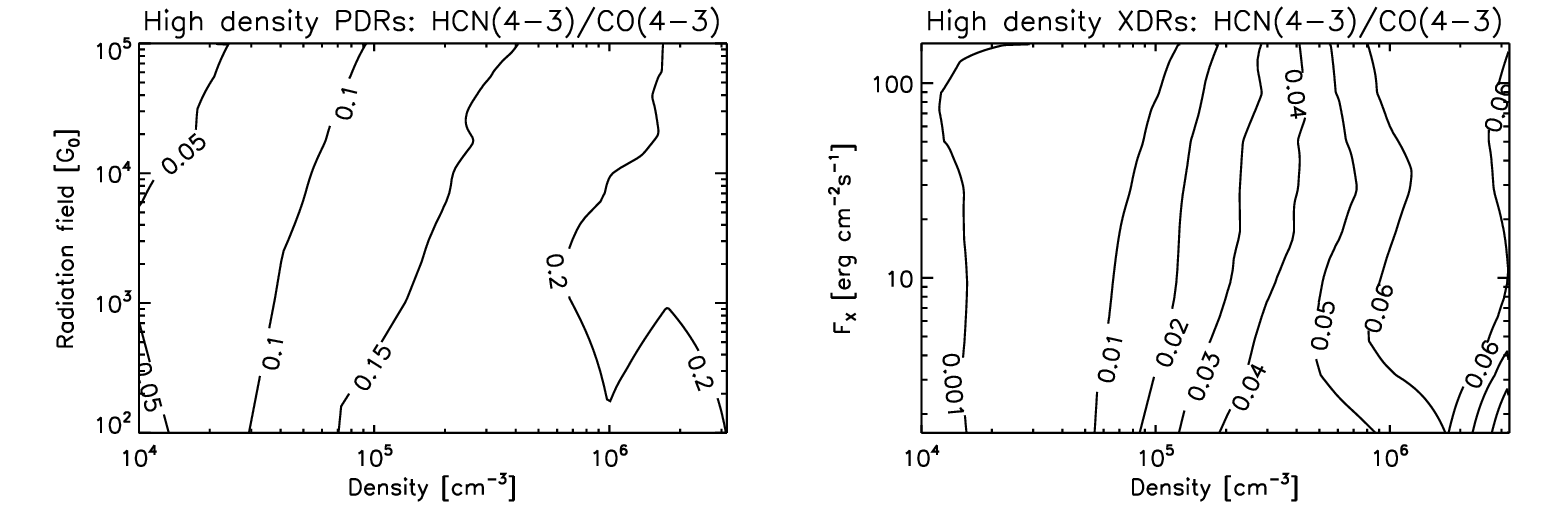}}
\caption{HCN(1-0)/CO(1-0) and HCN(4-3)/CO(4-3) ratios for PDR (left) and XDR (right) models.}
\label{ratio_HCN_CO}
\end{figure*}

\subsection{HCN/CO line intensity ratios}

Intensity ratios of lines from the same species, such as the
CO(4-3)/CO(1-0) ratio, vary with column density, due to the
temperature gradient throughout the cloud, optical depth effects,
varying abundance etc. Intensity ratios of lines from different
species, in addition vary because of abundance ratio differences,
complicating the interpretation of such line intensity ratios.  In
this section we turn our attention to the HCN/CO ratio, and start by
showing in Figs. \ref{Cum_dens_HCN10_CO10} and
\ref{Cum_dens_HCN43_CO43} the cumulative line intensity ratios for a
set of PDRs and XDRs at densities ranging from $n=10^4-10^6$~cm$^{-3}$
and incident fluxes between $G_0=10^3-10^5$
($F_X=1.6-160$~erg~s$^{-1}$~cm$^{-2}$). The cumulative line intensity
is the emergent intensity arising from the edge of the cloud to column
density $N_{\rm H}=n_{\rm H}z$:

\begin{equation}
I(z)=\frac{1}{2\pi}\int^z_0\Lambda(z')dz'.
\end{equation}

In the PDR, both the HCN(1-0)/CO(1-0) and the HCN(4-3)/CO(4-3) ratio
show a minimum. The HCN abundance shows a drop around the H/H$_2$
transition, while CO has its maximum abundance beyond the CO
transition. Deeper in the cloud, the HCN/CO abundance ratio is more or
less constant, but the CO line becomes optically thick and therefore
the ratio increases.

When $H_X/n$ is low in the XDR model, the gas is molecular, and the
HCN/CO abundance ratio is more or less constant. A slow rise in the
ratio is seen, as the CO line becomes optically thick. When the outer
part of the cloud is atomic (for high $H_X/n$), HCN shows a maximum,
before the H/H$_2$ transition. This also produces a maximum in the
HCN(1-0)/CO(1-0) ratio. In the PDR, the variation is not that large
for column densities $N_H > 10^{22}$~cm$^{-2}$. In the XDRs, however,
the variation between $N_H =10^{22}$ and $10^{23}$~cm$^{-2}$ can be
rather large.

In Fig. \ref{ratio_HCN_CO}, we show the HCN(1-0)/CO(1-0) and
HCN(4-3)/CO(4-3) line ratios for a fixed cloud size of one parsec
(cloud type A). The variation in the ratios is relatively large, due
to the high critical densities of the HCN transitions. The PDR models
produce the highest HCN(1-0)/CO(1-0) ratios, which are attained at
large densities ($n>10^6$~cm$^{-3}$) and may exceed unity. The
corresponding XDRs have ratios are only 0.1-0.2.  The interpretation
of the low J transitions is very difficult due to high opacities,
especially in the PDRs. For this reason, we also show the
HCN(4-3)/CO(4-3) ratio, which shows similar trends with density and
radiation field, but with somewhat lower absolute ratios.


\subsection{HCO$^+$ rotational lines and HCN/HCO$^+$ line intensity ratios}

In Fig. \ref{intensity_HCOp}, we show the HCO$^+$(1-0) and
HCO$^+$(4-3) line intensities, with critical densities
$n_{cr}\sim2\times10^5$ and $4\times10^6$~cm$^{-3}$,
respectively. These critical densities are significantly lower than
for HCN, causing a smaller spread in line intensities. Typically, the
HCO$^+$ lines are stronger in XDRs than in PDRs by a factor of at
least three. This is a direct consequence of the higher ionization
degree in XDRs (Meijerink \& Spaans 2005), leading to an enhanced
HCO$^+$ formation rate.

Note in this that Lepp \& Dalgarno (1996, their Fig.\ 2) find a rather
wide range of ionization rates for which the HCO$^+$ abundance is
large. As for the HCN discussed above, this is consistent with our
results since we integrate the depth dependent HCO$^+$ abundance
profile that results from the attenuation of the impinging X-ray
flux. The HCO$^+$ line-of-sight integral thus picks up a large
contribution and competes favorably with the PDR line emissivities.

\begin{figure*}[!ht]
\centering
\unitlength1cm
\begin{minipage}[b]{7cm}
\resizebox{7cm}{!}{\includegraphics*[angle=0]{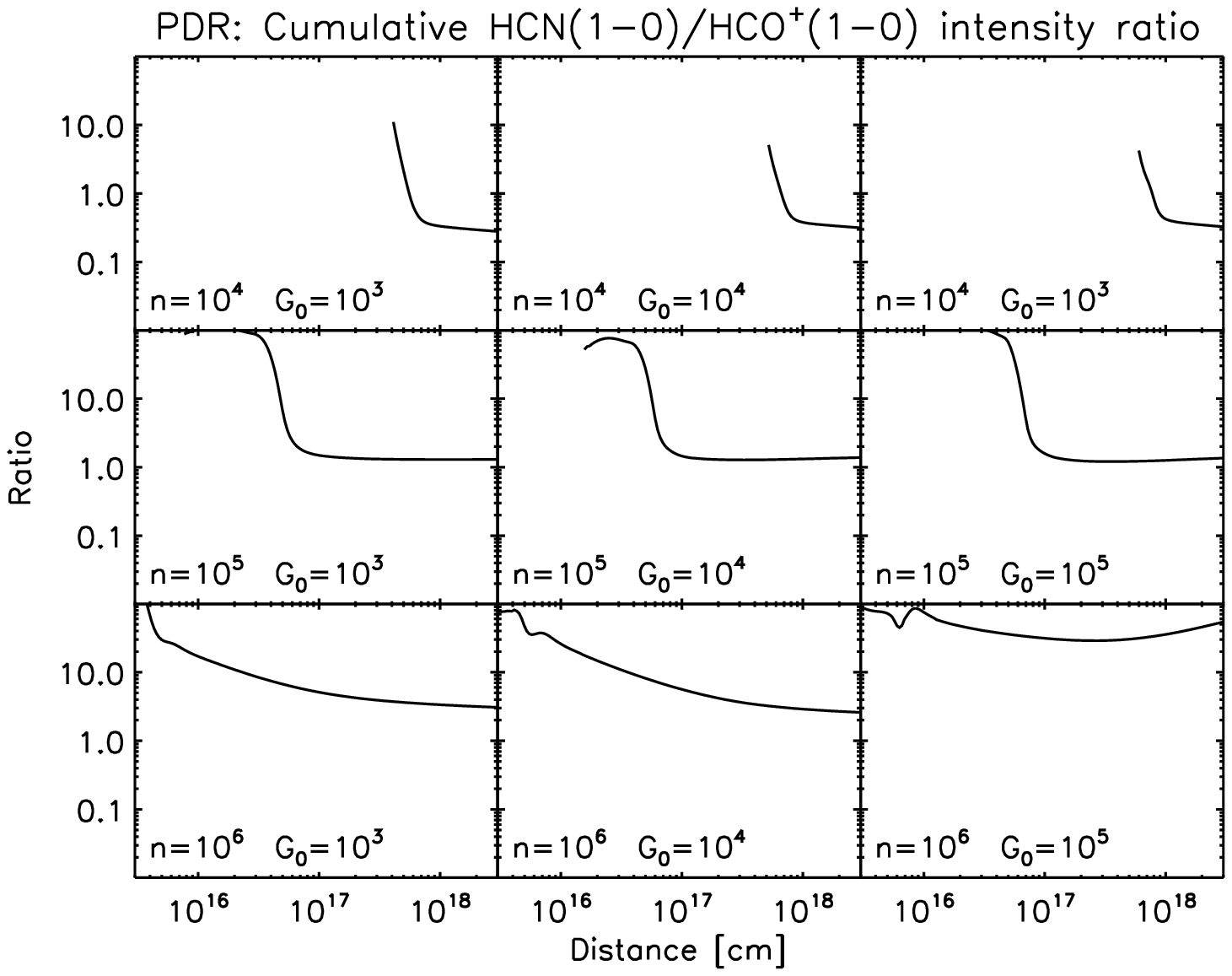}}
\end{minipage}
\begin{minipage}[b]{7cm}
\resizebox{7cm}{!}{\includegraphics*[angle=0]{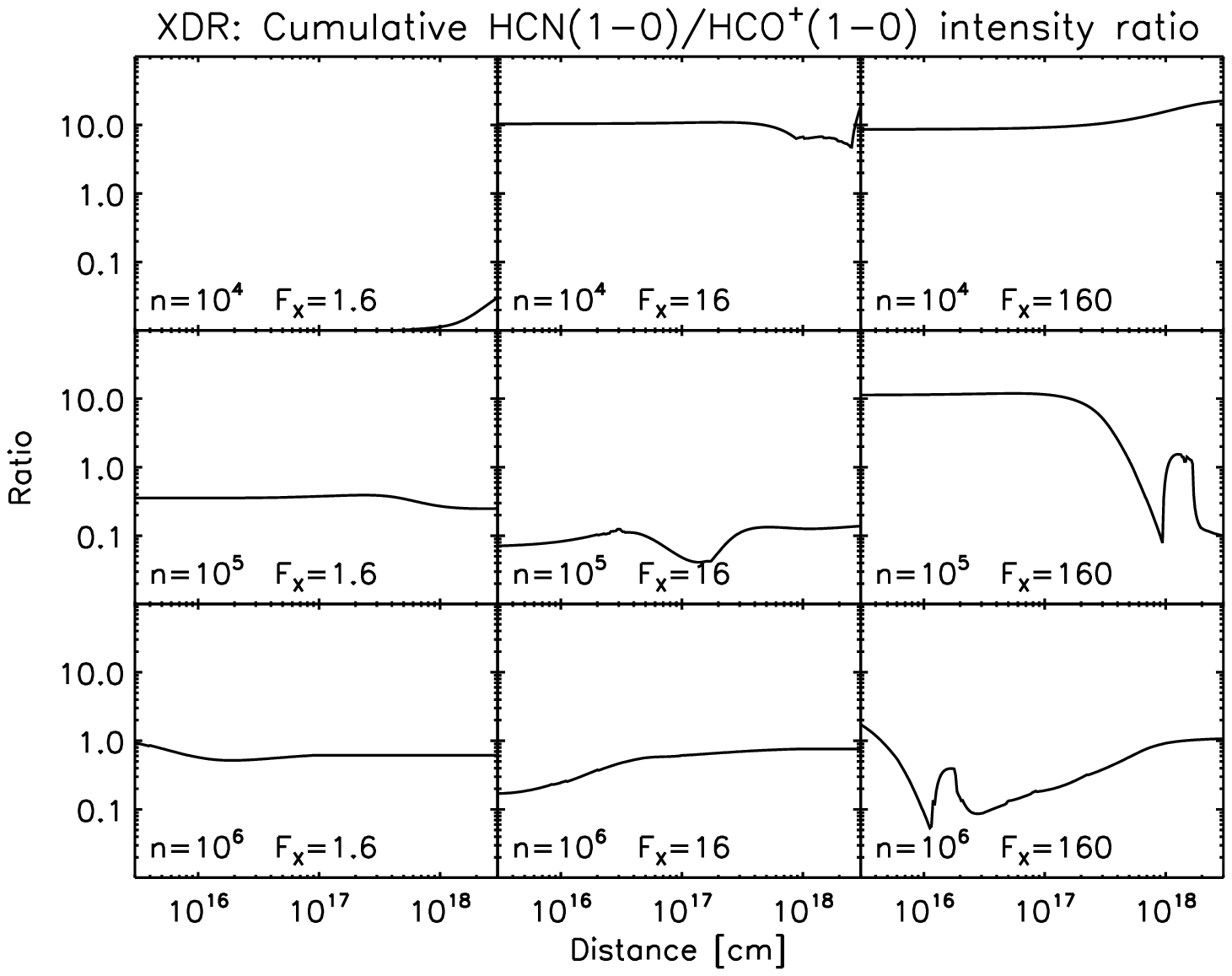}}
\end{minipage}
\caption[] {Cumulative HCN(1-0)/HCO$^+$(1-0) line intensity ratios for PDR
(left) and XDR(right).}
\label{cumul_ratio_HCN10_HCOp10}
\end{figure*}

\begin{figure*}[!ht]
\centerline{\includegraphics[height=50mm,clip=]{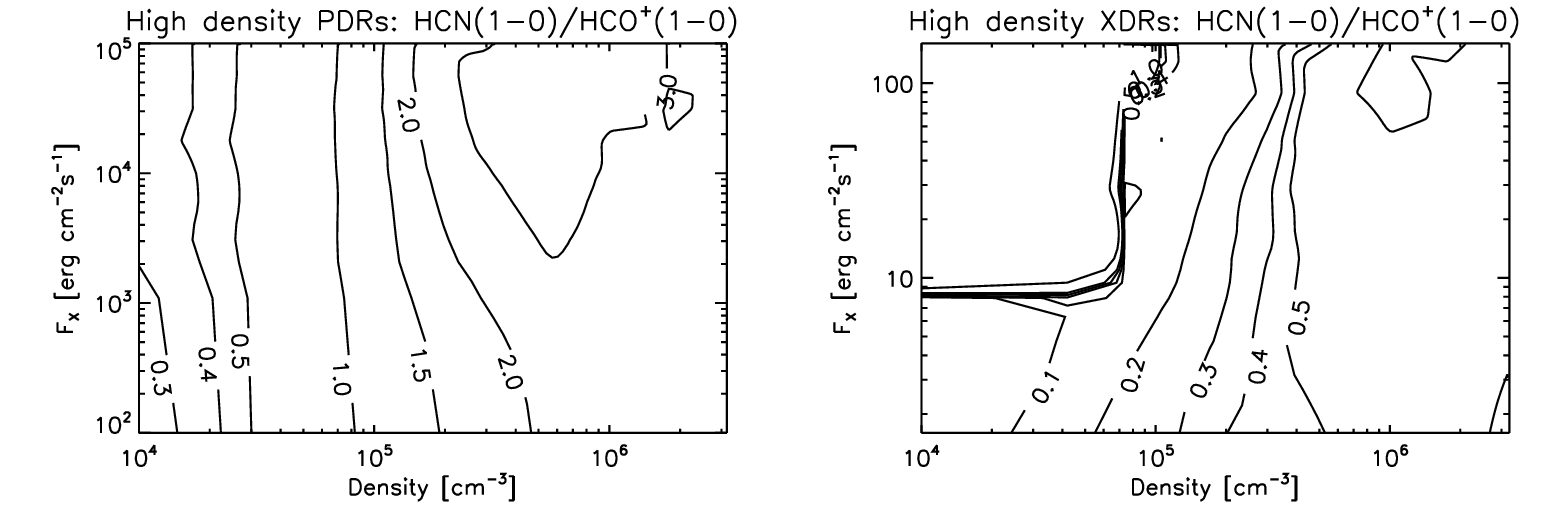}}
\centerline{\includegraphics[height=50mm,clip=]{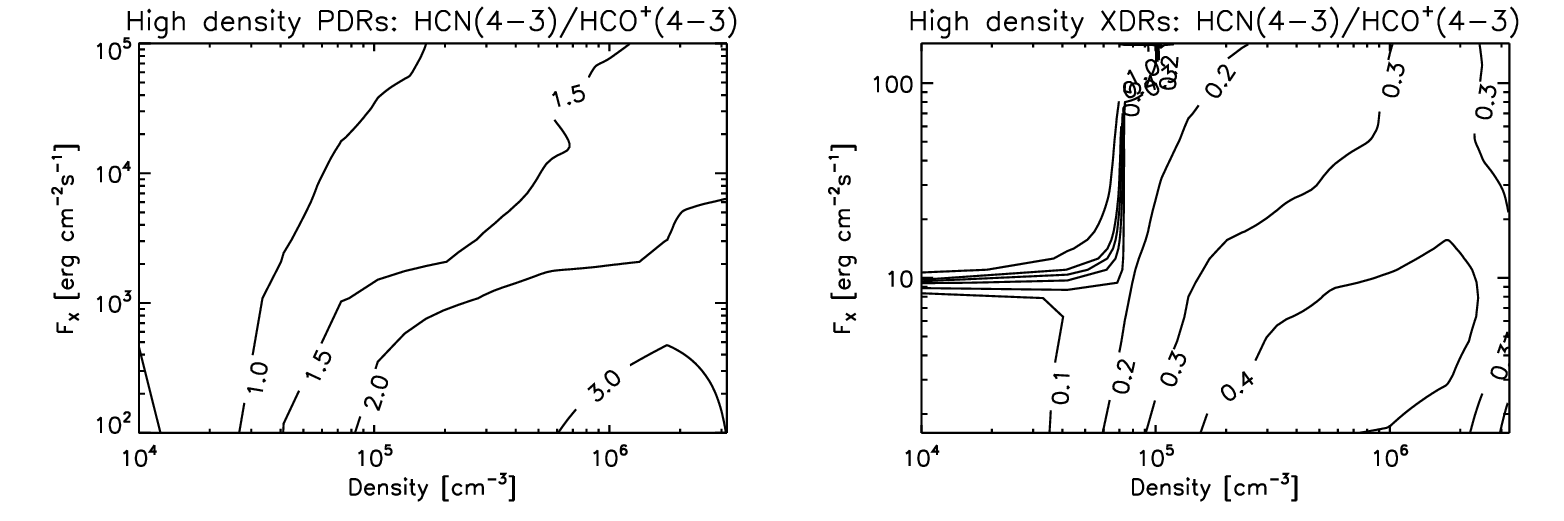}}
\caption{HCN(1-0)/HCO$^+$(1-0) and HCN(4-3)/HCO$^+$(4-3) ratios for PDR (left) and XDR (right) models.}
\label{ratio_HCN_HCOp}
\end{figure*}

In Figs. \ref{cumul_ratio_HCN10_HCOp10} and
\ref{cumul_ratio_HCN43_HCOp43}, we show the cumulative
HCN(1-0)/HCO$^+$(1-0) and HCN(4-3)/HCO$^+$(4-3) line intensity ratios,
for the same PDR and XDR models as in Section 5.5. Depending on the
incident radiation field, HCN or HCO$^+$ is more abundant at the PDR
edge of the cloud. Around the H/H$_2$ transition a minimum in the
HCO$^+$ abundance is seen in the PDR. Deeper into the cloud, the
HCO$^+$ abundance increases again, and is then constant. This is also
the case for HCN, and the HCN/HCO$^+$ abundance ratio is larger than
unity. Therefore, at sufficiently large columns and densities, the
HCN(1-0)/HCO$^+$(1-0) line intensity ratio becomes larger than one.

In the XDR models, HCO$^+$ is chemically less abundant than HCN for
very large $H_X/n$ (Meijerink \& Spaans 2005).  For larger columns
HCO$^+$ becomes more abundant, however, and eventually the cumulative
column density of HCO$^+$ becomes larger than HCN (see specifically
Fig.\ 10 in Paper I). This follows directly from the fact that the
HCO$^+$ abundance is high over a much wider range of ionization rates
than HCN (Lepp \& Dalgarno 1996, their Figs.\ 2 and 3).

Fig. \ref{ratio_HCN_HCOp} clearly shows that the HCN/HCO$^+$ ratio
discriminates between PDRs and XDRs in the density range between
$n=10^5$ and $10^{6.5}$~cm$^{-3}$ (cloud type A). The
HCN(1-0)/HCO$^+$(1-0) and HCN(4-3)/HCO$^+$(4-3) line ratios are both
much larger in the PDR models than XDR models, for columns of
$10^{23}$ cm$^{-2}$ and larger (Paper I). The difference ranges
between a factor of 4-10, depending on the density. The XDR
HCN(1-0)/HCO$^+$(1-0) ratio becomes larger than unity for more modest
columns of $10^{22.5}$ cm$^{-2}$ and less.


\begin{figure*}[!ht]
\centering
\unitlength1cm
\begin{minipage}[b]{7cm}
\resizebox{7cm}{!}{\includegraphics*[angle=0]{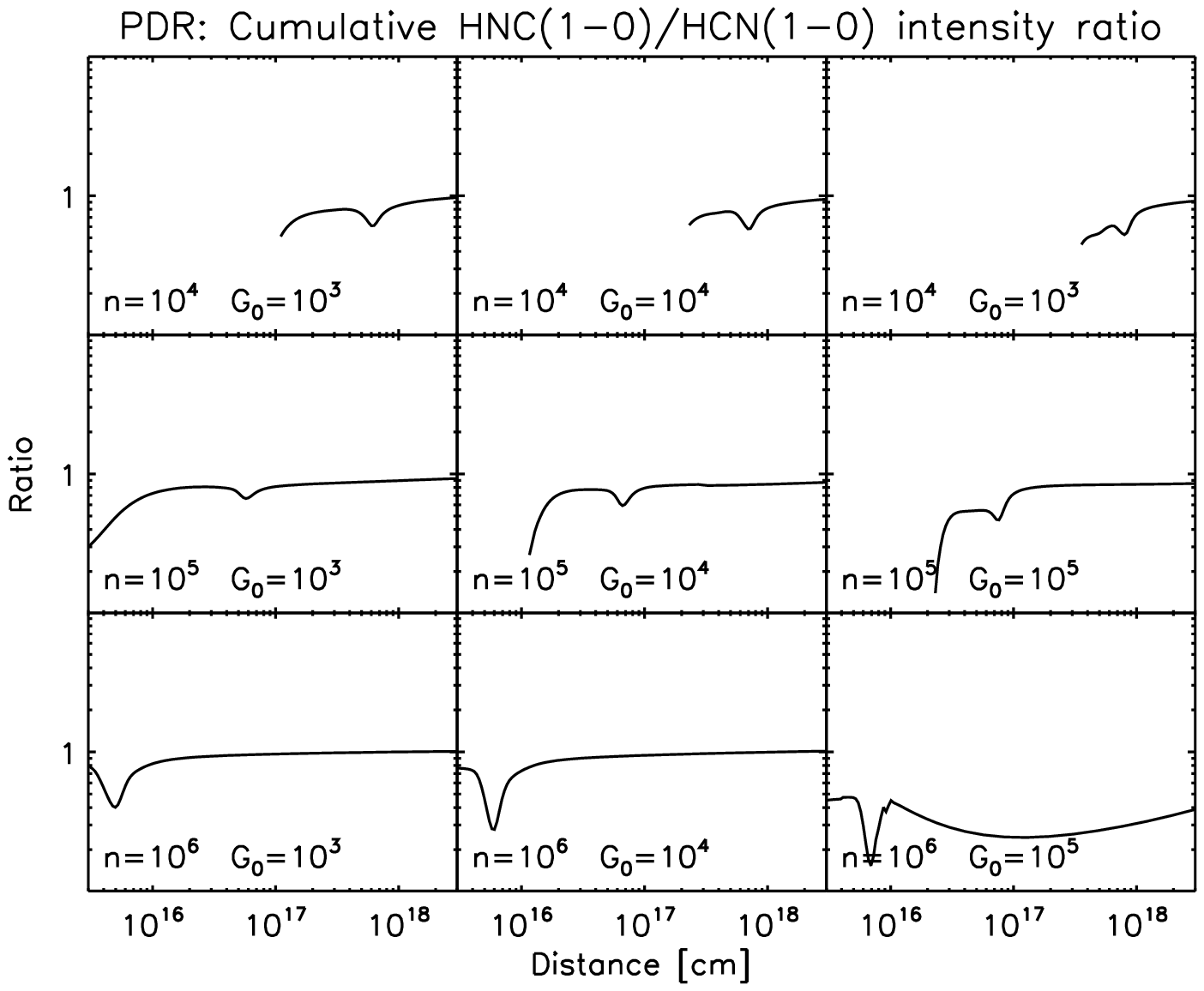}}
\end{minipage}
\begin{minipage}[b]{7cm}
\resizebox{7cm}{!}{\includegraphics*[angle=0]{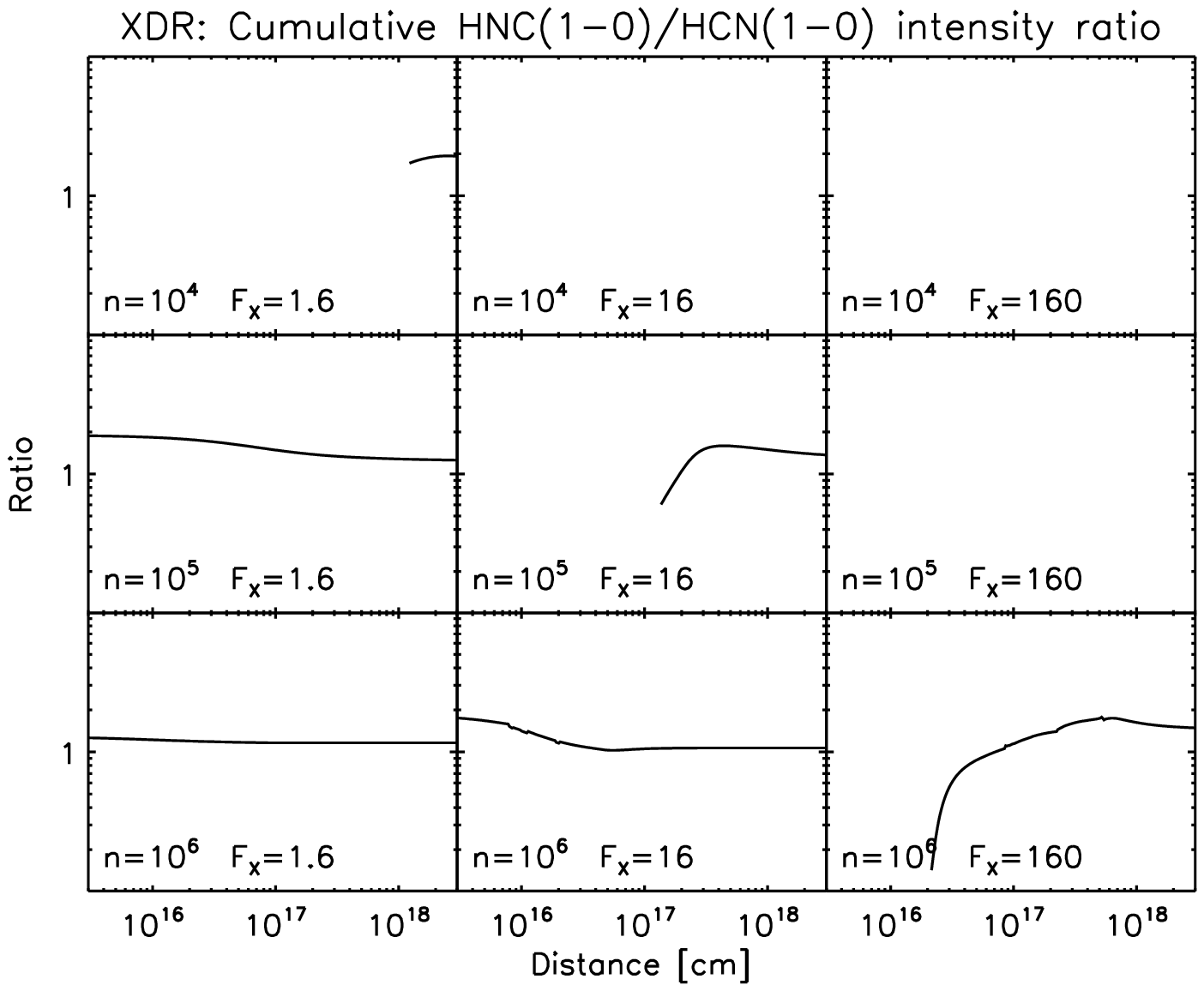}}
\end{minipage}
\caption[] {Cumulative HNC(1-0)/HCN(1-0) line intensity ratios for PDR
(left) and XDR(right).}
\label{cumul_ratio_HNC10_HCN10}
\end{figure*}

\begin{figure*}[!ht]
\centerline{\includegraphics[height=50mm,clip=]{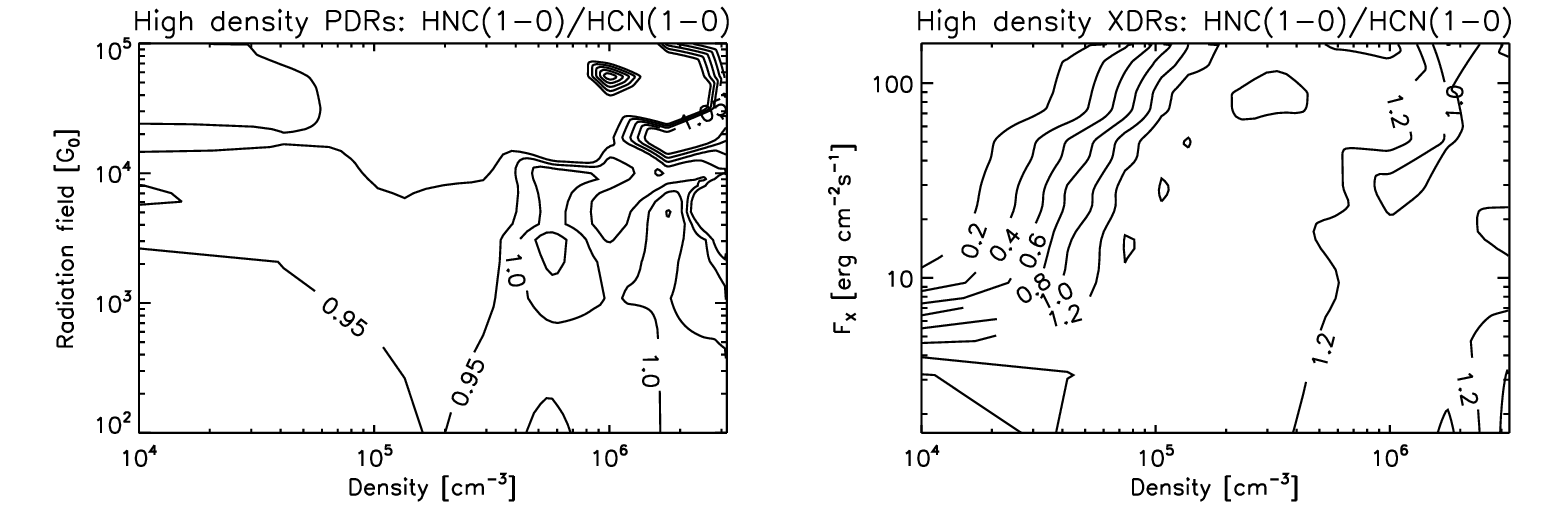}}
\centerline{\includegraphics[height=50mm,clip=]{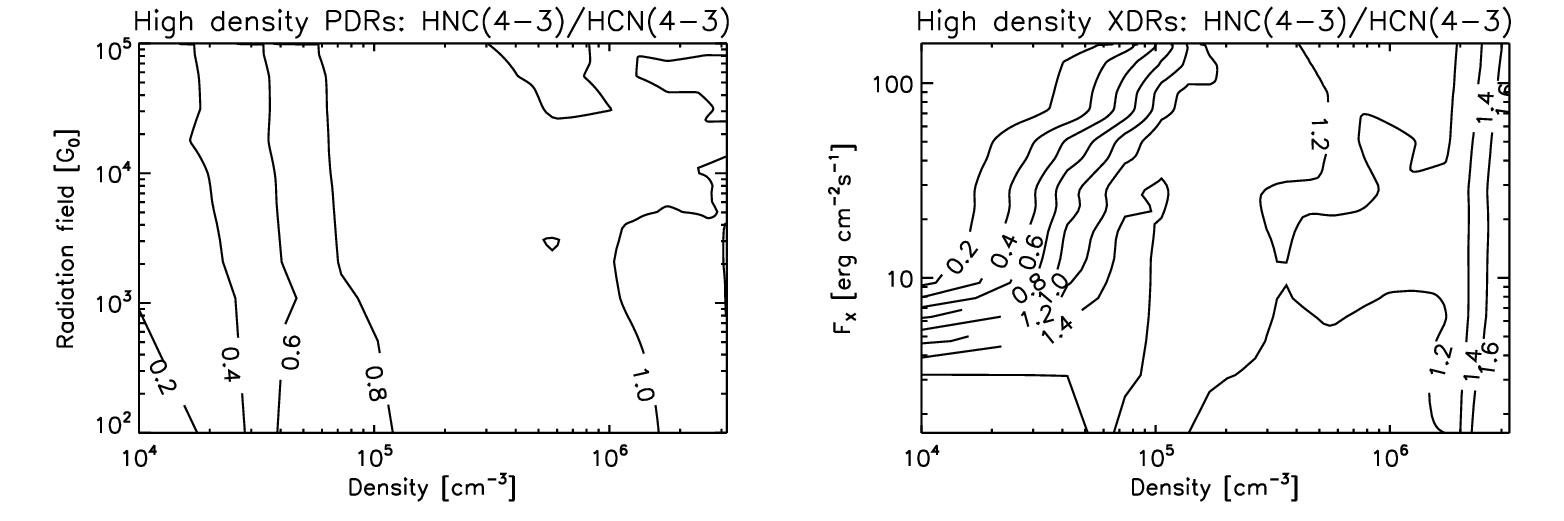}}
\caption{HNC(1-0)/HCN and HNC(4-3)/HCN(4-3) ratios for PDR (left) and XDR (right) models.}
\label{ratio_HNC_HCN}
\end{figure*}

\subsection{HNC/HCN ratios}

The critical densities of HCN and HNC are almost identical, so that
the only differences in line ratio should be due to differences in the
abundances. In PDRs, HCN is more abundant in the radical region, but
deeper in the cloud the abundance ratio approaches unity. In XDRs, HCN
is more abundant in the highly ionized part of the cloud. However, HNC
is equally or even more abundant than HCN deep into the cloud. As a
result, the HNC(1-0)/HCN(1-0) line intensity ratio is around one for
the PDRs if the column density is larger than $10^{22}$ cm$^{-2}$,
while the HNC(1-0)/HCN(1-0) ratio is less than unity for $N_H<10^{22}$
cm$^{-2}$. The XDR models, however, show low ratios for the low, $\sim
10^4$ cm$^{-3}$, densities and strong, $>10$ erg s$^{-1}$ cm$^{-2}$,
radiation fields. The ratios increase for lower incident radiation
fields, and at highest densities ($n=10^{6.5}$~cm$^{-3}$) the line
ratios are always larger than one, irrespective of irradiation.

In the PDRs, the HNC(4-3)/HCN(4-3) ratio quickly drops below unity at
densities below $n=10^5$~cm$^{-3}$. This density is far below the
critical densities of the lines, and therefore high temperatures are
needed to excite them. Such high temperatures are indeed found in the
radical regions of the PDRs, but there the HNC abundance is much lower
than the HCN abundance, which explains the drop in the ratio. In the
XDRs, the HNC(4-3)/HCN(4-3) ratios are quite similar to the
HNC(1-0)/HCN(1-0) ratios, except for densities $n>10^6$~cm$^{-3}$,
where they are even high than these.


\begin{figure*}[!ht]
\centerline{\includegraphics[height=50mm,clip=]{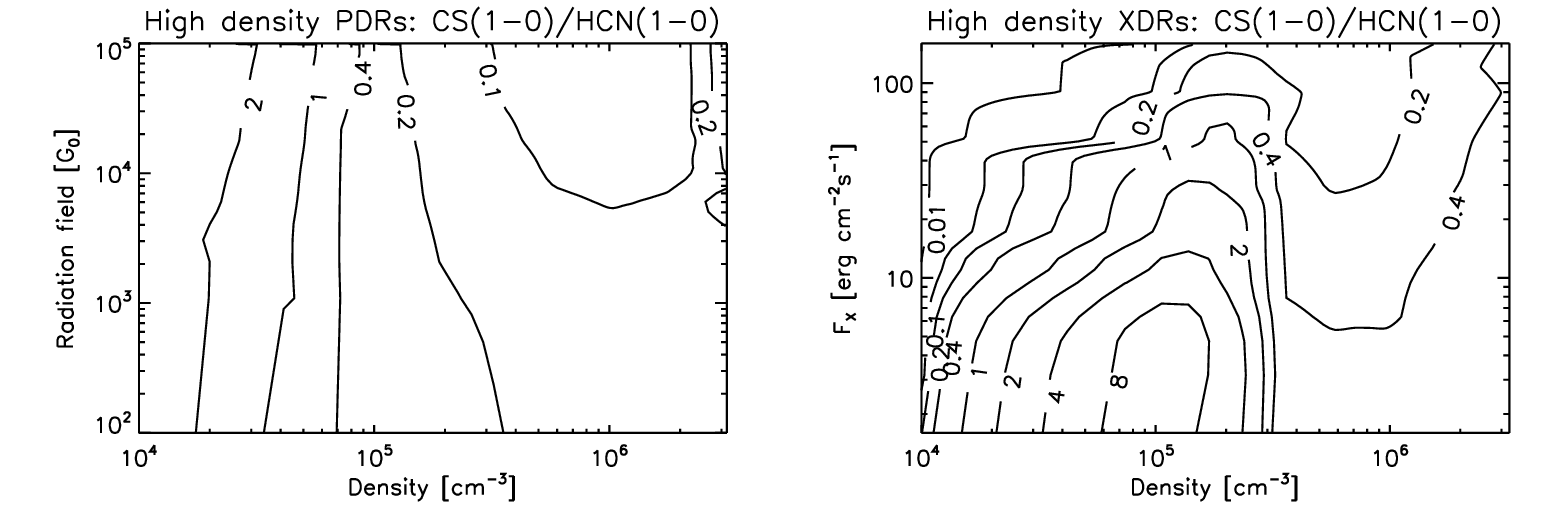}}
\centerline{\includegraphics[height=50mm,clip=]{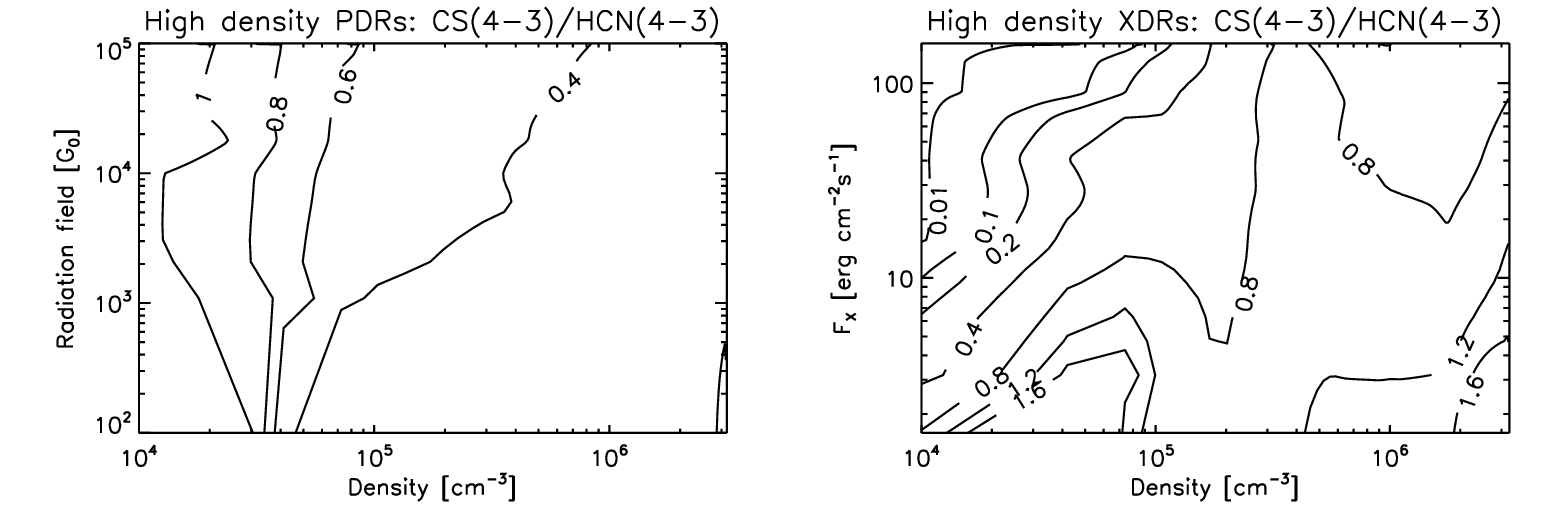}}
\caption{CS(1-0)/HCN and CS(4-3)/HCN(4-3) ratios for PDR (left) and XDR (right) models.}
\label{ratio_CS_HCN}
\end{figure*}

\subsection{SiO and CS}

Although SiO is usually considered to be a good tracer of shocks, we
do find that the SiO(1-0)/CO(1-0) ratio is typically larger in XDRs
than in PDRs by a factor of 2-3, for densities around $10^{5.5}$
cm$^{-3}$.  For the higher excitation J=4-3 lines, the effect
disappears because CO is generally warmer in XDRs compared to PDRs.

The CS(1-0)/HCN(1-0) ratio is a factor of two larger (smaller) in XDRs
for densities above (below) $10^5$ cm$^{-3}$. Interestingly, the
corresponding 4-3 ratio continues this trend but changes in the ratio
from $10^4$ to $10^6$ cm$^{-3}$ are now as large as a factor of 10.

\section{Column density ratios}

Unfortunately, for many molecular species of interest no reliable
collisional cross sections are available. For these species we are
unable to accurately predict line intensities, but we can still
calculate the column density ratios. In this section, we discuss
column density ratios for a number of species that are of potential
interest in attempts to discriminate between PDRs and XDRs.

\begin{figure*}[!ht]
\centerline{\includegraphics[height=50mm,clip=]{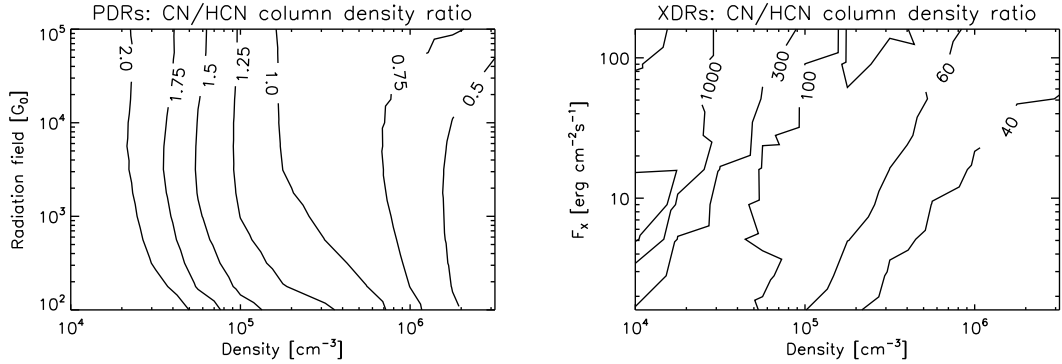}}
\caption{CN/HCN column density ratios for PDR (left) and XDR (right) models.}
\label{column_ratio_CN_HCN}
\end{figure*}


\subsection{CN/HCN column density ratio}

In Fig. \ref{column_ratio_CN_HCN}, we show the CN/HCN column density
ratios. There is an enormous difference between the ratios for PDRs
and XDRs. In the PDRs, the cloud type A ratios range from 0.5
($n\sim10^6$~cm$^{-3}$) to 2.0 ($n\sim10^4$~cm$^{-3}$), while in the
XDR models the same ratio varies from 40 ($n\sim10^6$~cm$^{-3}$) to
over a 1000 ($n\sim10^4$~cm$^{-3}$). We find higher CN/HCN ratios at
lower densities, where the chemical rates are lower, making it more
difficult to form large molecules. The PDR ratios are only dependent
on density, which is explained by the fact that most CN and HCN
molecules are formed beyond the H/H$_2$ transition. This part of the
cloud is shielded from FUV photons and the chemistry is dominated by
the cosmic ray ionization rate, which is the same in every model. In
the XDR models, there is much less variation in the CN and HCN
abundance throughout the cloud. The variations do not exceed more than
two to three orders of magnitude, while this is over ten orders of
magnitude in the PDR models. All parts of the cloud contribute almost
equally to the column density ratio, including the region with very
high $H_X/n$. $H_X/n$ is a major factor in the resulting ratio, and
therefore the XDRs also show a large dependence on incident radiation
field.

On the website, we show the PDR and XDR cumulative CN/HCN ratios for a
few specific densities and radiation fields. The variation in the
cumulative column density ratio is much less in the XDR models than in
the PDR models. At PDR edges, the gas is highly ionized (as in the
XDRs), and here we find ratios resembling those of XDRs. Abundances,
however, are very low here because of the high photo-dissociation
rate.

\subsection{CH/HCN column density ratios}

In Fig. \ref{column_ratio_CH_HCN}, we show the CH/HCN column density
ratio. The differences between the ratios in PDRs and XDRs are even
larger than in the case of CN/HCN. While the PDR ratios increase from
0.2 ($n\sim10^6$~cm$^{-3}$) to 0.9 ($n\sim10^4$~cm$^{-3}$), the XDR
ratios range from 20 to more than 10000. The PDR ratio does not depend
on density only. At relatively low densities
($n=1-3\times10^4$~cm$^{-3}$), we find a dependency on incident
radiation field as well. CH reaches its maximum abundance at greater
depths than HCN, which means that the cumulative ratio is still
increasing at large column densities. Our model cloud sizes are not
large enough to allow the ratio to converge to a constant ratio (see
website). For the XDR models, we can roughly state that the ratio
increases toward higher $H_X/n$. However, at $n\sim10^5$~cm$^{-3}$ the
lowest ratio is not found at the lowest incident radiation field
strength. At such densities, the limited cloud sizes are comparable to
the depth of the H/H$_2$ transition.  This transition is included in
models with the lowest radiation field strengths, but not in those
with the highest radiation field strengths.

\subsection{CH$^+$/HCN column density ratios}

In Fig. \ref{column_ratio_CHp_HCN}, we show the CH$^+$/HCN ratios for
both PDR and XDR models. They range from $10^{-6}$
($n\sim10^6$~cm$^{-3}$) to $6\times10^{-4}$ ($n\sim10^4$~cm$^{-3}$) in
the PDRs, and from $10^{-3}$ ($n\sim10^6$~cm$^{-3}$) to over 1000
($n\sim10^4$~cm$^{-3}$) in the XDRs. In the PDR models, the highest
CH$^+$ abundance is seen close to the edge of the cloud but it
decreases very quickly beyond the H/H$_2$ transition. HCN, on the
contrary, reaches its highest abundance beyond this transition. This
explains the decrease of the PDR cumulative column density ratios with
increasing depth (see website). The PDR models show a dependence on
both density and incident radiation field for this ratio. The decrease
in the ratio with increasing density is caused by the fact that on the
one hand CH$^+$ is more easily destroyed (due to higher recombination
rates) and on the other hand HCN more easily formed when densities are
higher. The ratios do not converge to a constant value deep into the
cloud. At identical densities, the ratio increases with increasing
radiation field, since the HCN column densities become smaller while
the CH$^+$ column densities become larger for the cloud-size
considered here. In the XDR models, the largest ratios are seen for
the highest $H_X/n$ (low density and high incident radiation
field). The fluctuations in the CH$^+$ and HCN abundances are more
gradual. The HCN abundance increases and the CH$^+$ abundance
decreases when the X-ray photons are gradually absorbed. Therefore,
the XDR cumulative column densities ratios show less variation than
the PDR ratios (see website).

\subsection{HCO/HCO$^+$ column density ratios}

Fig. \ref{column_ratio_HCO_HCOp} shows the HCO/HCO$^+$ column density
ratios. For the PDRs, we find much larger ratios than for the XDRs.
The PDRs show ratios between 0.1 ($n\sim10^4$~cm$^{-3}$) and 10.0
($n\sim10^6$~cm$^{-3}$), while the ratios in the XDR range from
$10^{-5}$ ($n\sim10^4$~cm$^{-3}$) to $10^{-3}$
($n\sim10^6$~cm$^{-3}$). In the PDRs, the ratios depend only on
density for radiation fields $G_0<10^4$. With larger incident
radiation fields, the ratio becomes dependent on the radiation field
strength as well. The HCO abundance reaches its maximum and more or
less constant abundance somewhat deeper into the cloud than HCO$^+$.
As column density increases, the HCO/HCO$^+$ ratio slowly converges to
a constant value (see website). For large radiation fields
($G_0>10^4$), the cloud size considered here is too small to allow
column densities to converge to a constant ratio.  In the XDR models,
the ratio depends on both density and radiation field in all
regimes. The lowest ratios are seen for high $H_X/n$
($n=10^4$~cm$^{-3}$ and $F_X=160$).

\subsection{HOC$^+$/HCO$^+$ column density ratios}

In Fig. \ref{column_ratio_HOCp_HCOp}, we show the HOC$^+$/HCO$^+$
column density ratios. The XDRs show larger ratios, ranging from
$10^{-4}$ ($n\sim10^6$~cm$^{-3}$) to 0.6 ($n\sim10^4$~cm$^{-3}$), than
PDRs, where the ratios range from $10^{-7}$ to $10^{-5}$. The XDR
ratios increase for larger $H_X/n$ (with maximum at
$n\sim10^4$~cm$^{-3}$ and $F_X=160$). The behavior is more complex for
the PDR models. The ratios depend on density only for $G_0>10^4$, but
below this value there is also a dependency on radiation field
strength. This is explained by the fact that ratios drop very fast,
when the gas becomes molecular (see website), but abundances are still
significant. A large fraction of the gas at the edge of the cloud is
molecular for densities $n>10^{5.5}$~cm$^{-3}$ and radiation fields
$G_0<10^4$. Here we find a fast decrease in the ratio for lower
radiation fields.

\subsection{NO/CO column density ratios}

In Fig \ref{column_ratio_NO_CO}, we show the NO/CO column density
ratios. The ratios in XDRs are much larger, $10^{-4}$
($n\sim10^6$~cm$^{-3}$) to $10^{-3}$ ($n\sim10^4$~cm$^{-3}$), than for
PDRs ($10^{-6}-10^{-5}$). In the PDR the ratios are largely determined
in the molecular part of the cloud and depend more or less on density
only. The ratios in the XDR depend mostly on $H_X/n$. The largest
ratios are seen for the largest $H_X/n$ ($n\sim10^4$~cm$^{-3}$ and
$F_X=160$).

CO(1-0) lines have optical depths of $\tau({\rm CO(1-0)})\sim
100$. Therefore, it could be possible to observe NO(1-0)/CO(1-0) line
intensity ratios as large as 0.1, while the maximum column density
ratios are only $10^{-3}$.

\subsection{N$_2$H$^+$/CO column density ratios}

In Fig. \ref{column_ratio_N2Hp_CO}, we show the N$_2$H$^+$/CO column
density ratio. In both PDRs and XDRs the model ratios are rather low.
They decrease with density in the PDR ($10^{-7}-10^{-9}$), which is
opposite to the behavior in XDRs. In the XDRs, we find a decrease
with increasing $H_X/n$. There are collisional data available for
N$_2$H$^+$, but since the abundances are so low, the line intensities
are too small to be observable.

\section{Summary and outlook}

We have presented a large set of PDR and XDR models that can be used
to determine the physical conditions that pertain to irradiated gas
clouds. This grid spans a large range in densities ($n_{\rm
H}=10^2-10^{6.5}$~cm$^{-3}$), irradiation ($G_0=10^{0.5}-10^5$ and
$F_X=1.6\times10^{-2}-160$~erg~cm$^{-2}$~s$^{-1}$) and column
densities ($N_{\rm H}=1.5\times10^{22}-1\times10^{25}$~cm$^{-2}$). We
have used the results to make predictions for the intensities and
ratios of the most important atomic fine-structure lines, e.g., [CII],
[OI], [CI], [SiII], and [FeII], rotational lines for molecular species
such as HCO$^+$, HCN, HNC, CS and SiO (up to $J=4$), CO and $^{13}$CO
up to $J=16$, and for column densities for CN, CH, CH$^+$, HCO,
HOC$^+$, NO, and N$_2$H$^+$. It is not possible to to present
all the results, but they are available on-line at the following URL:
{\small \tt http://www.strw.leidenuniv.nl/$\sim$meijerin/grid/}. Here
we summarize the most important conclusions:

\begin{enumerate}

\item The surface temperatures are higher (lower) in PDRs compared to
XDRs for densities $n>10^4$~cm$^{-3}$ ($n<10^4$~cm$^{-3}$). Two
opposing effects play a major role in determining the resulting
surface temperature: (1) The heating efficiency, which is much higher
in XDRs (up to 70 percent) than in PDRs (0.5-3.0 percent); (2) The
absorption cross sections which are much smaller for X-rays than for
FUV photons.

\item For the atomic lines, we find that the fine-structure line
ratios of [SiII] 35~$\mu$m/[CII] 158~$\mu$m,
[OI] 63~$\mu$m/[CII] 158~$\mu$m, [FeII] 26~$\mu$m/[CII] 158~$\mu$m, and
[CI] 369~$\mu$m/[CI] 609~$\mu$m are higher in XDRs than in PDRs, for a
given density, column, and irradiation strength. Whereas PDR ratios
depend on density and irradiation strength only, XDR depend on column
density as well. In PDRs, fine-structure line emission is only
produced at the edge of the cloud, while in XDRs almost all parts of
the cloud contribute.

\item We find higher CO line ratios for XDRs. In PDRs, CO is formed
beyond the H/H$_2$ transition and typically has temperatures in the
range $T\sim20-50$~K. In XDRs, CO is present throughout the cloud in
significant abundances, even in the highly ionized part. When using CO
line ratios, the best way to distinguish between PDRs and XDRs is to
consider ratios such as CO(16-15)/CO(1-0), where the differences are
largest (Fig. \ref{high_J_CO_ratio}).

\item HCN/HCO$^+$ ratios discriminate well between PDRs and XDRs, even
in the lower rotational lines, and especially when densities are as
high as $n>10^5$~cm$^{-3}$. At such densities, the
HCN(1-0)/HCO$^+$(1-0) line intensity ratios are $<1$ in XDRs, while
PDRs have ratios $>1$ for column densities $N_{\rm
H}>10^{23}$~cm$^{-2}$. Although the HCN/HCO$^+$ line ratio in an XDR
may become even larger in clouds of relatively modest density
($10{^4}$ cm$^{-3}$) subjected to high radiation field strengths
($>100$~erg~s$^{-1}$~cm$^{-2}$), we find that the line intensities in
this part of parameter space are too low to be detectable (see also
Meijerink et al.\ \citeyear{Meijerink2006}).

\item For densities between $10^4$ and $10^{6.5}$~cm$^{-3}$,
HNC(1-0)/HCN(1-0) ratios in PDRs are of order 1 ($<1$) for columns
larger (smaller) than $10^{22}$ cm$^{-2}$, while the ratios range
between 0.2-1.2 for XDRs. PDR HNC(4-3)/HCN(4-3) ratios never exceed
unity, while we find XDR ratios up to 1.6 for densities
$n>10^6$~cm$^{-3}$.

\item HCN/CO ratios are typically smaller for in XDRs than in PDRs,
for two reasons: (1) The HCN abundance is boosted only in high
(column) density gas, with columns in excess of $10^{23}$~cm$^{-2}$
and densities larger than $10^4$~cm$^{-2}$; (2) CO is warmer in XDRs,
which leads to stronger emission, and this suppresses the HCN/CO
ratios as well. In PDRs, the very high HCN(1-0)/CO(1-0) ratios of
order $>1$ are only obtained at very high column densities ($N_{\rm
H}>10^{23}$~cm$^{-2}$). High column densities $N_H$ are needed to
obtain a large HCN column density and to obtain optically thick HCN
lines. This happens much faster for CO, due to its higher fractional
abundance.

\item We find that CN/HCN, NO/CO, and HOC$^+$/HCO$^+$ column density
ratios are discriminant between PDRs and XDRs. Molecules such as CH
and CH$^+$ also look very promising. However, we urgently need
reliable collisional cross section data in order to make predictions
for observed line ratios.

\end{enumerate}

\noindent
We conclude that both atomic fine-structure and molecular rotational
lines have significant diagnostic value to allow us to distinguish
between clouds irradiated by star-bursts (FUV) and by active galactic
nuclei (X-ray) in the centers of galaxies. Note, however, that the
line intensities (and ratios) presented in this work do not take the,
likely complex, kinematics of nuclear gas into account. For example,
gas rotating in an accretion disk around an AGN will cause the line
widths of, say, HCN, HNC and HCO$^+$ transitions to differ, depending
on where their chemical abundances peak with depth.

The XDR/AGN contribution will typically be of a much smaller (possibly
beam diluted) angular scale than that of a PDR/Starburst. A 10-25\%
PDR contribution may already suppress our ability to recognize XDR
excitation from HCN/HCO$^+$ and HNC/HCN line ratios. A solution to
this can be found in the very high J CO lines (e.g., CO(16-15)), that
are excellent indicators of an XDR contribution. These very high
rotational lines will be observable with the ESA Herschel (HIFI) space
observatory scheduled for launch in the near future; they can be seen
in absorption in the near-infrared with Subaru. Currently available
(sub)millimeter facilities lack the spatial resolution to separate the
PDR/stellar and XDR/AGN contributions in distant active galaxies. Up
to now, these components can only be spatially resolved in the Milky
Way. However, the resolving power of ALMA will bring this possibility
within reach for external galaxy centers as well.

\begin{acknowledgements} 
We thank Dieter Poelman for making his radiation transfer code {\it
beta3D} available to us.
\end{acknowledgements}


\begin{appendix}

\section{On-line figures}



\begin{figure*}[!ht]
\centerline{\includegraphics[height=150mm,clip=]{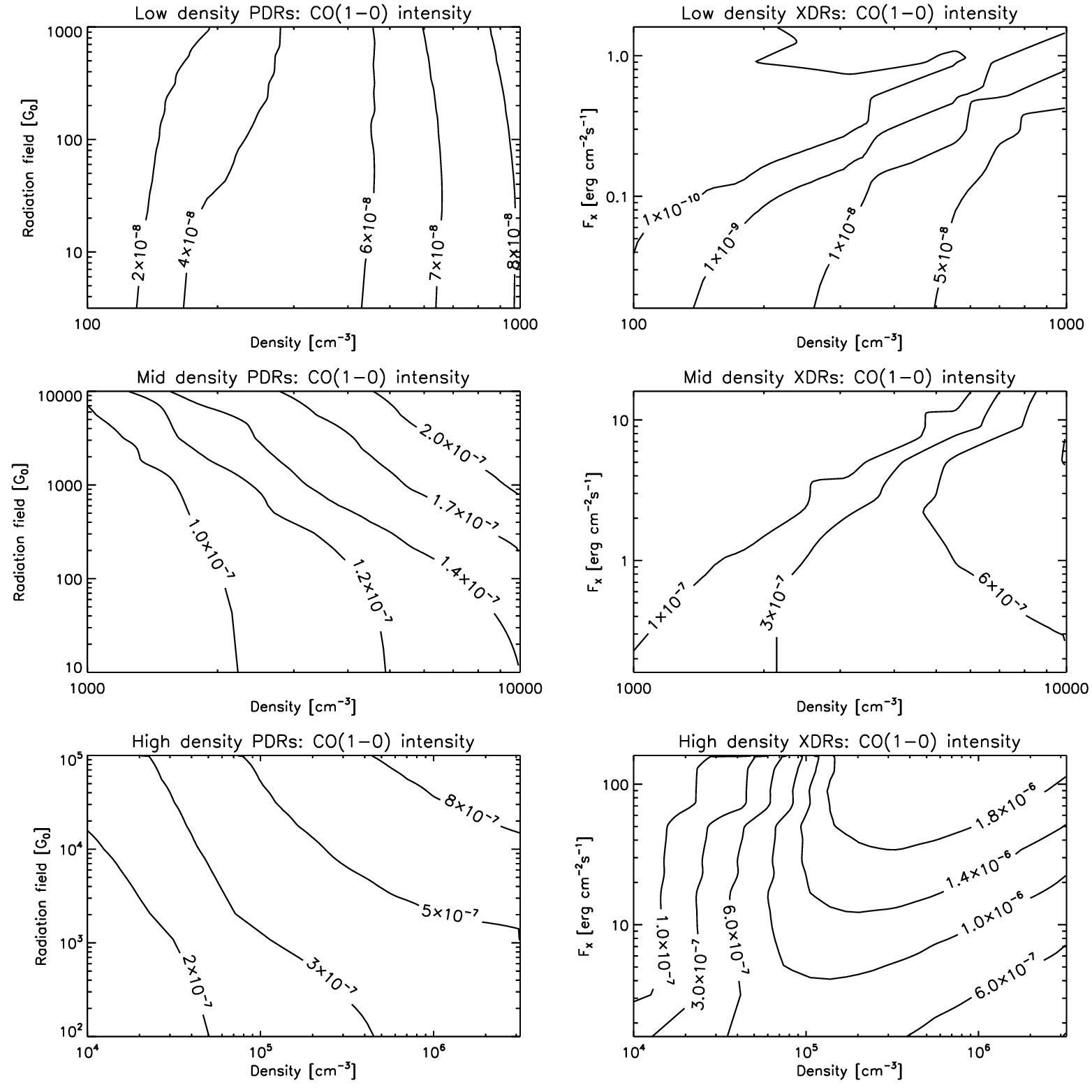}}
\caption{CO(1-0) intensity for PDR (left) and XDR (right) models.}
\label{intensity_CO10}
\end{figure*}

\begin{figure*}[!ht]
\centerline{\includegraphics[height=150mm,clip=]{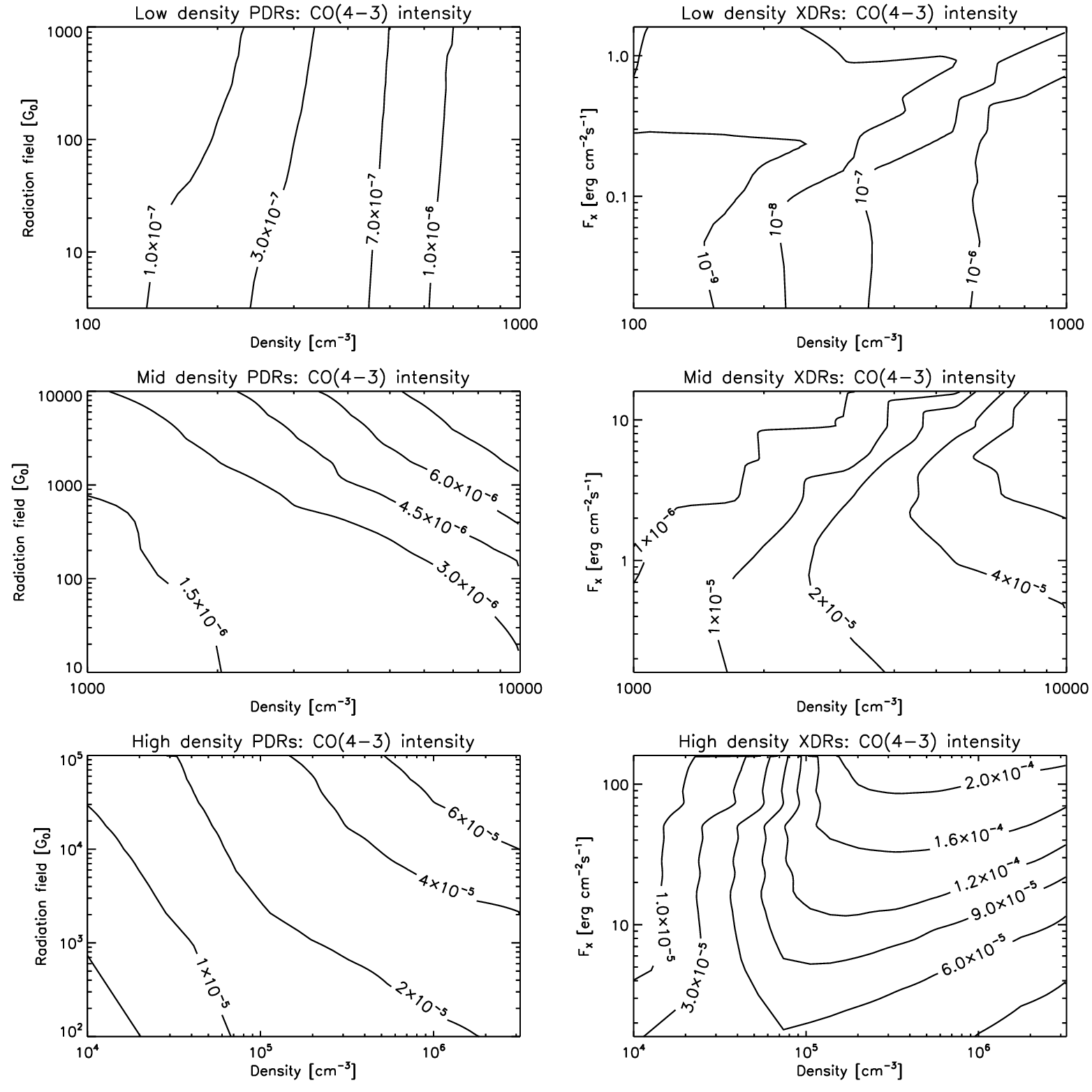}}
\caption{CO(4-3) intensity for PDR (left) and XDR (right) models.}
\label{intensity_CO43}
\end{figure*}

\begin{figure*}[!ht]
\centerline{\includegraphics[height=50mm,clip=]{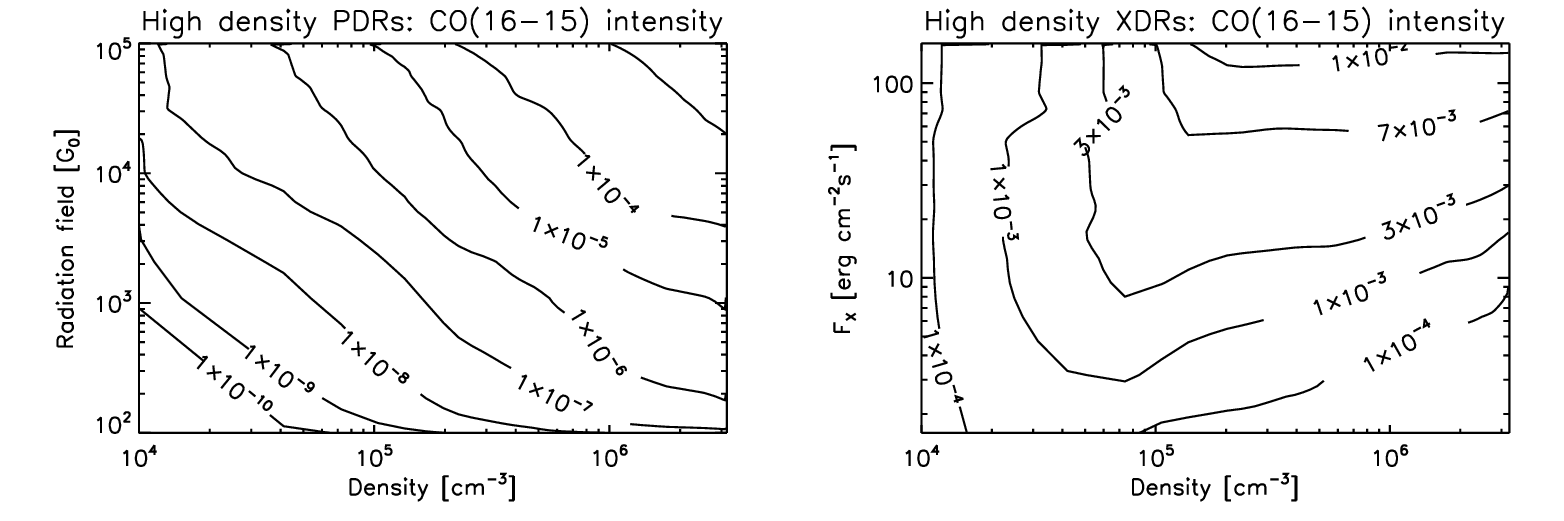}}
\centerline{\includegraphics[height=50mm,clip=]{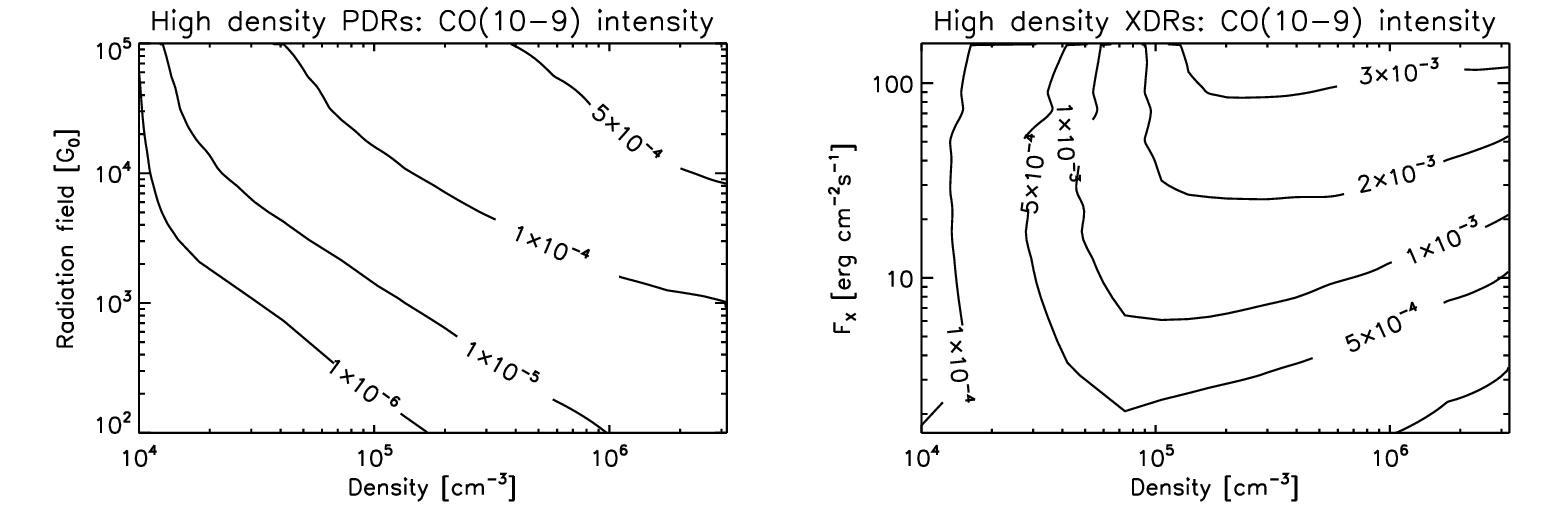}}
\centerline{\includegraphics[height=50mm,clip=]{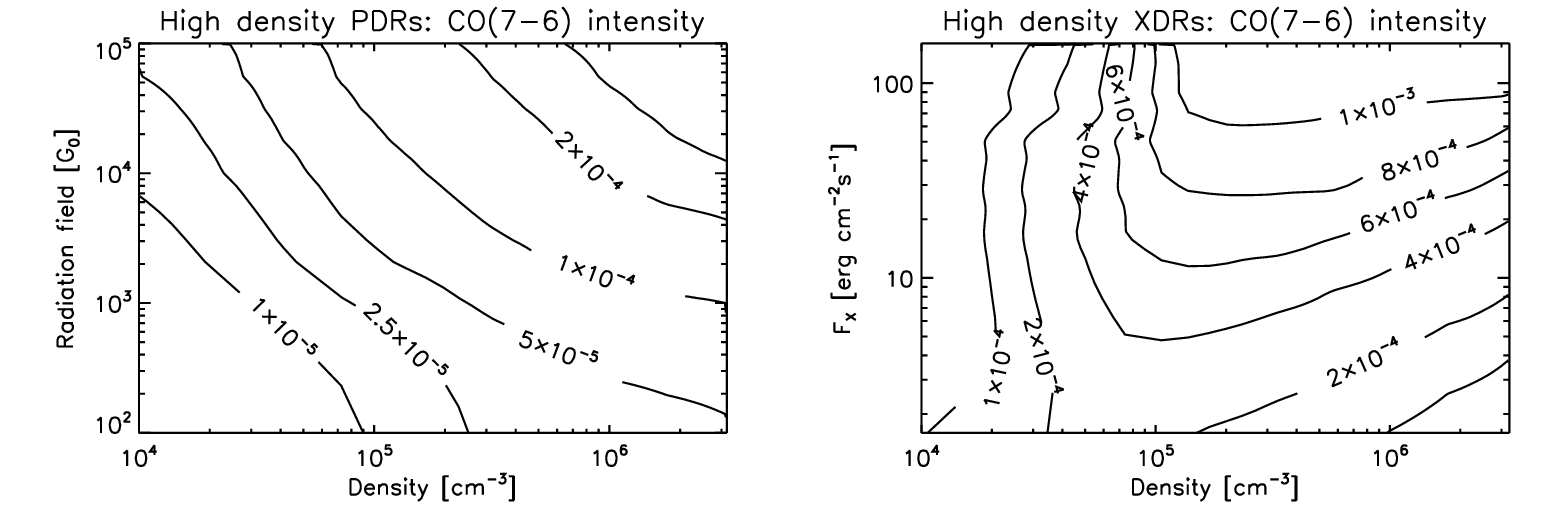}}
\caption{CO(16-15), CO(10-9), and CO(7-6) intensity for PDR (left) and XDR (right) models.}
\label{high_J_CO_intens}
\end{figure*}

\begin{figure*}[!ht]
\centerline{\includegraphics[height=150mm,clip=]{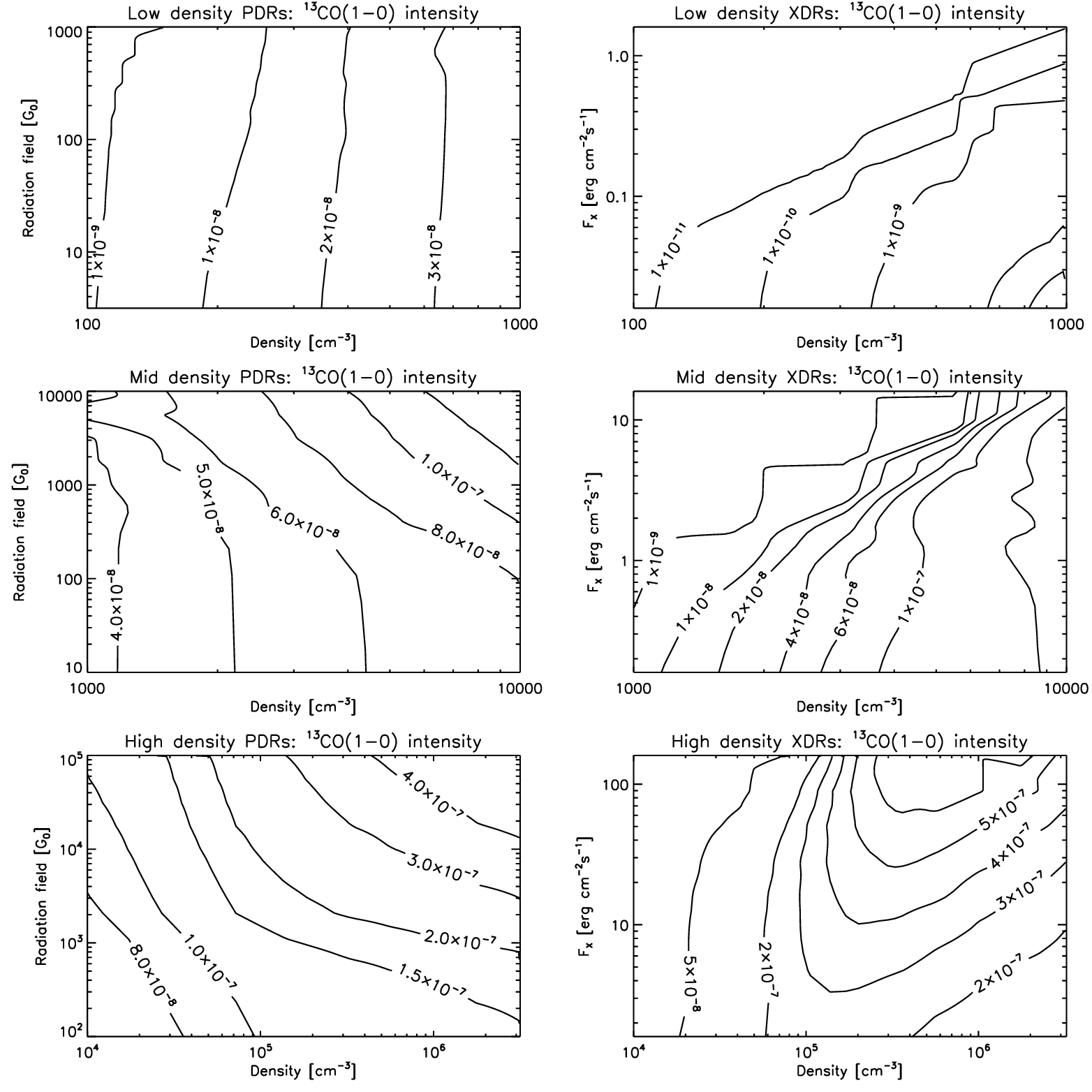}}
\caption{$^{13}$CO(1-0) intensity for PDR (left) and XDR (right) models.}
\label{intensity_13CO10}
\end{figure*}

\begin{figure*}[!ht]
\centerline{\includegraphics[height=150mm,clip=]{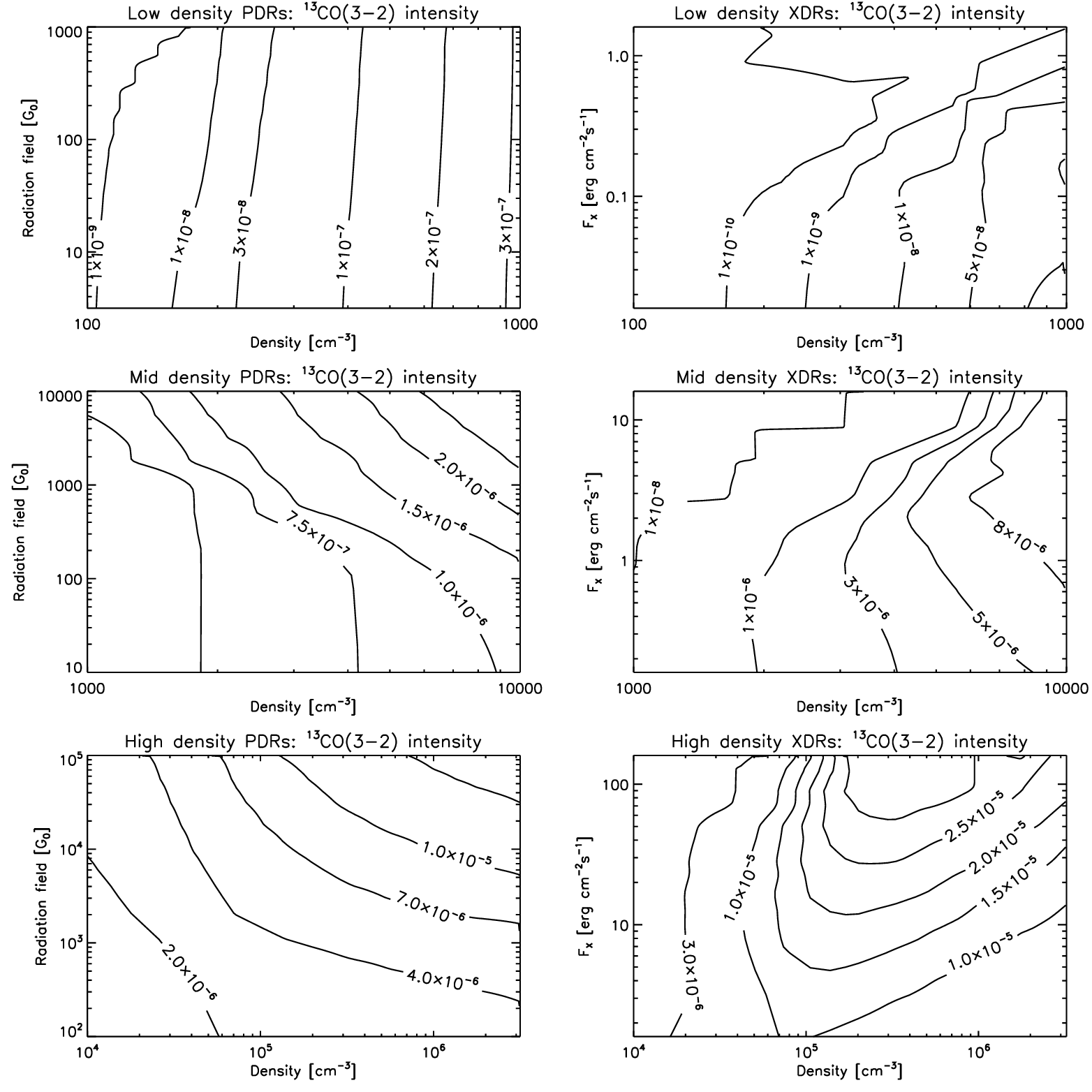}}
\caption{$^{13}$CO(3-2) intensity for PDR (left) and XDR (right) models.}
\label{intensity_13CO32}
\end{figure*}

\begin{figure*}[!ht]
\centerline{\includegraphics[height=150mm,clip=]{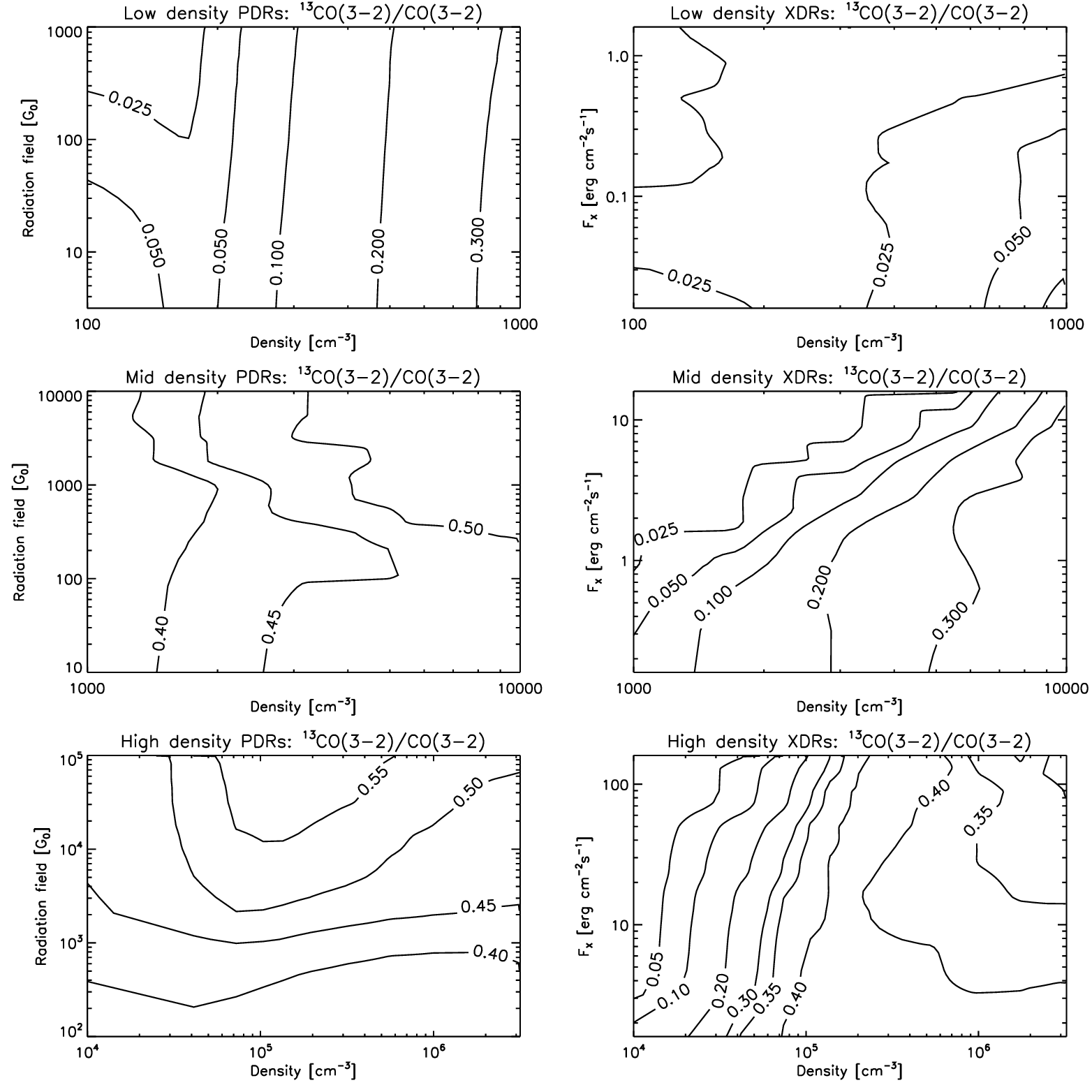}}
\caption{$^{13}$CO(3-2)/CO(3-2) ratio for PDR (left) and XDR (right) models.}
\label{ratio_13CO32_CO32}
\end{figure*}

\begin{figure*}[!ht]
\centerline{\includegraphics[height=50mm,clip=]{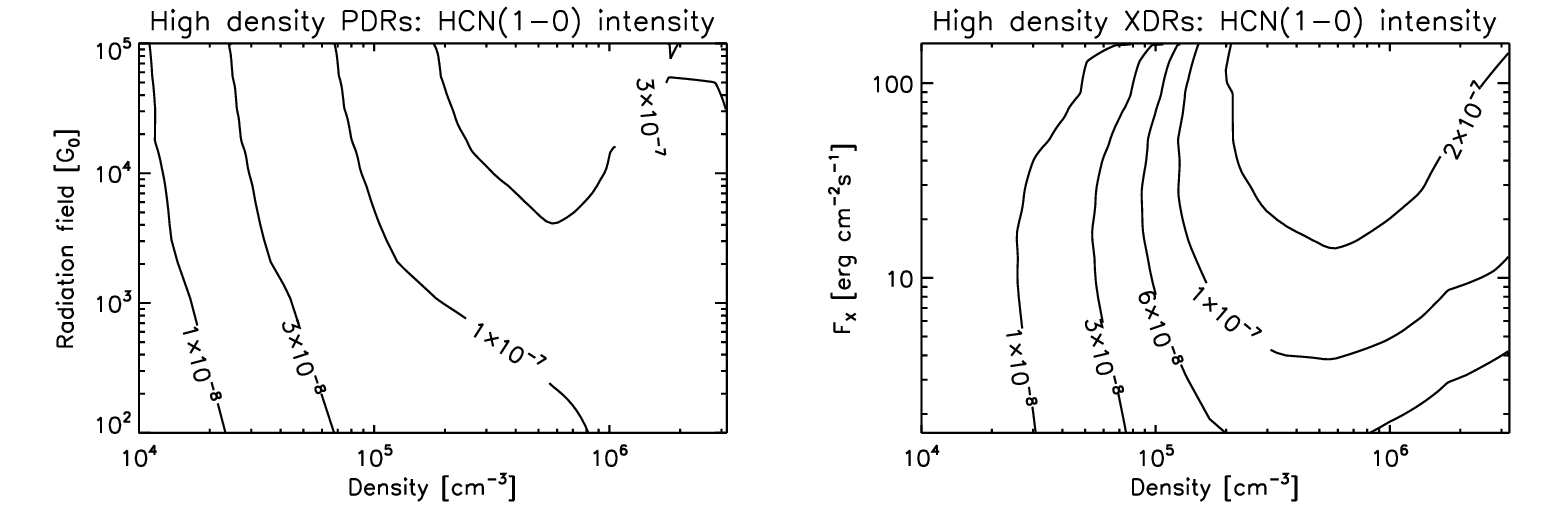}}
\centerline{\includegraphics[height=50mm,clip=]{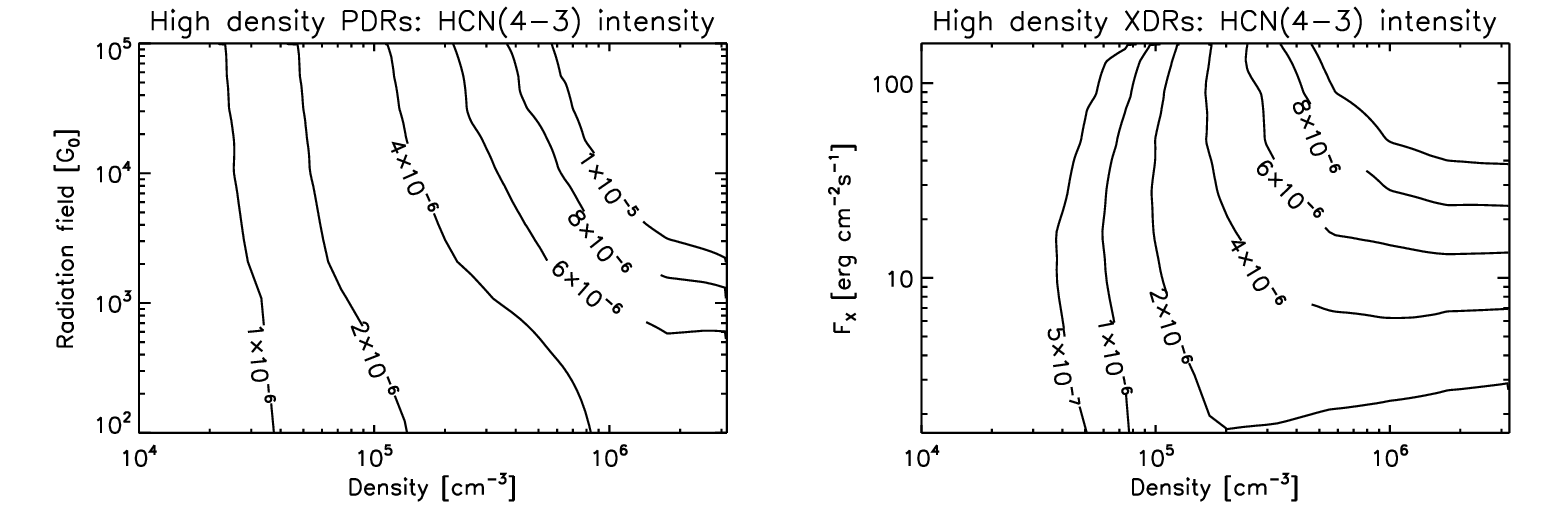}}
\caption{HCN(1-0) and HCN(4-3) intensity for PDR (left) and XDR (right) models.}
\label{intensity_HCN}
\end{figure*}

\begin{figure*}[!ht]
\centering
\unitlength1cm
\begin{minipage}[b]{7cm}
\resizebox{7cm}{!}{\includegraphics*[angle=0]{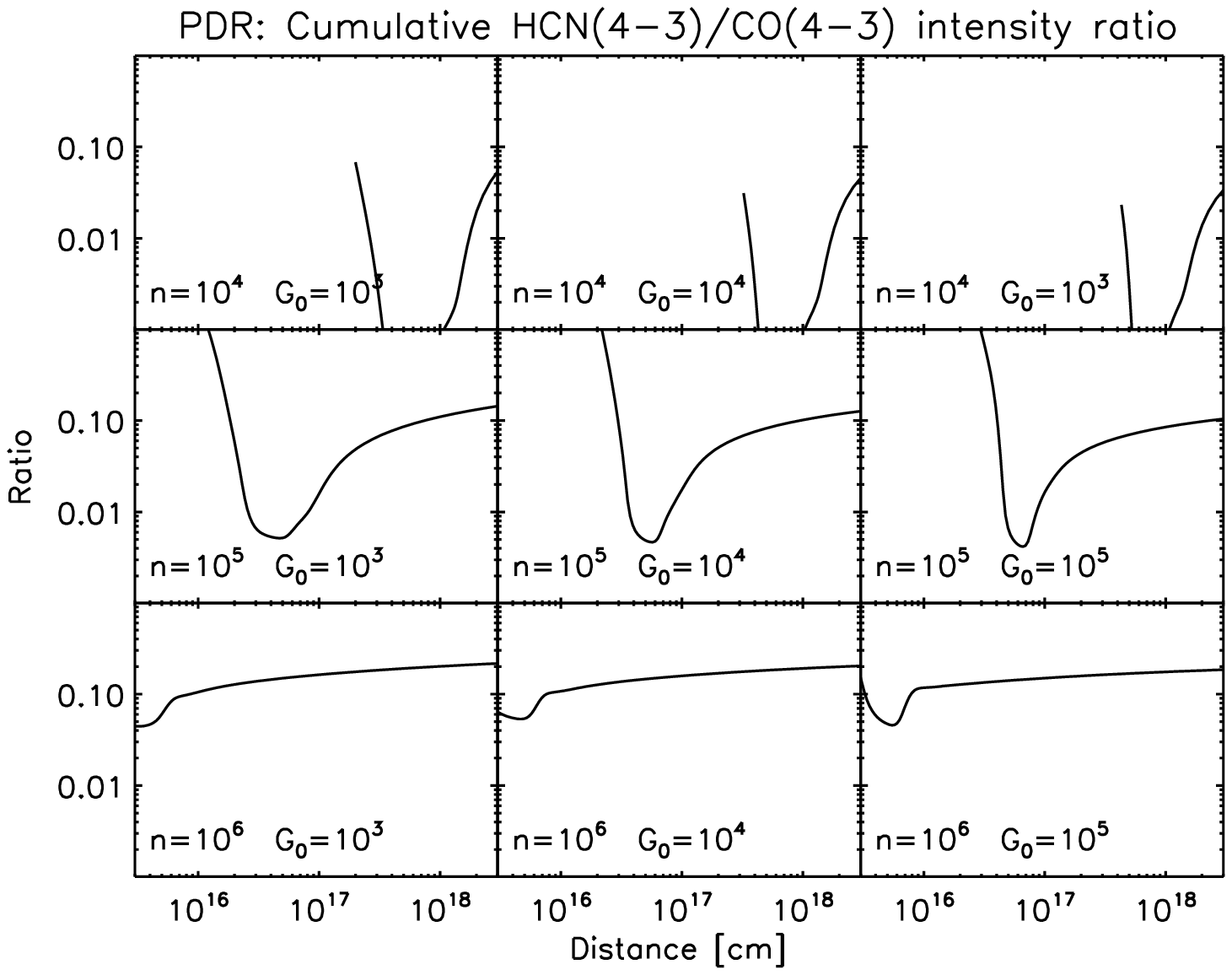}}
\end{minipage}
\begin{minipage}[b]{7cm}
\resizebox{7cm}{!}{\includegraphics*[angle=0]{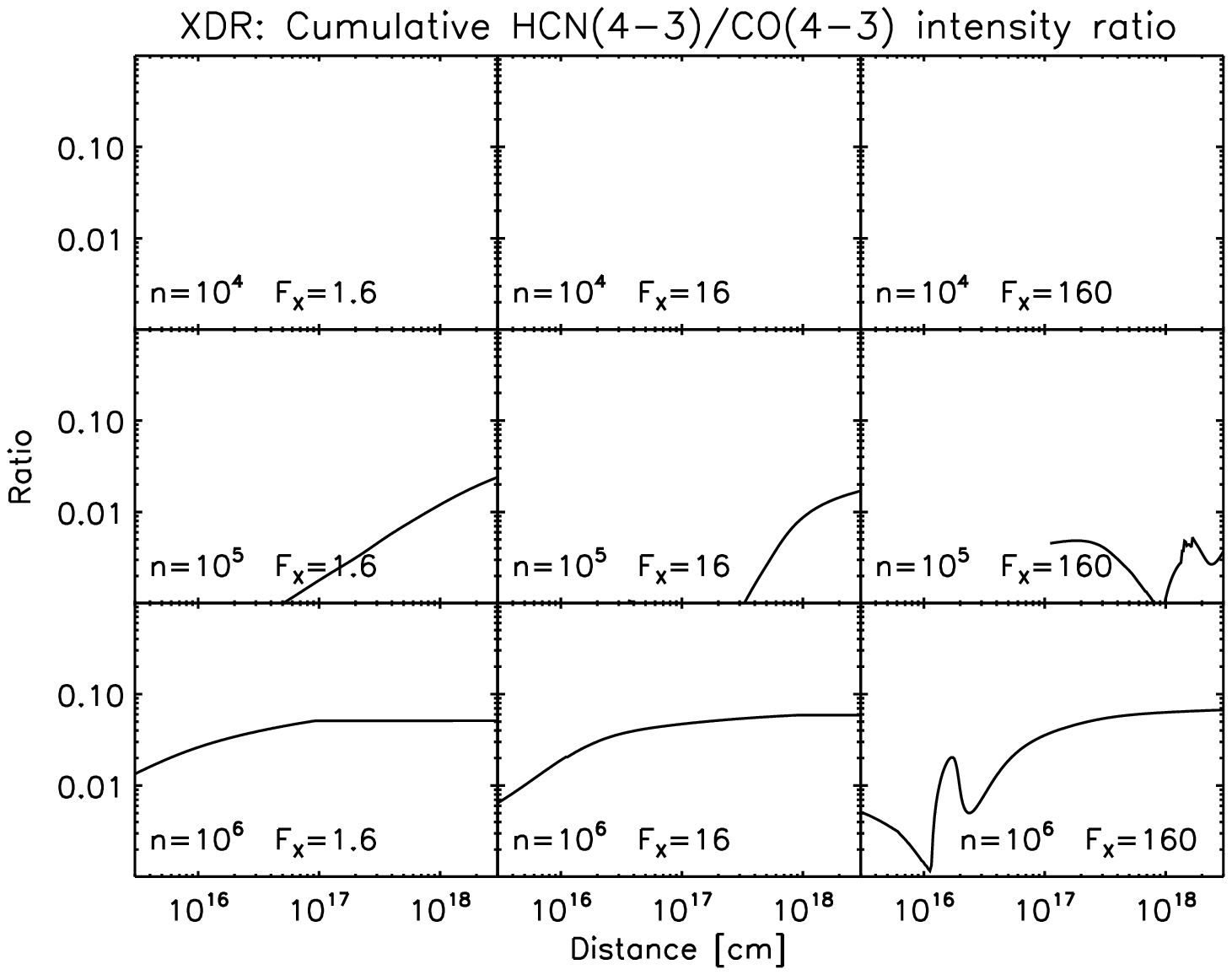}}
\end{minipage}
\caption[] {Cumulative HCN(4-3)/CO(4-3) line intensity ratios for PDR
(left) and XDR(right).}
\label{Cum_dens_HCN43_CO43}
\end{figure*}

\begin{figure*}[!ht]
\centerline{\includegraphics[height=50mm,clip=]{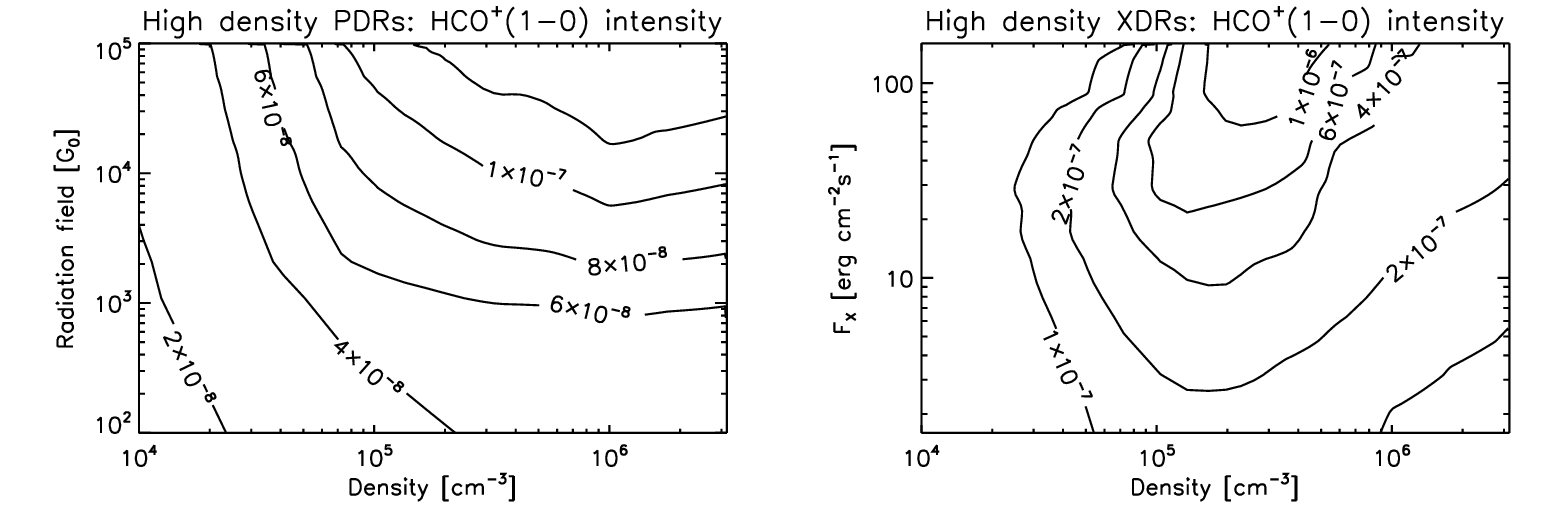}}
\centerline{\includegraphics[height=50mm,clip=]{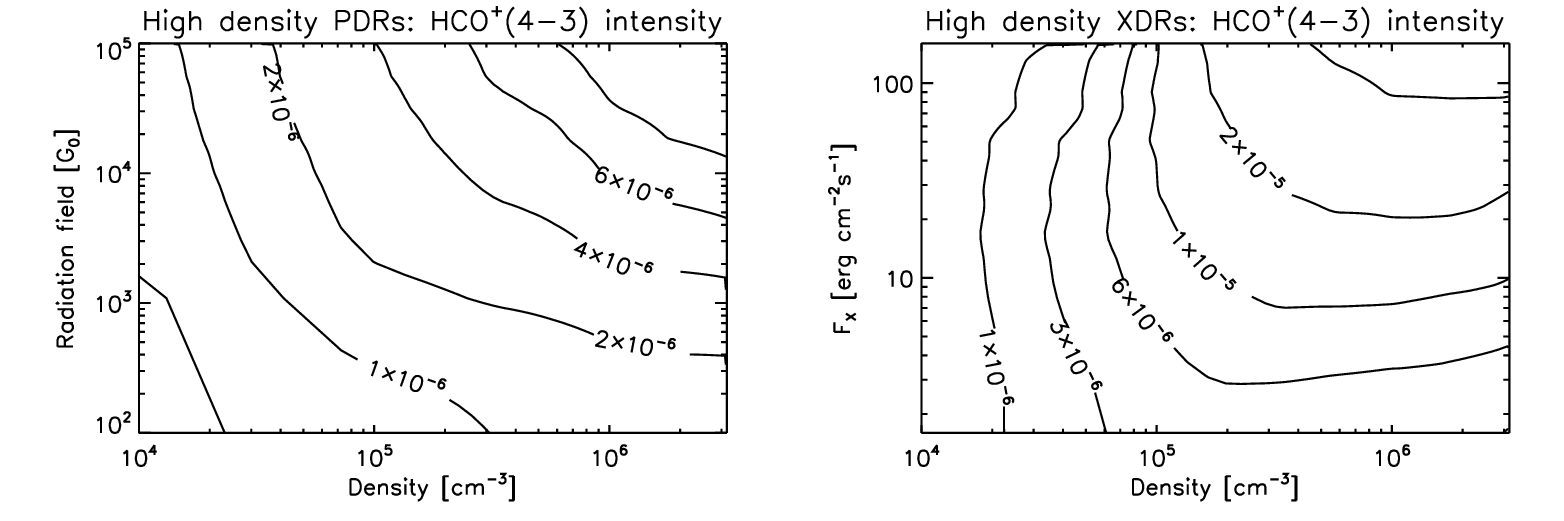}}
\caption{HCO$^+$(1-0) and HCO$^+$(4-3) intensity for PDR (left) and XDR (right) models.}
\label{intensity_HCOp}
\end{figure*}

\begin{figure*}[!ht]
\centering
\unitlength1cm
\begin{minipage}[b]{7cm}
\resizebox{7cm}{!}{\includegraphics*[angle=0]{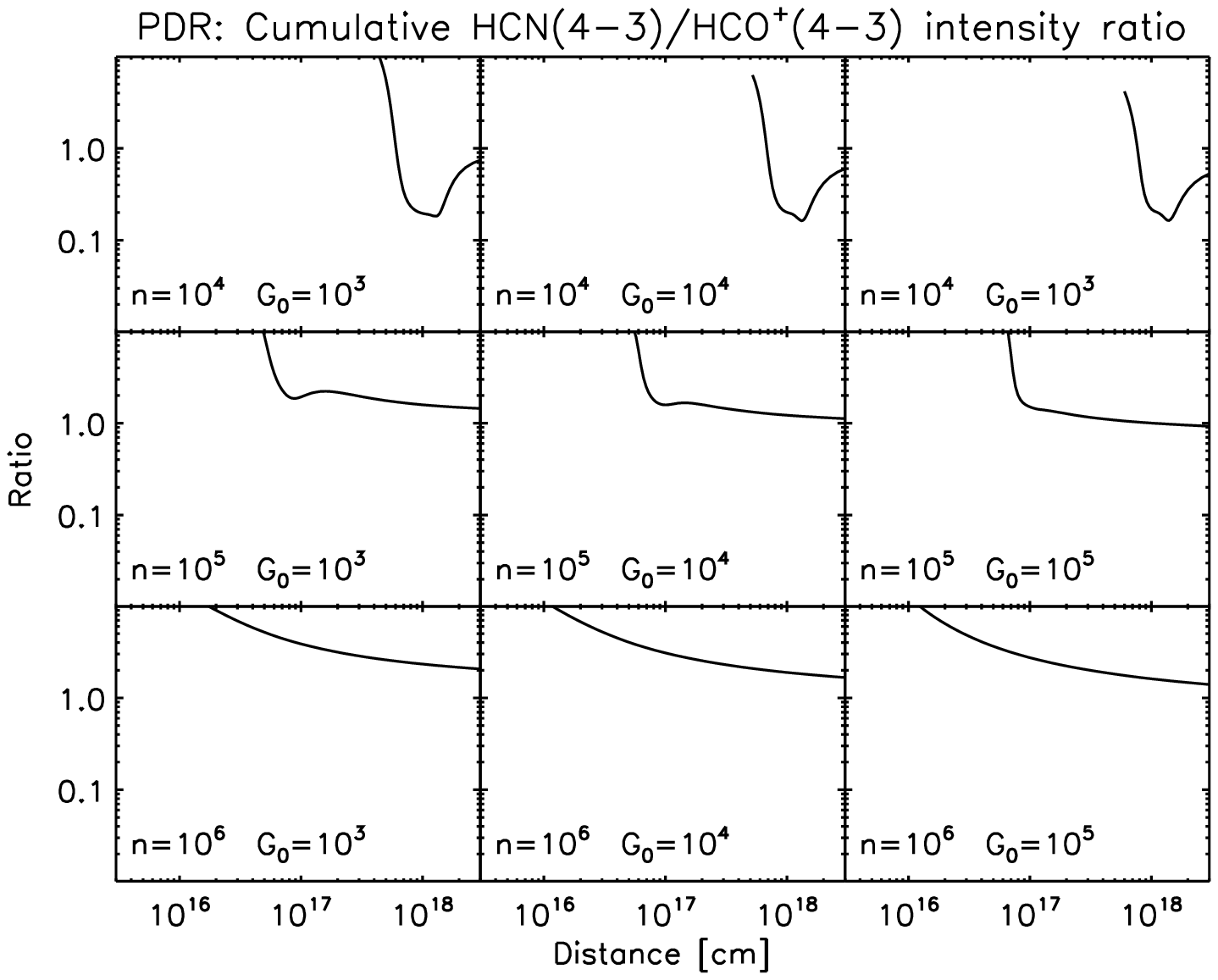}}
\end{minipage}
\begin{minipage}[b]{7cm}
\resizebox{7cm}{!}{\includegraphics*[angle=0]{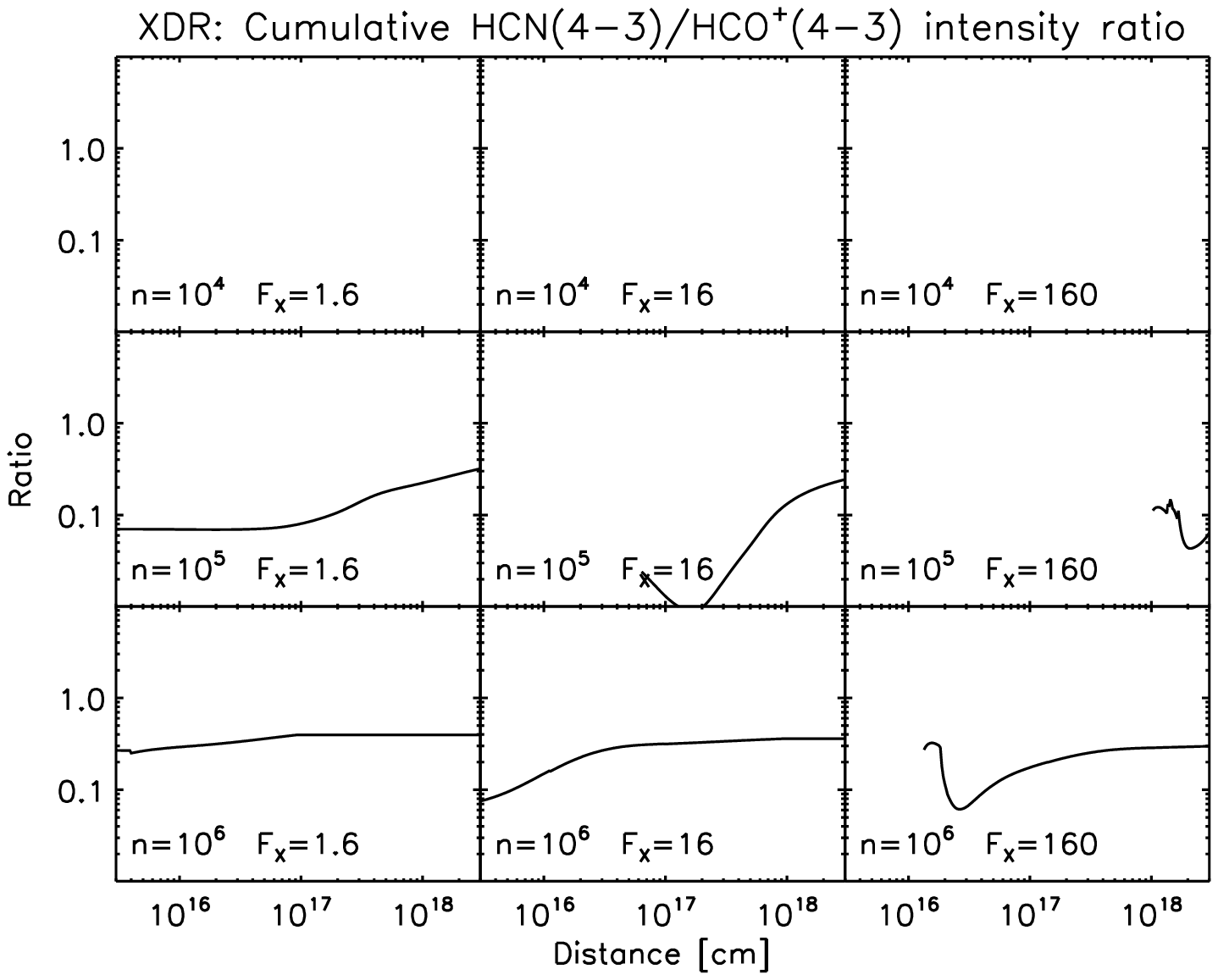}}
\end{minipage}
\caption[] {Cumulative HCN(4-3)/HCO$^+$(4-3) line intensity ratios for PDR
(left) and XDR(right).}
\label{cumul_ratio_HCN43_HCOp43}
\end{figure*}

\begin{figure*}[!ht]
\centering
\unitlength1cm
\begin{minipage}[b]{7cm}
\resizebox{7cm}{!}{\includegraphics*[angle=0]{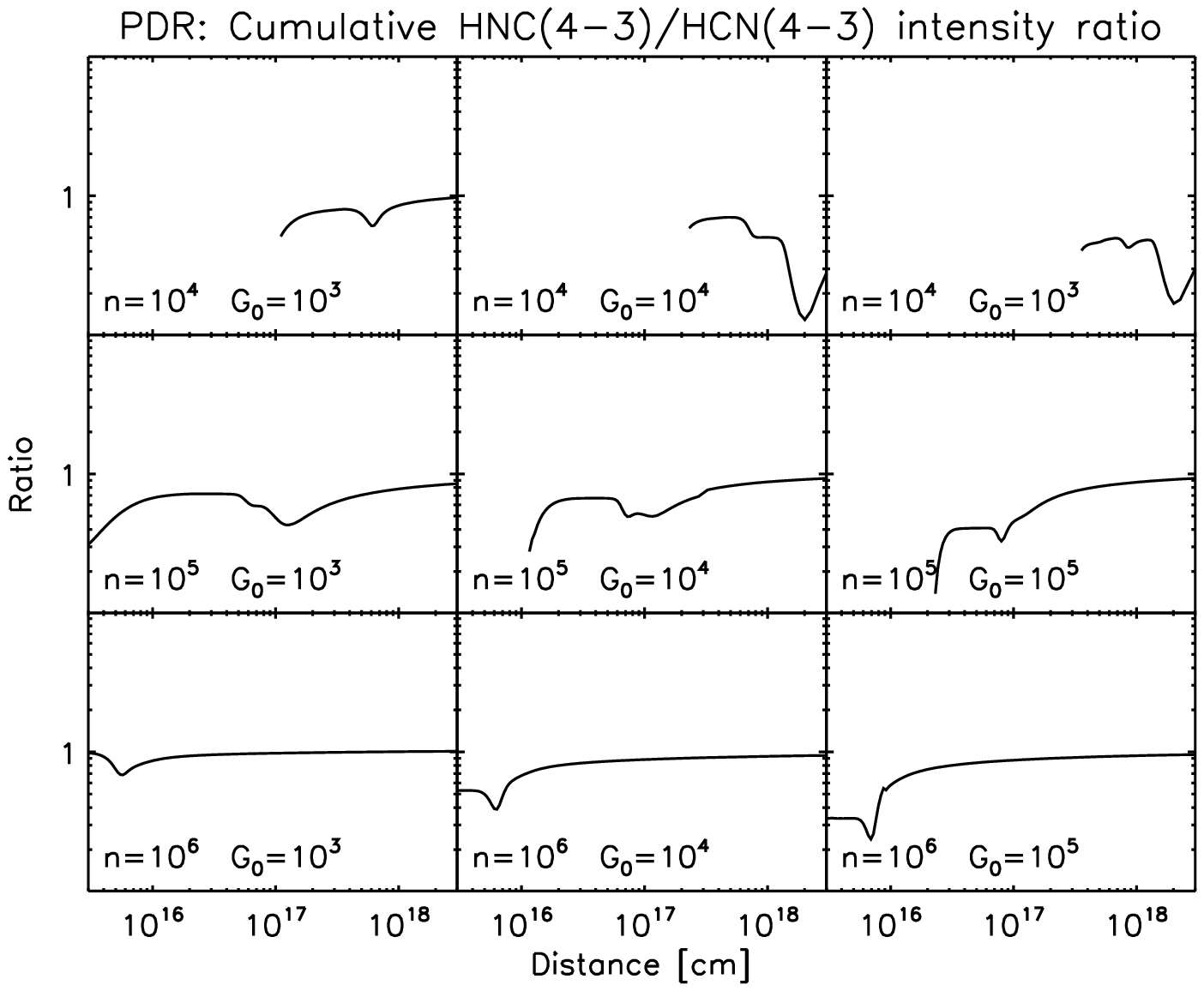}}
\end{minipage}
\begin{minipage}[b]{7cm}
\resizebox{7cm}{!}{\includegraphics*[angle=0]{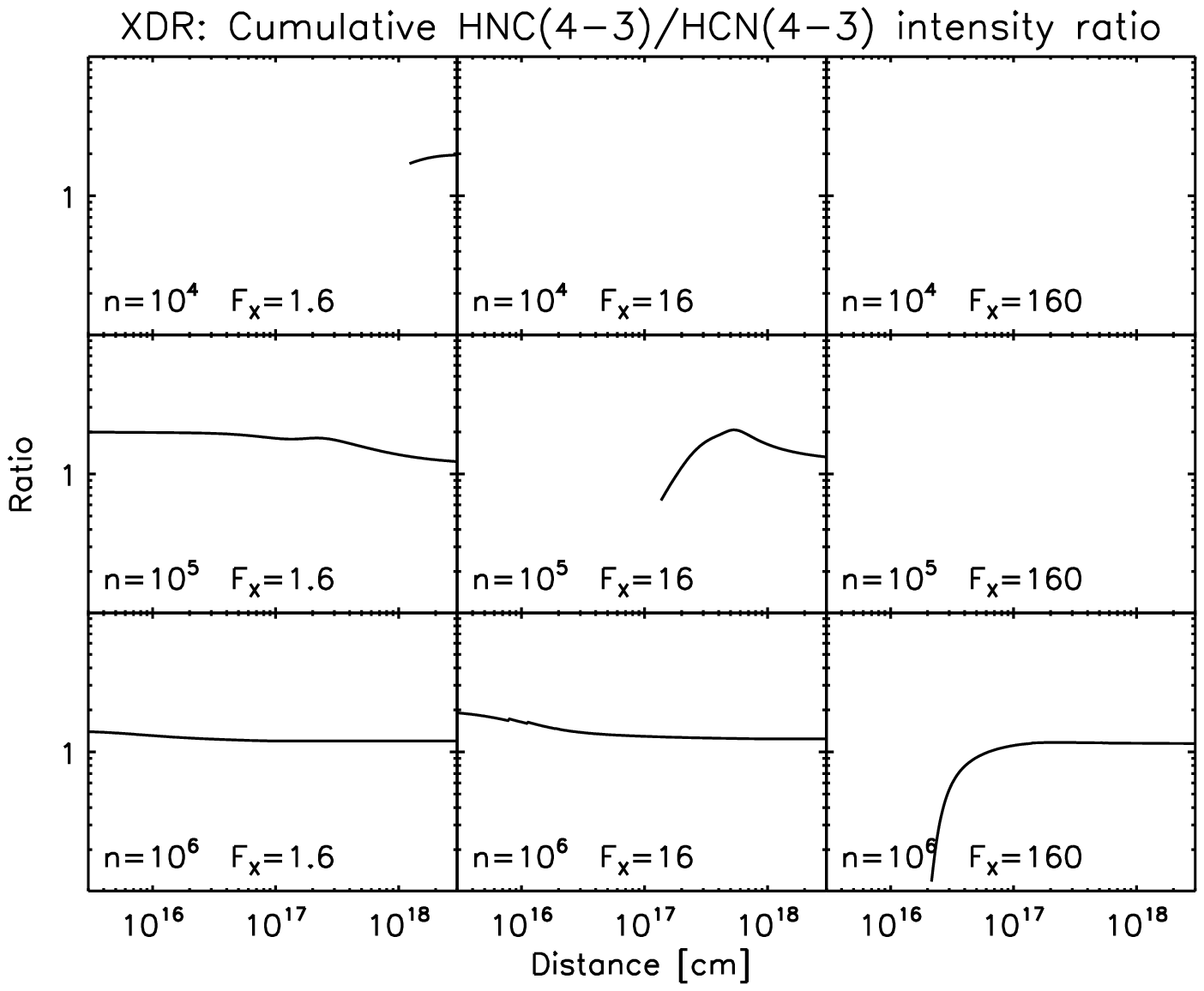}}
\end{minipage}
\caption[] {Cumulative HNC(4-3)/HCN(4-3) line intensity ratios for PDR
(left) and XDR(right).}
\label{cumul_ratio_HNC43_HCN43}
\end{figure*}

\begin{figure*}[!ht]
\centerline{\includegraphics[height=50mm,clip=]{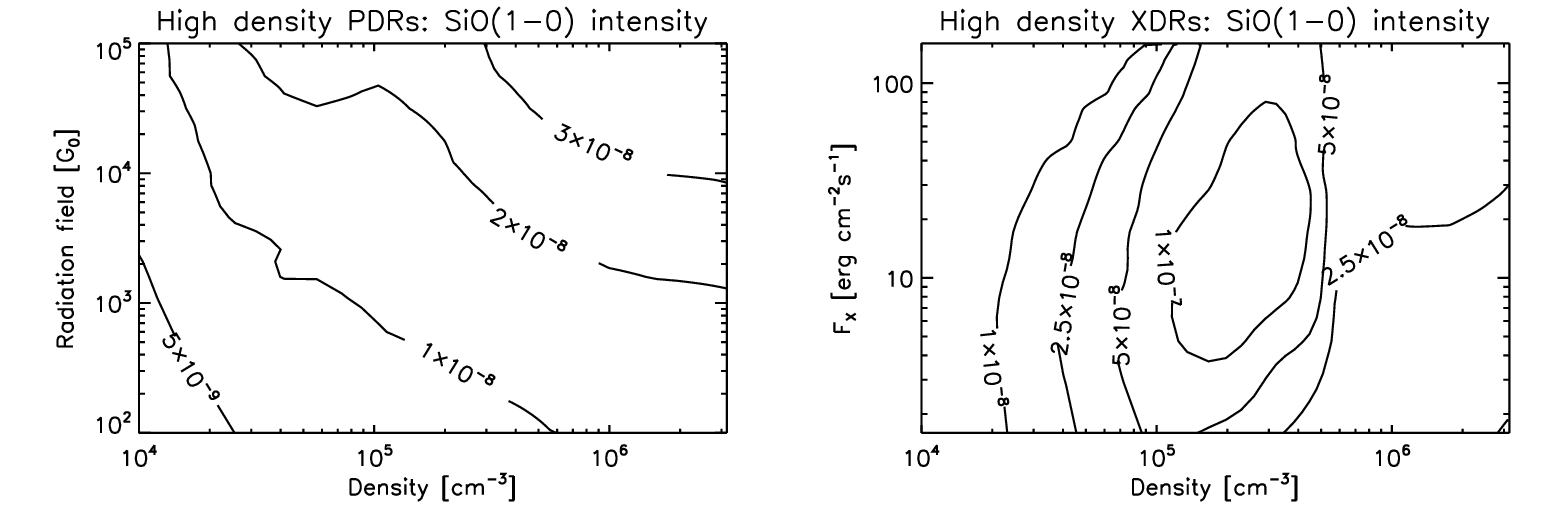}}
\centerline{\includegraphics[height=50mm,clip=]{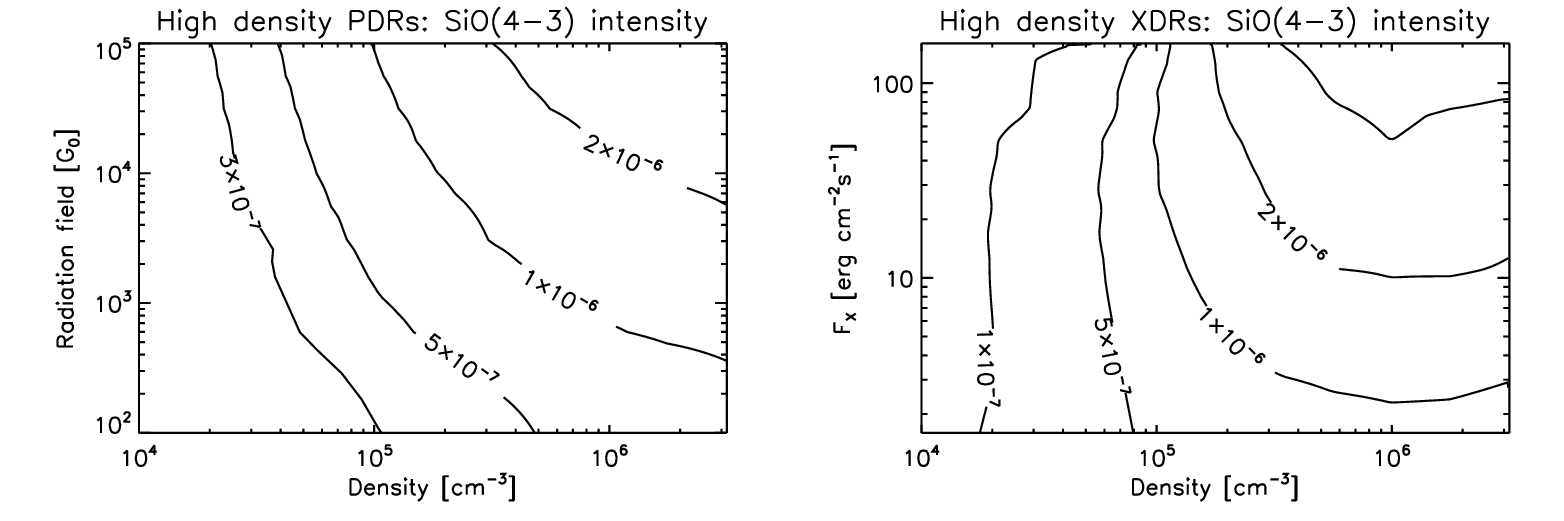}}
\caption{SiO(1-0) and SiO(4-3) intensity for PDR (left) and XDR (right) models.}
\label{intensity_SiO}
\end{figure*}

\begin{figure*}[!ht]
\centerline{\includegraphics[height=50mm,clip=]{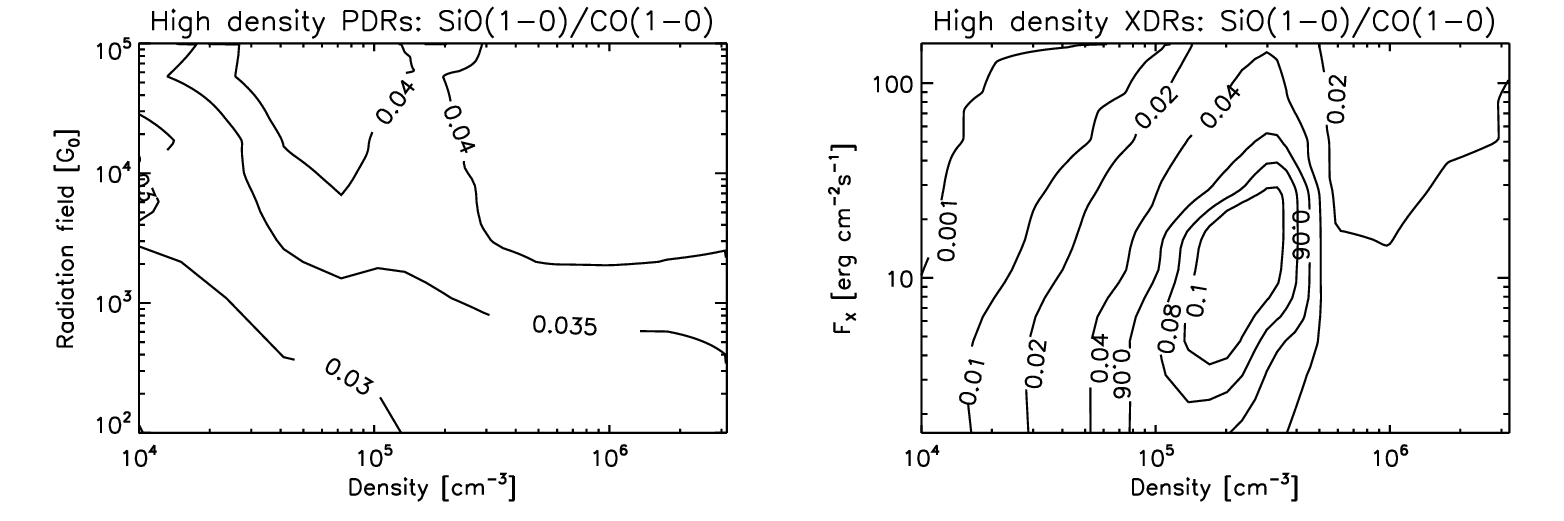}}
\centerline{\includegraphics[height=50mm,clip=]{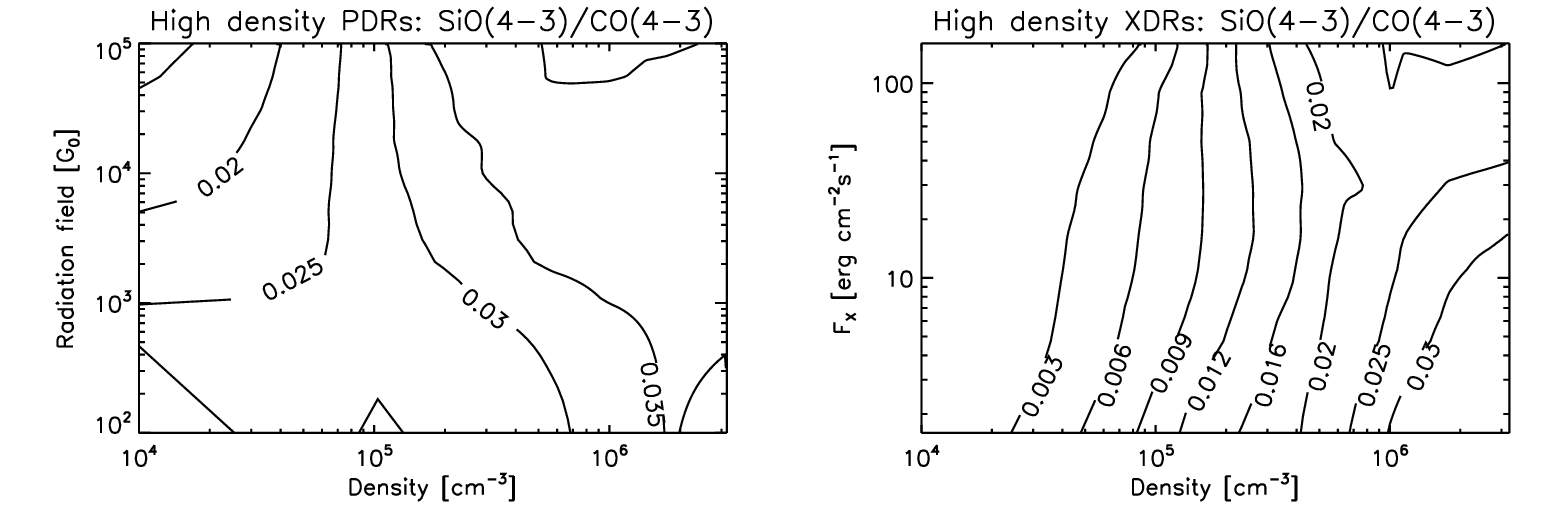}}
\caption{SiO(1-0)/CO and SiO(4-3)/CO(4-3) ratios for PDR (left) and XDR (right) models.}
\label{ratio_SiO_CO}
\end{figure*}

\begin{figure*}[!ht]
\centerline{\includegraphics[height=50mm,clip=]{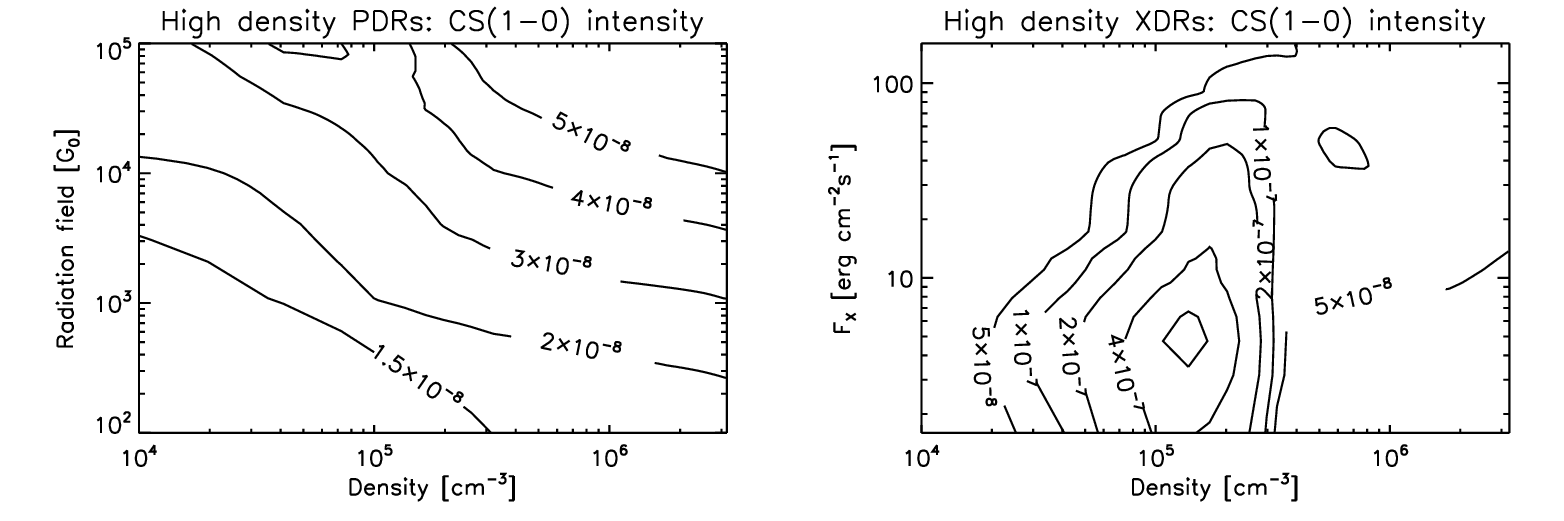}}
\centerline{\includegraphics[height=50mm,clip=]{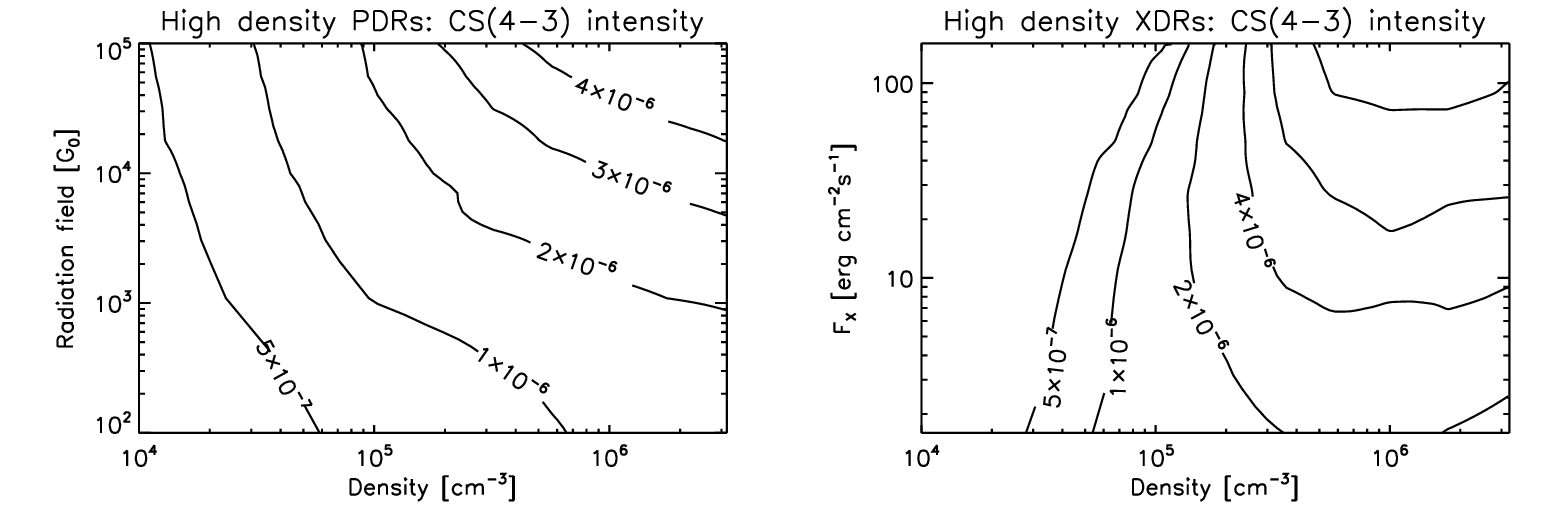}}
\caption{SiO(1-0) and SiO(4-3) intensity for PDR (left) and XDR (right) models.}
\label{intensity_CS}
\end{figure*}

\begin{figure*}[!ht]
\centerline{\includegraphics[height=50mm,clip=]{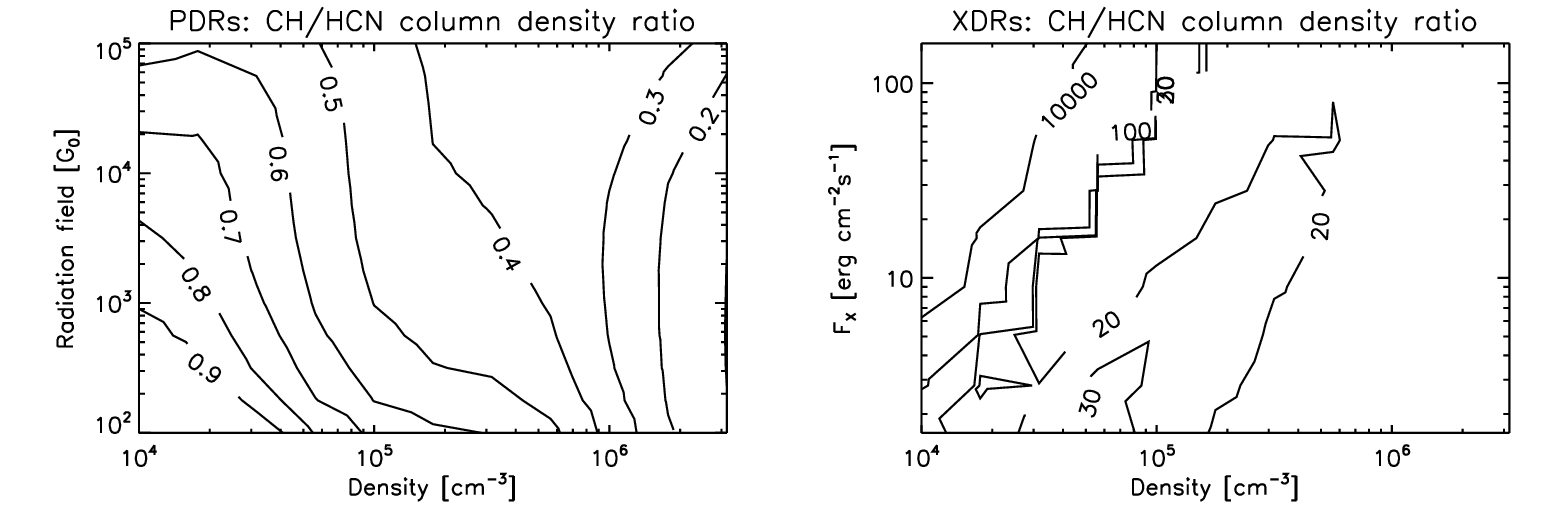}}
\caption{CH/HCN column density ratios for PDR (left) and XDR (right) models.}
\label{column_ratio_CH_HCN}
\end{figure*}


\clearpage

\begin{figure*}[!ht]
\centerline{\includegraphics[height=50mm,clip=]{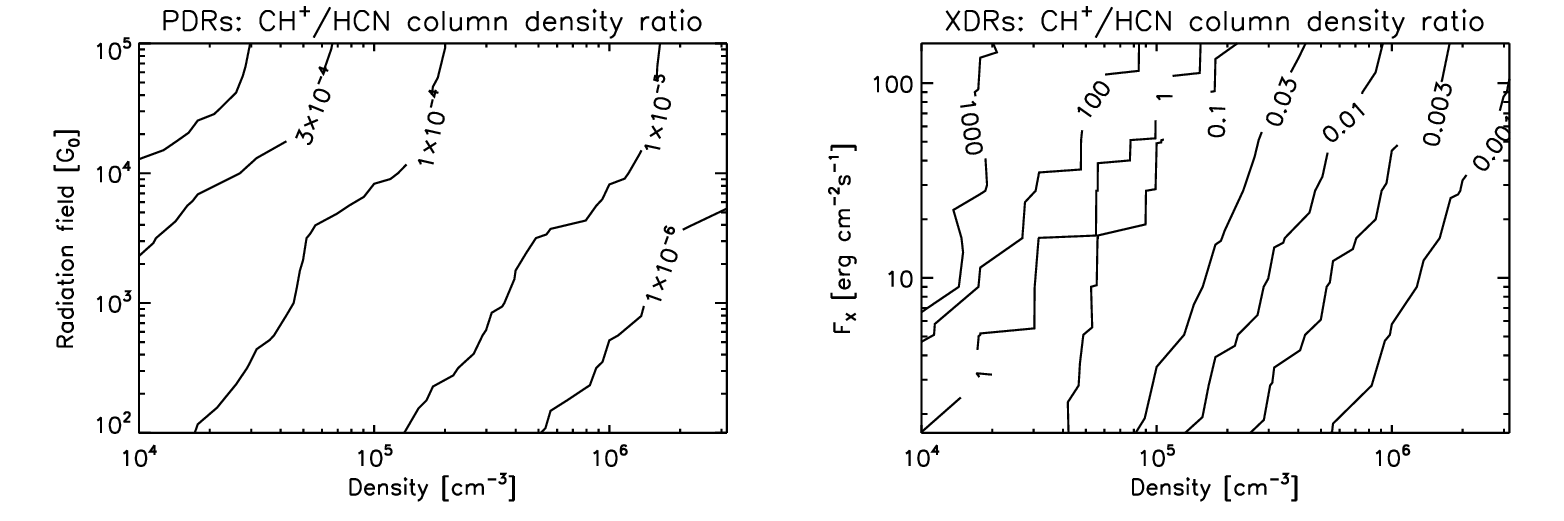}}
\caption{CH$^+$/HCN column density ratios for PDR (left) and XDR (right) models.}
\label{column_ratio_CHp_HCN}
\end{figure*}


\begin{figure*}[!ht]
\centerline{\includegraphics[height=50mm,clip=]{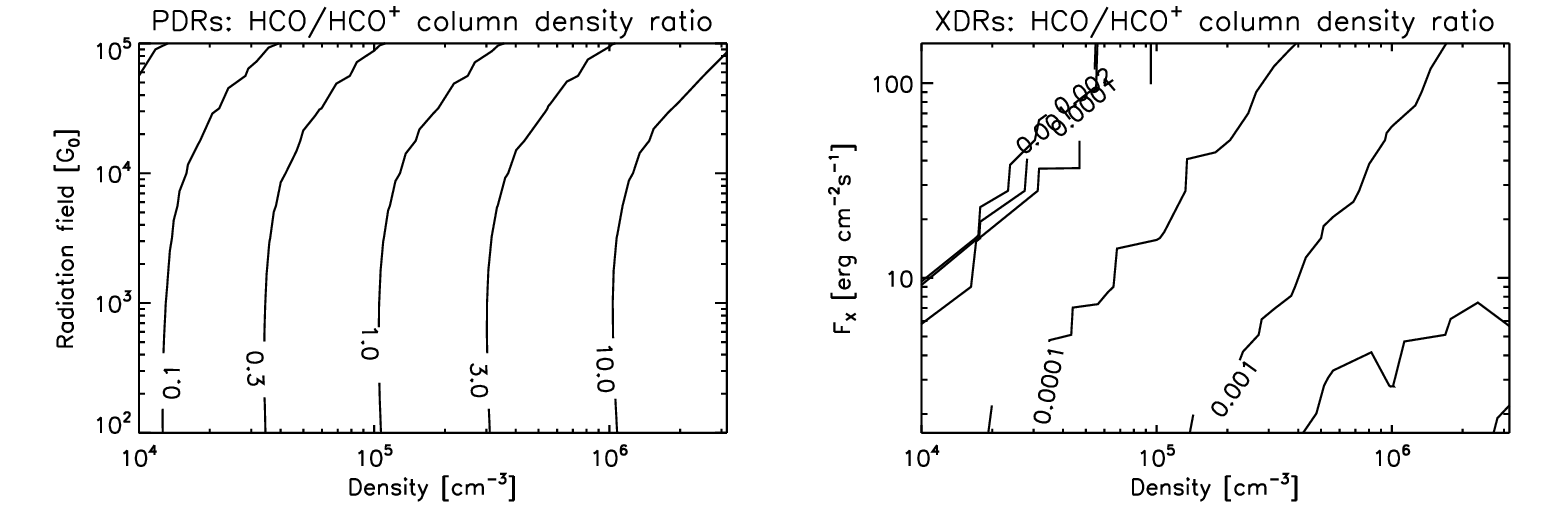}}
\caption{HOC/HCO$^+$ column density ratios for PDR (left) and XDR (right) models.}
\label{column_ratio_HCO_HCOp}
\end{figure*}


\begin{figure*}[!ht]
\centerline{\includegraphics[height=50mm,clip=]{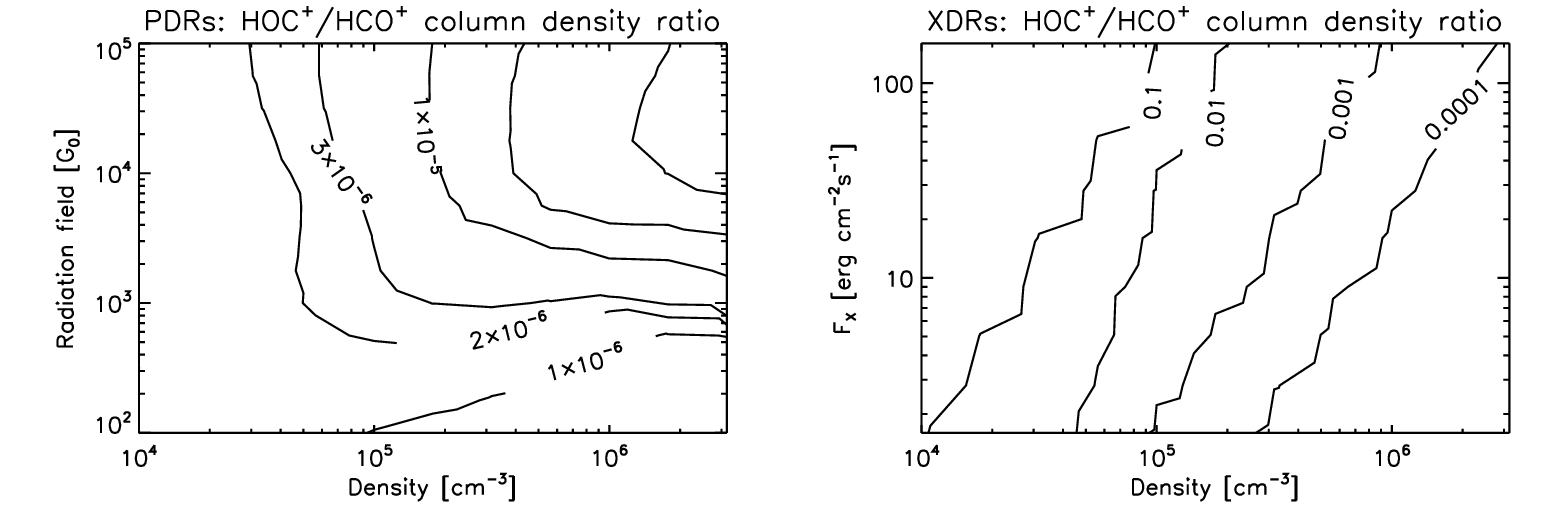}}
\caption{HOC$^+$/HCO$^+$ column density ratios for PDR (left) and XDR (right) models.}
\label{column_ratio_HOCp_HCOp}
\end{figure*}


\begin{figure*}[!ht]
\centerline{\includegraphics[height=50mm,clip=]{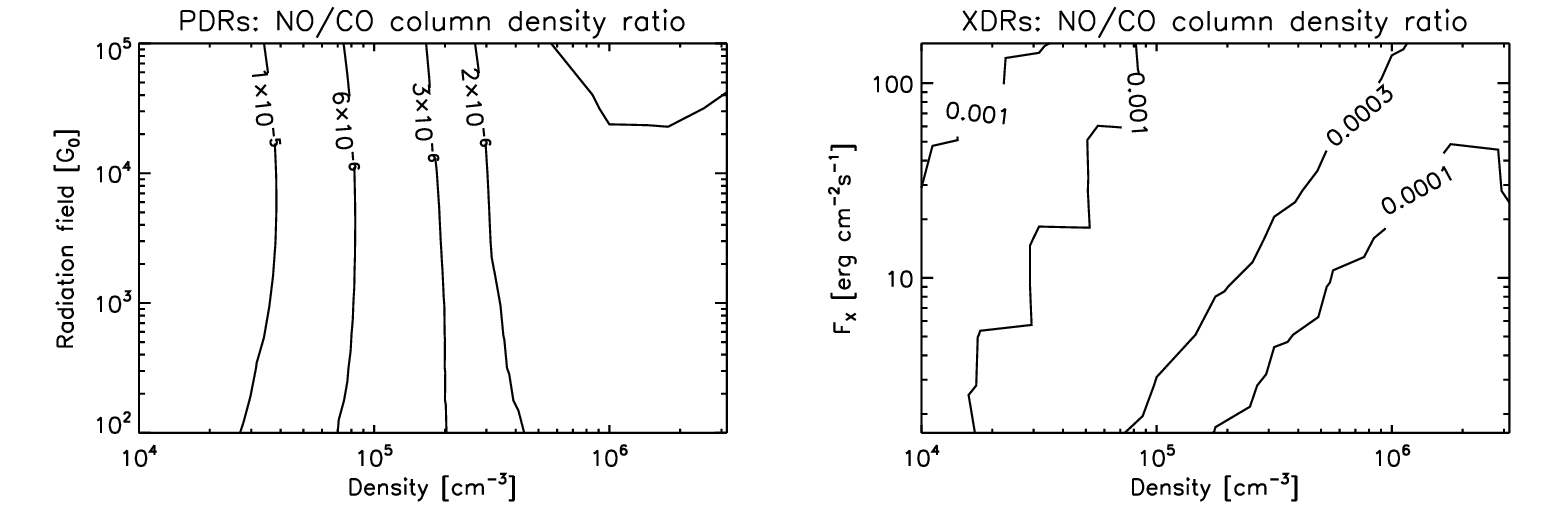}}
\caption{NO/CO column density ratios for PDR (left) and XDR (right) models.}
\label{column_ratio_NO_CO}
\end{figure*}

\clearpage


\begin{figure*}[!ht]
\centerline{\includegraphics[height=50mm,clip=]{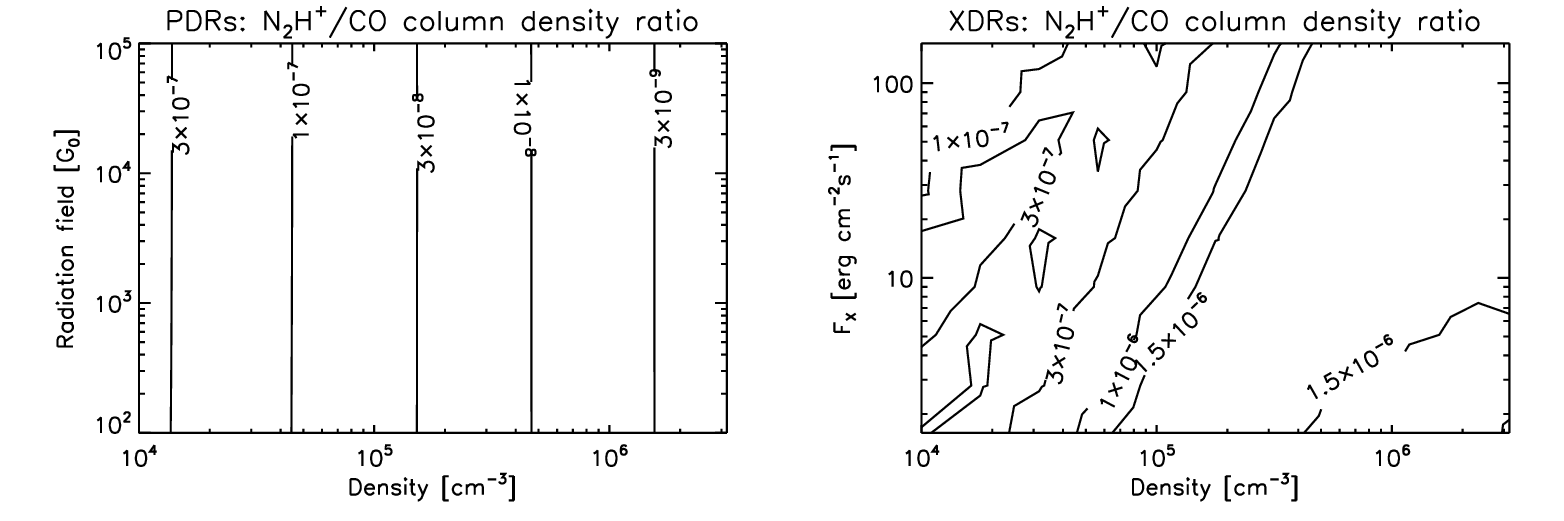}}
\caption{N$_2$H$^+$/CO column density ratios for PDR (left) and XDR (right) models.}
\label{column_ratio_N2Hp_CO}
\end{figure*}


\end{appendix}

\end{document}